\documentclass[aps,prx,twocolumn,superscriptaddress,nofootinbib,nolongbibliography,nobalancelastpage,10pt,floatfix]{revtex4-2}
\usepackage{standalone}
\usepackage{xcolor,amsthm,amsmath,amssymb,amsxtra,amsfonts,dsfont,graphicx,bm,bbm,bbold,braket,natbib,booktabs,mathtools}
\usepackage[colorlinks,citecolor=blue,linkcolor=blue,urlcolor=blue]{hyperref}
\usepackage[capitalize]{cleveref}
\crefname{appendix}{App.}{App.}

\usepackage[normalem]{ulem}
\usepackage[utf8]{inputenc}
\usepackage{microtype}
\usepackage{comment}
\usepackage{multirow}

\usepackage[percent]{overpic}
\usepackage{placeins}

\usepackage[caption=false]{subfig}
\def\Id{{\openone}}

\newcommand*\dd{\mathop{}\!\mathrm{d}}
\DeclareMathOperator{\tr}{tr}

\newif\ifshowcomments
    \showcommentstrue 

\usepackage{tikz}
\usetikzlibrary{quantikz2,backgrounds,fit,positioning,shapes.geometric,calc,decorations.pathreplacing}

\DeclareRobustCommand{\criticize}[3]{%
  \ifmmode
    \text{\textcolor{%
      \ifnum#2=5 red\else\ifnum#2=4 orange\else\ifnum#2=3 yellow!80!red\else\ifnum#2=2 yellow!60!black\else\ifnum#2=0 green!60!black\else gray\fi\fi\fi\fi\fi
      !%
      \ifnum#3<3 30\else\ifnum#3<5 70\else 100\fi\fi
      !black
    }{[#1 (S#2/C#3)]}}%
  \else
    \textcolor{%
      \ifnum#2=5 red\else\ifnum#2=4 orange\else\ifnum#2=3 yellow!80!red\else\ifnum#2=2 yellow!60!black\else\ifnum#2=0 green!60!black\else gray\fi\fi\fi\fi\fi
      !%
      \ifnum#3<3 30\else\ifnum#3<5 70\else 100\fi\fi
      !black
    }{[#1 (S#2/C#3)]}%
  \fi
}

\pdfoutput=1

\begin{document}

\title{Benchmark of quantum algorithms for ground state preparation in the presence of noise}

\author{Daniel Molpeceres}
\email{daniel.molpeceres@tum.de}
\affiliation{Munich Center for Quantum Science and Technology (MCQST), Schellingstrasse 4, D-80799 M\"unchen, Germany}
\affiliation{Technical University of Munich, TUM School of Natural Sciences, Physics Department, James-Franck-Str. 1, D-85748 Garching, Germany}

\author{Sirui Lu}
\affiliation{Munich Center for Quantum Science and Technology (MCQST), Schellingstrasse 4, D-80799 M\"unchen, Germany}
\affiliation{Max-Planck-Institut f\"ur Quantenoptik, Hans-Kopfermann-Strasse 1, D-85748 Garching, Germany}

\author{J. Ignacio Cirac}
\affiliation{Munich Center for Quantum Science and Technology (MCQST), Schellingstrasse 4, D-80799 M\"unchen, Germany}
\affiliation{Max-Planck-Institut f\"ur Quantenoptik, Hans-Kopfermann-Strasse 1, D-85748 Garching, Germany}

\author{Barbara Kraus}
\affiliation{Munich Center for Quantum Science and Technology (MCQST), Schellingstrasse 4, D-80799 M\"unchen, Germany}
\affiliation{Technical University of Munich, TUM School of Natural Sciences, Physics Department, James-Franck-Str. 1, D-85748 Garching, Germany}

\date{\today}

\begin{abstract}
    We compare the performance of representative cooling, adiabatic, and optimization algorithms for ground-state preparation in the presence of noise. Using an exactly solvable family of quadratic fermionic Hamiltonians subject to depolarizing noise, we derive the scaling of the achievable relative energy as a function of the noise rate and support these results with numerical simulations. The Hamiltonian exhibits two phases, separated by a quantum phase transition. As expected, the performance of the different algorithms depends on the phase: adiabatic evolution is favorable in the trivial phase, while a multi-frequency cooling algorithm, as proposed in~\cite{Molpeceres2025Quantum}, becomes competitive or superior in the topological phase, where gap-closing limits adiabatic protocols. We further present numerical results for the quantum approximate optimization algorithm~\cite{Farhi2014Quantum}, showing that it performs competitively with cooling in the trivial phase but is typically outperformed in the topological regime. Finally, we show that for this model the cooling protocol exhibits enhanced robustness to parameter imperfections, highlighting its potential advantage for realistic implementations of noisy quantum state preparation.
    The analytical approach developed here, in conjunction with numerical validation, establishes an extendable approach to benchmarking ground-state preparation algorithms.
\end{abstract}

\maketitle


\section{Introduction}
\label{sec:intro}

Preparing ground states of many-body Hamiltonians is a central task in quantum simulation and computation.
Various strategies have been developed, including adiabatic evolution~\cite{Albash2018Adiabatic}, variational algorithms, such as the Quantum Approximate Optimization Algorithm (QAOA)~\cite{Farhi2014Quantum,Zhou2020Quantum,Cerezo2021Variational}, and dissipative or cooling protocols~\cite{Kraus2008Preparation,Verstraete2009Quantum,Diehl2008Quantum,Terhal2000Problem,Molpeceres2025Quantum,Lloyd2025Quasiparticle,Lloyd2025Quantum,Zhan2026Rapid,Ding2025EndtoEnd,Mi2024Stable,Hahn2025Efficient,Langbehn2025Universal,Ramon-Escandell2025Thermal,Wang2026Lindblad,Motlagh2024Ground,Chen2025Efficient,Schlomer2026Modelagnostic}.
In current experimental settings, all these approaches are limited by the presence of noise, including decoherence and imperfections, which degrades the quality of the prepared states and imposes limits on the reachable energies. Thus, the following questions naturally arise: how does the achievable energy scale with the noise rate, and which strategy performs best for given hardware constraints?

The adiabatic algorithm~\cite{Albash2018Adiabatic} prepares the target ground state by slowly evolving the system from the ground state of a simple initial Hamiltonian to the target Hamiltonian, relying on the adiabatic theorem to keep the state close to the instantaneous ground state throughout the evolution. Its performance is often limited by the minimum spectral gap along the interpolation path, which closes at quantum phase transitions and demands impractically long runtimes~\cite{Zurek2005Dynamics,Dziarmaga2005Dynamics}.
Variational algorithms such as the Quantum Approximate Optimization Algorithm (QAOA)~\cite{Cerezo2021Variational,Farhi2014Quantum} circumvent some of those issues by optimizing the parameters of a fixed-depth quantum circuit in order to minimize the energy expectation value. Their scalability is often limited by barren plateaus at large circuit depth~\cite{Cerezo2021Variational,McClean2018Barren}, or by the expressivity of the chosen family of gates.
Cooling algorithms take a qualitatively different approach: they engineer a coupling between the system and a cold auxiliary bath and extract energy through repeated cycles of joint evolution and bath reset~\cite{Terhal2000Problem}.
Closely related are dissipative state preparation (DSP) protocols~\cite{Kraus2008Preparation,Verstraete2009Quantum,Diehl2008Quantum,Poyatos1996Quantum}, which engineer the system--bath coupling so that the dissipative dynamics drives the system to a desired target state. Both cooling and dissipative algorithms are described in terms of a channel which is repeatedly applied. Their performance is limited, among other factors, by the spectral gap of that channel.

The last two algorithms mentioned above have the advantage that the state of interest is prepared as the stationary state of the evolution (i.e. the fixed point of a linear map). Several complementary approaches have been put forward in that context. For instance, dissipative state preparation (DSP) schemes have been proposed to prepare ground states of frustration-free Hamiltonians~\cite{Verstraete2009Quantum,Kraus2008Preparation}.
Ding et al. and Zhan et al.~\cite{Ding2024Singleancilla,Zhan2026Rapid} proposed a filtered Lindbladian approach to prepare the ground or the thermal state of arbitrary (local) Hamiltonians (see also~\cite{Chen2025Efficient}).
Lloyd et al.~\cite{Lloyd2025Quasiparticle,Lloyd2025Quantum} introduced both a cooling scheme that uses time-dependent system--bath couplings and a quasiparticle cooling framework underlying a recent quantum processor experiment~\cite{Mi2024Stable}. In~\cite{Molpeceres2025Quantum} a cooling protocol which uses static couplings that are instantaneously switched on and off, combined with randomized cycle times and multiple bath frequencies was introduced.

In this work, we analyze how noise affects some of the ground state preparation algorithms.
We compare various algorithms in the presence of noise, with the goal of identifying the fundamental mechanisms that limit their performance and determining their optimal behavior. Our approach combines analytical derivations, where we obtain the achievable energy as a function of the noise rate $\kappa$, with numerical optimization of the control parameters for each algorithm. Prior work has studied noise effects on individual algorithms, including the robustness of adiabatic computation to decoherence~\cite{Childs2001Robustness,Sarovar2013Error,Schiffer2024Quantum} and noise-induced errors in quantum simulation. However, those works do not directly compare the performance of different algorithms.

We compare the plain adiabatic algorithm, QAOA and the cooling protocol introduced in~\cite{Molpeceres2025Quantum}. As in that work, we choose a simple Hamiltonian that contains two different phases, one trivial and the other one topological, connected by a quantum phase transition \cite{Pfeuty1970The,Greiter2014The,Mbeng2024Quantum}. This allows us to derive analytical expressions and to perform numerical simulations in sufficiently large system sizes. The initial state of all the protocols is a trivial one (product state, or a more realistic state close to the trivial one, see below) and we choose as a figure of merit the relative energy, which, as opposed to the global fidelity, is an intensive quantity which is well suited in the context of quantum simulation.

We divide our analysis in two parts. The first one compares analytically the adiabatic algorithm with the cooling protocol in the weak coupling regime, where one can derive an effective master equation describing the dynamics. In that case, we optimize the evolution time in both methods to reach the lowest energy. As expected, the relative performance of the methods depends on the phase of the Hamiltonian. We show that both, the energy scaling with the noise rate $\kappa$ and the prefactor in front of it are relevant for analyzing each method.

In the second part we compare two algorithms in which several control parameters are optimized: QAOA and the optimizable cooling algorithm of~\cite{Molpeceres2025Quantum}. Note that, thanks to the variational parameters, both outperform the adiabatic algorithm and the multifrequency cooling algorithm explained previously. Thus, we only concentrate on the optimized QAOA and cooling algorithm. In order to enable a fairer comparison, we optimize in both the same number of parameters. We find that both algorithms perform comparably in the trivial phase but cooling outperforms QAOA in the topological regime. Additionally, we observe that the cooling algorithm is more resilient to parameter imperfections. We emphasize that this paper only compares the final performance of the protocols, since our aim is to understand the fundamental limits of the different algorithms.

The remainder of the paper is organized as follows.
\Cref{sec:summary} provides an overview of the main results and their physical interpretation.
\Cref{sec:model} defines the model, the bath, and the noise channel.
\Cref{sec:algs} describes the three algorithms we consider here.
\Cref{sec:analytics_cooling_vs_adiabatic} derives the analytic scaling laws with noise strength for cooling and adiabatic evolution and compares them across the phase diagram.
\Cref{sec:numerics_qaoa_vs_cooling} presents a numerical comparison of QAOA and optimized cooling, including an analysis of robustness to parameter imperfections.
\Cref{sec:conclusion} summarizes our conclusions and outlook.

\section{Summary of results}
\label{sec:summary}

In this section we present and discuss the main results derived in this work. We briefly introduce the model Hamiltonian, the noise model, and the figure of merit, as well as the quantum algorithms we are going to analyze. We then consider the weak coupling regime and compare the cooling and the adiabatic protocols. Then, we relax the condition of weak coupling and present the results for QAOA and the cooling methods.

\subsection{Model}

The goal is to prepare the ground state of a family of translationally invariant quadratic fermionic Hamiltonians, $H_S(\theta)$, parametrized by $\theta\in [0,\pi/2]$. This is the same textbook model as the one used in Ref.~\cite{Molpeceres2025Quantum}. Due to translational invariance, $H_S(\theta)$ can be analytically block-diagonalized via a Fourier transformation, leading to a direct sum of Hamiltonians acting only on the mode pairs $(k,-k)$ \cite{Mbeng2024Quantum}. Each of these conserves the fermionic parity, which allows us to use a tensor product formulation of the fermionic states. In particular, we will call $\ket{0}^{\otimes N}$ the vacuum state.
For $\theta=\pi/2$, the Hamiltonian is local and the ground state is the vacuum. For $\theta=0$ the Hamiltonian is at the topological point of the Majorana chain with zero correlation length \cite{Kitaev2001Unpaired,Ortiz2014Manybody}. The model possesses a phase transition at $\theta =\pi/4$ that separates the trivial ($\theta >\pi/4$) and the topological ($\theta<\pi/4$) phases. At the critical point, the gap for the mode with $k=N/2$ closes.

As a figure of merit for the different algorithms, we consider the relative energy
\begin{equation}
    e=\left|\frac{E-E_{\rm GS}}{E_{\rm GS}}\right|,
    \qquad \mbox{ with }
    E=\tr(H_S\rho)
    \label{eq:e_definition}
\end{equation}
where $E_{\rm GS}$ denotes the ground-state energy of $H_S$ and $\rho$ the output state of the corresponding algorithm.

As in~\cite{Molpeceres2025Quantum}, the noise is modeled through a depolarizing master equation given in Sec.~\ref{subsec:noise}, characterized by a noise rate $\kappa$. Note that it commutes with the dynamics generated by the Hamiltonian, which simplifies the analysis.

As mentioned in the introduction, we will consider three quantum algorithms: adiabatic, QAOA and cooling. There are many different ways of performing each of those algorithms, and it is out of the scope of this work to analyze every one of them. This is why we take some of the standard approaches used in the adiabatic and QAOA algorithms. In the first one, we start with the Hamiltonian at $H(\pi/2)$ and transform it linearly with time to $H(\theta)$. In the QAOA, we consider pairs of layers that contain the evolution dictated by $H(\pi/2)$ and $H(0)$, and optimize with respect to the evolution times in each layer independently. Regarding the cooling algorithm, we select the one considered in our previous work~\cite{Molpeceres2025Quantum} with a slight modification. Despite the fact that we consider a specific cooling algorithm, we expect that other cooling and dissipative algorithms will share many of the features that we obtain in our analysis.

A key distinction between the algorithms is the role of the initial state. Whereas in a cooling protocol the state of interest is the stationary state, in the adiabatic and QAOA approaches the relevant state is obtained via a unitary evolution. Consequently, the initial state plays an important role in the latter two algorithms, while in a cooling process the final state is essentially independent of it. As for the initial state, one could simply consider the vacuum state. However, this is already the ground state for $\theta=\pi/2$, so that in that case both the adiabatic and the QAOA algorithm will give the perfect state (without performing any operation), whereas the cooling algorithm will give an imperfect one since it will have to reach a steady state. To enable a meaningful comparison in the presence of noise, it is crucial to adopt a common framework in which all algorithms are evaluated under the same noise model and their performance is optimized with respect to their respective control parameters. In particular, this requires considering realistic initial states that can be prepared under this noise, rather than idealized states such as $\ket{0}^{\otimes N}$, thereby ensuring that the comparison reflects experimentally accessible conditions. Due to that, we use as initial state $\rho^{\rm cool}(\pi/2)$, obtained from a simple cooling protocol at $\theta=\pi/2$. Although this does not change the scalings that we find, it may slightly influence the numerical results.

\subsection{Comparison between adiabatic and cooling algorithms}

To understand how the adiabatic and cooling algorithms compare in the presence of noise, it is useful to analyze both methods analytically. Our central analytical result is the derivation of scaling laws for the achievable relative energy as a function of the noise rate $\kappa$.
We are interested in the limit of low noise rate, and in the energy scaling with $\kappa$ in that limit. When comparing two algorithms, we will say that one has better scaling with $\kappa$ if the relative energy increases more slowly in terms of that parameter.
For instance, a scaling with $\kappa^{1/3}$ is better than one with $\kappa^{2/3}$, despite the fact that $\kappa\ll1$. As we show, this scaling depends on both the algorithm under consideration (adiabatic versus cooling) and on the phase being studied (trivial versus topological).

\begin{table}[ht]
    \caption{Summary of the optimal scaling laws for the relative energy $e_{\rm opt}(\kappa)$ for the cooling and adiabatic algorithms.
        Prefactors: $F=3(A/(gt))^{2/3}/C$ with $A,C$ defined in \cref{eq:e_cool_scaling};
        $\tilde{A} \approx \cos^{2} \theta_f/8$;
        $\tilde{B}$ depends on $\theta_f$ via the Landau--Zener coefficient~$G$ (Appendix~\ref{app:adiabatic_derivations}), and
        $D = 9(gt)^{-4/3}/4$ (see Appendix~\ref{subapp:few-frequency_corrections_cooling}).}
    \label{tab:scaling_summary}
    \begin{ruledtabular}
        \small
        \begin{tabular}{lcc}
            Algorithm           & \shortstack{Trivial                                                                                \\($\theta_f>\pi/4$)} & \shortstack{Topological\\($\theta_f<\pi/4$)}                  \\
            \hline
            Cooling             & \multicolumn{2}{c}{$F\kappa^{1/3}$}                                                                                                                                                      \\
            Adiab.\ (ideal)     & $3\tilde{A}^{1/3}\kappa^{2/3}$                                                                                           & $\tfrac{3}{2^{1/3}}\tilde{B}^{2/3}\kappa^{1/3}$               \\
            Adiab.\ (realistic) & $(3\tilde{A}^{1/3}+D)\kappa^{2/3}$                                                                                       & $\tfrac{3}{2^{1/3}}\tilde{B}^{2/3}\kappa^{1/3}+D\kappa^{2/3}$ \\
        \end{tabular}
    \end{ruledtabular}
\end{table}

\paragraph{Cooling.}
We consider the multi--frequency, randomized-time cooling protocol developed in Ref.~\cite{Molpeceres2025Quantum} (see also \cref{subsec:cooling_alg} and Appendix~\ref{app:analytic_cooling}), where a bath is locally weakly coupled to the system with a coupling constant $g$ and extracts energy from the system iteratively.
We analyze the state after the concatenation of $\ell$ cooling cycles where in each cycle we choose a random $\Delta$ and $\tau$
from certain distributions. We consider a uniform distribution for $\tau\in[0,2t]$, whereas for $\Delta$ we take both continuous and discrete distributions in the interval $\Delta\in[\Delta_m,\Delta_M]$, which contains the whole single-particle spectrum of energies $\{\epsilon_m,...,\epsilon_M\}$. We further constrain our analysis to the weak-coupling ($(gt)^2 \ll 1$), cooling ($\epsilon_m t \gg 1$), and low-noise ($\kappa t \ll 1$) limits. Note that this requires $\kappa \ll \epsilon_m$.
As shown in Appendix~\ref{app:analytic_cooling}, in the limit of $\ell \rightarrow \infty$, the state obtained by such a cooling process is close
to the stationary state of an averaged map for any distribution of $\tau$ and $\Delta$.

Let us briefly comment on the choice of distributions of the bath frequencies considered here. 
As shown in~\cite{Molpeceres2025Quantum}, for a single bath frequency $\Delta$ (i.e., the distribution of the bath frequency is a delta function), the cooling rate as a function of the energy is described by a Lorentzian function proportional to $t$, centered at $\Delta$, and with a width proportional to $1/t$. Thus, only the modes with energies close to that frequency $\Delta$ will be effectively cooled. As $t$ increases, the energy of the corresponding mode becomes lower, but fewer and fewer modes are cooled. This is why it is necessary to choose many frequencies $\Delta$ in an interval that includes all mode energies. One possibility is to consider a continuous distribution $\Delta \in [\Delta_m, \Delta_M]$. This is analyzed in Appendix~\ref{app:analytic_cooling} and allows us to obtain analytical expressions for the cooling rate and the final energy. Alternatively, one can consider a distribution containing $R$ discrete frequencies, equispaced in $[\Delta_m, \Delta_M]$. This is the approach adopted in Ref.~\cite{Molpeceres2025Quantum}, and is also used here for the numerical computations\footnote{Note that in contrast to Ref.~\cite{Molpeceres2025Quantum}, we do not apply a long sequence of cooling maps with randomly chosen cycle times for each frequency, but instead vary both the cycle time and the bath frequency randomly.}. In order to cover the whole single-particle spectrum, we have to choose $R \propto t$. In the limit $R \to \infty$, we recover the continuum case, but the discrete distribution also allows us to tackle special cases where only a few, or even just one frequency is better. For the remainder of the paper, we will focus on this discrete distribution.

As the cooling rate increases with $t$, longer cycle times lead to lower attainable energies. However, in the presence of noise, longer $t$ introduces more noise into the system. Therefore, for a given noise rate, there exists an optimal time $t$ that minimizes the final energy. The results of this optimization are summarized in Table~\ref{tab:scaling_summary}. We obtain that the optimal energy achievable under depolarizing noise scales as a power law with the noise rate $\kappa$. These results are derived analytically in the limits described above. We first derive the relative energy and then optimize it under those constraints (see Table~\ref{tab:scaling_summary}). This result shows that for the cooling algorithm, the energy scales with the noise rate as $\kappa^{1/3}$ and does not depend on the quantum phase (see Table~\ref{tab:scaling_summary}). We also derive results on the achievable fidelity with this cooling process (see  Appendix~\ref{subapp:weak_coupling_map}).

\paragraph{Adiabatic evolution.}
As is common in adiabatic algorithms, we consider a linear interpolation between the initial (simple) and the target Hamiltonian (see \cref{subsec:adiabatic_alg}). As before, increasing the total evolution time increases the performance of the algorithm. However, the effect of noise also increases with time. This interplay will dictate the optimal time and energy, leading to the results presented in Table~\ref{tab:scaling_summary}. Additionally, the performance of the adiabatic algorithm will strongly depend on whether the adiabatic path crosses the critical point at $\theta=\pi/4$ or not. In case the target Hamiltonian still lies in the trivial phase (no crossing), the optimal time can be shown to scale like $\kappa^{-1/3}$ and the achievable relative energy to scale like $\kappa^{2/3}$ (see Table~\ref{tab:scaling_summary}). In the topological phase ($\theta < \pi/4$), the gap is closed during the adiabatic evolution. Using Landau--Zener theory and including noise, we show that the relative energy scales as in \cref{eq:e_adiab_ent}: we achieve for the optimal time a scaling proportional to $\kappa^{-2/3}$ and the achievable energy as $\kappa^{1/3}$ (see Table~\ref{tab:scaling_summary}).

It should be noted that scalings alone are not enough to determine the performance of the algorithm. The prefactors in front of them play an important role in the comparison between energy densities at finite values of $\kappa$. Taking them into account, it can be seen that the performance of the adiabatic algorithm in the selected range of $\kappa$ is better in the trivial phase than in the topological phase. That is, even though the rate at which relative energy grows with $\kappa$ is slower in the topological phase, the total relative energy is lower in the trivial phase. This can be understood from the absence of a level crossing in the trivial phase.

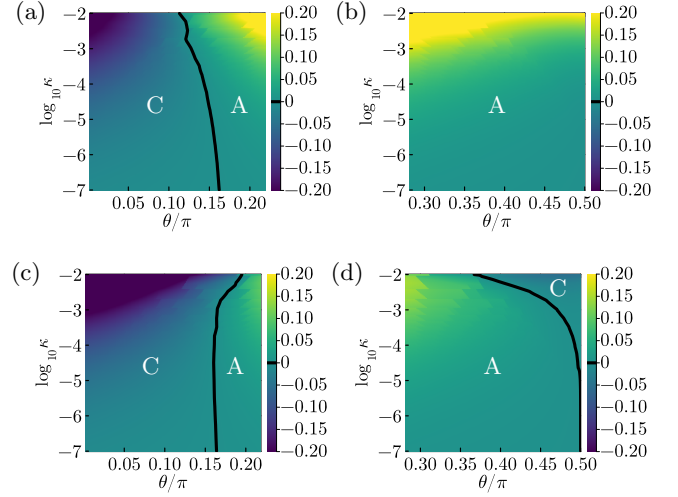
\begin{figure}
    \begin{minipage}{0.23\textwidth}
        \centering
        \begin{overpic}[width = 1\textwidth]{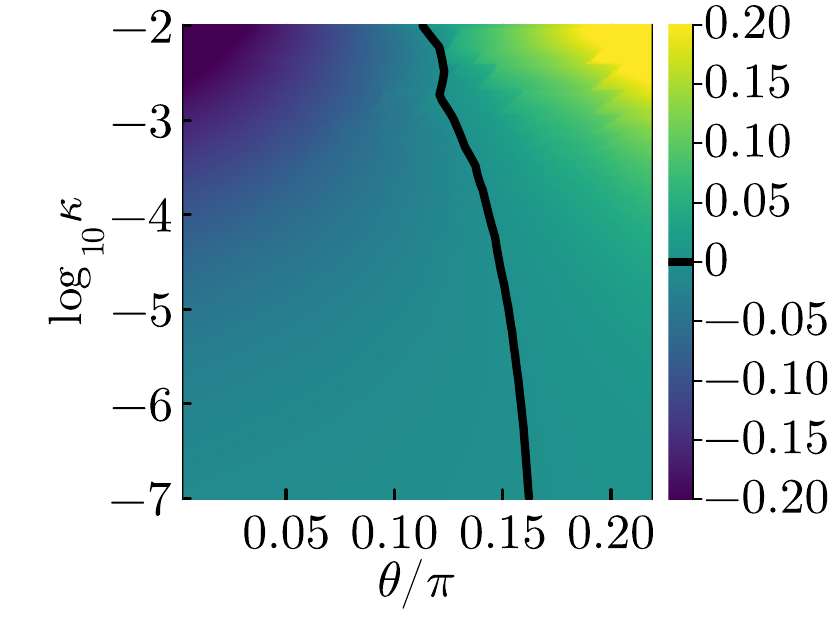}
            \put(-2,70){{(a)}}
            \put(40,40){{\color{white}{C}}}
            \put(67,40){{\color{white}{A}}}
        \end{overpic}
        \label{fig:cool_adiab_ent}
    \end{minipage}
    \begin{minipage}{0.23\textwidth}
        \centering
        \begin{overpic}[width = 1\textwidth]{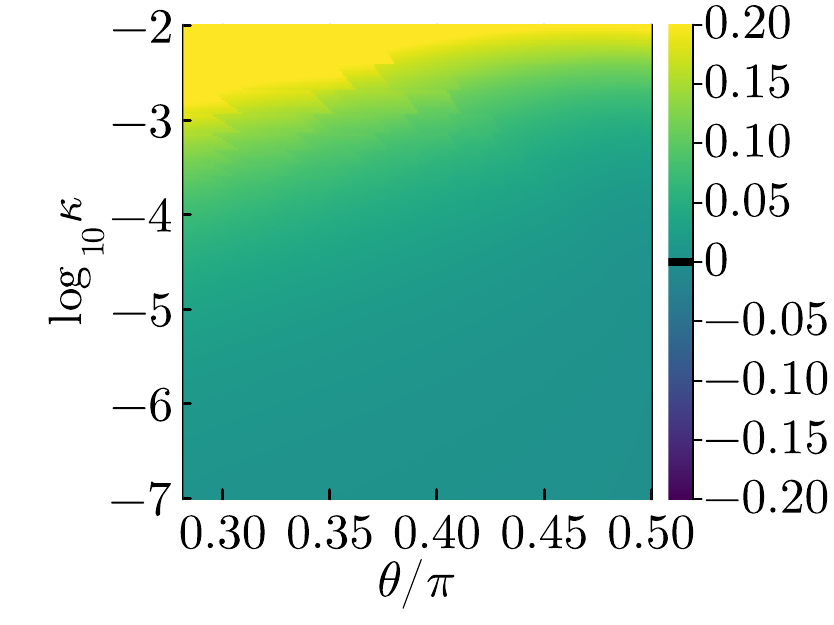}
            \put(-2,70){{(b)}}
            \put(47,40){{\color{white}{A}}}
        \end{overpic}
        \label{fig:cool_adiab_prod}
    \end{minipage}
    \begin{minipage}{0.23\textwidth}
        \centering
        \begin{overpic}[width = 1\textwidth]{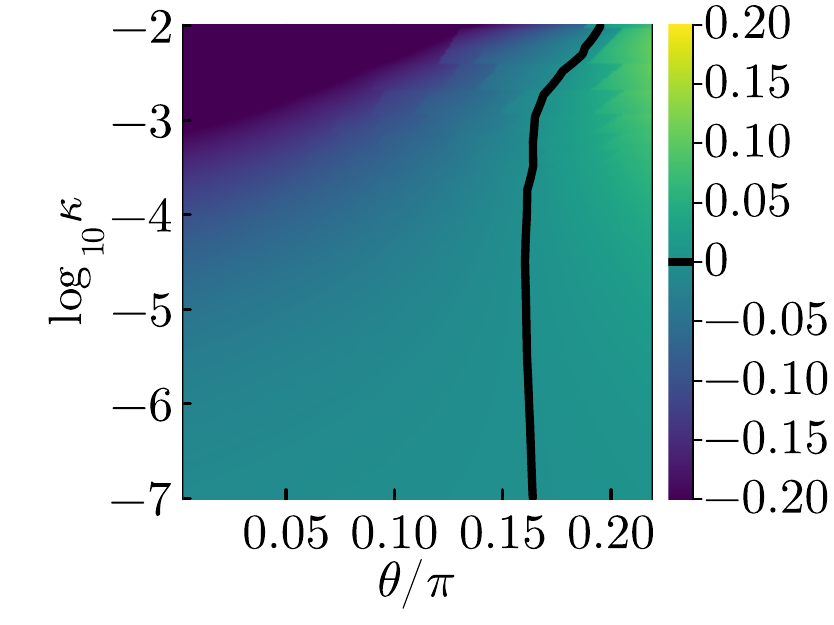}
            \put(-2,70){{(c)}}
            \put(40,40){{\color{white}{C}}}
            \put(67,40){{\color{white}{A}}}
        \end{overpic}
        \label{fig:cool_realadiab_ent}
    \end{minipage}
    \begin{minipage}{0.23\textwidth}
        \centering
        \begin{overpic}[width = 1\textwidth]{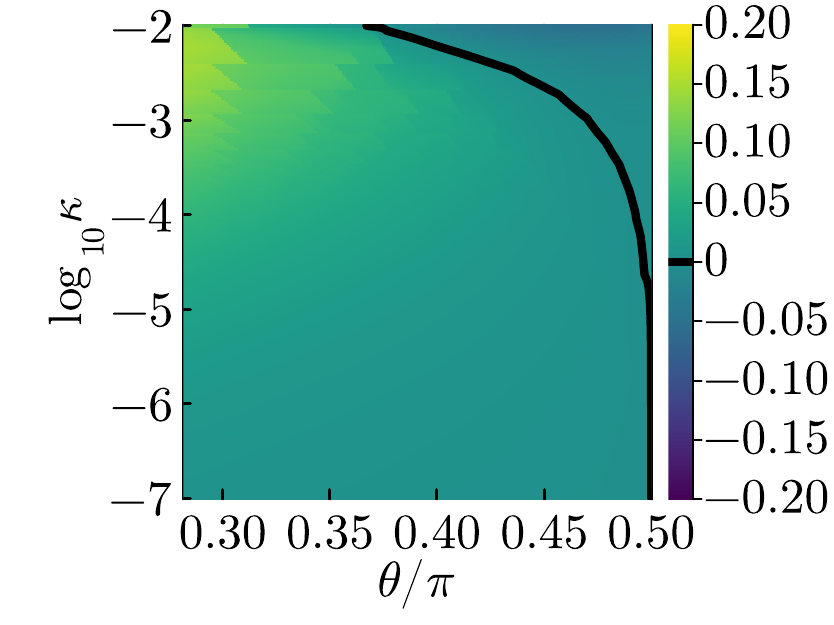}
            \put(-2,70){{(d)}}
            \put(47,40){{\color{white}{A}}}
            \put(68,65){{\color{white}{C}}}
        \end{overpic}
        \label{fig:cool_realadiab_prod}
    \end{minipage}
    \vspace{-2em}
    \caption{Difference $e^{\rm cool}_{\rm opt}-e^{\rm adiab}_{\rm opt}$ between the optimal relative energies achieved by the cooling and the adiabatic algorithms, plotted as a function of $\theta/\pi$ (horizontal axis) and $\log_{10}\kappa$ (vertical axis) for $N=100$ systems.
        Negative values (blue/dark) indicate that cooling gives lower energy (region labeled ``C''); positive values (yellow/bright) indicate that the adiabatic algorithm gives lower energy (region labeled ``A''). The black contour marks $e_{\rm cool}^{\rm opt}=e_{\rm adiab}^{\rm opt}$.
        (a)~Topological phase, adiabatic algorithm starting from $\ket{0}^{\otimes N}$ (ideal): cooling gives lower energy throughout most of this phase.
        (b)~Trivial phase, ideal adiabatic algorithm: the adiabatic algorithm gives lower energy everywhere due to its small non-adiabatic prefactor and the $\kappa^{2/3}$ power law in this gapped regime.
        (c)~Topological phase, realistic adiabatic algorithm starting from $\rho^{\rm cool}(\pi/2)$: the region where cooling yields lower energy expands.
        (d)~Trivial phase, realistic adiabatic: the imperfect initial state reduces the adiabatic advantage, and cooling yields lower energy for $\kappa\gtrsim 10^{-3}$.
        For cooling the relative energies are computed numerically at $N=100$, whereas for relative energies achieved by the adiabatic algorithm we use the large-$N$ analytic formulas to suppress the finite-size effect. The deviations at high $\kappa$ are due to the fact that we enter a regime where the here chosen discretization of $R$ matters (see App. \ref{subapp:few-frequency_corrections_cooling}).}
    \label{fig:cooling_vs_adiabatic_heatmaps}
\end{figure}

\paragraph{Numerical comparison between adiabatic and cooling algorithms.}
To complement the analytical results, we numerically optimize the runtime of the algorithms. This allows us not only to confirm the analytical predictions, but also to assess their performance beyond the assumptions of the analytical treatment. In \cref{fig:cooling_vs_adiabatic_heatmaps} we present the difference between the numerically optimized (over the runtime) relative energies achieved with the adiabatic and the cooling algorithms as a function of the noise rate $\kappa$ and the Hamiltonian parameter $\theta$. We perform this comparison for the initial state $\ket{0}^{\otimes N}$ and the more realistic initial state $\rho^{\rm cool}(\pi/2)$. As can be clearly seen, the adiabatic algorithm outperforms the cooling algorithm in the trivial phase. This is expected, as for $\theta=\pi/2$ the ground state coincides with the initial state (or is close to it in the realistic scenario). However, cooling is more effective in the topological phase due to the gapless transition in the adiabatic path.

\subsection{Comparison between numerically optimized algorithms}

In addition to cooling and adiabatic evolution, we consider QAOA as a third approach to ground-state preparation. To include it on equal footing in our comparison, we compare it to a numerically optimized version of our cooling algorithm (see \cite{Molpeceres2025Quantum}). Its performance is evaluated under the same noise model and both are optimized over the same number of parameters, ensuring a consistent and fair comparison. This allows us to assess how a finite-depth variational protocol competes with cooling across different phases.

\paragraph{QAOA.}
As is common in QAOA, we consider $p$ layers, each consisting of two unitaries generated by non-commuting Hamiltonians $H_M$ and $H_P$, which are both translationally invariant and quadratic. Consequently, the unitaries can again be decomposed into $2 \times 2$ unitaries acting on each mode pair $(k,-k)$. Each unitary is then determined by a single parameter.

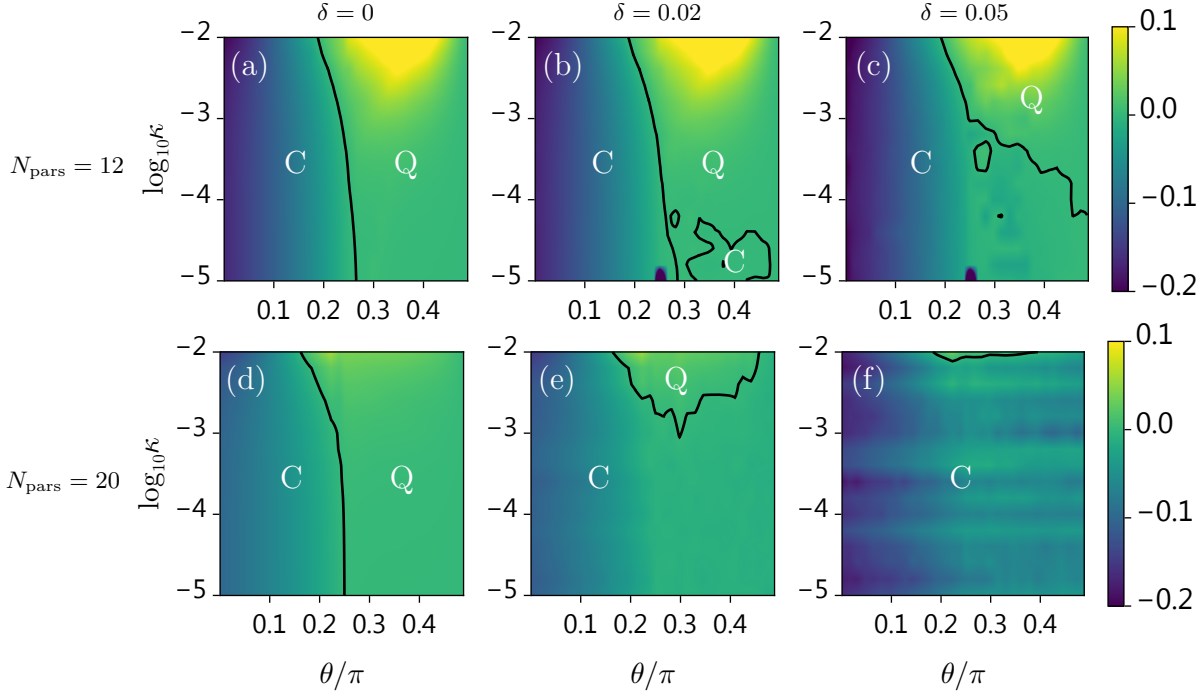
\begin{figure*}
    \begin{minipage}{0.9\textwidth}
        \centering
        \begin{overpic}[width=0.9\textwidth]{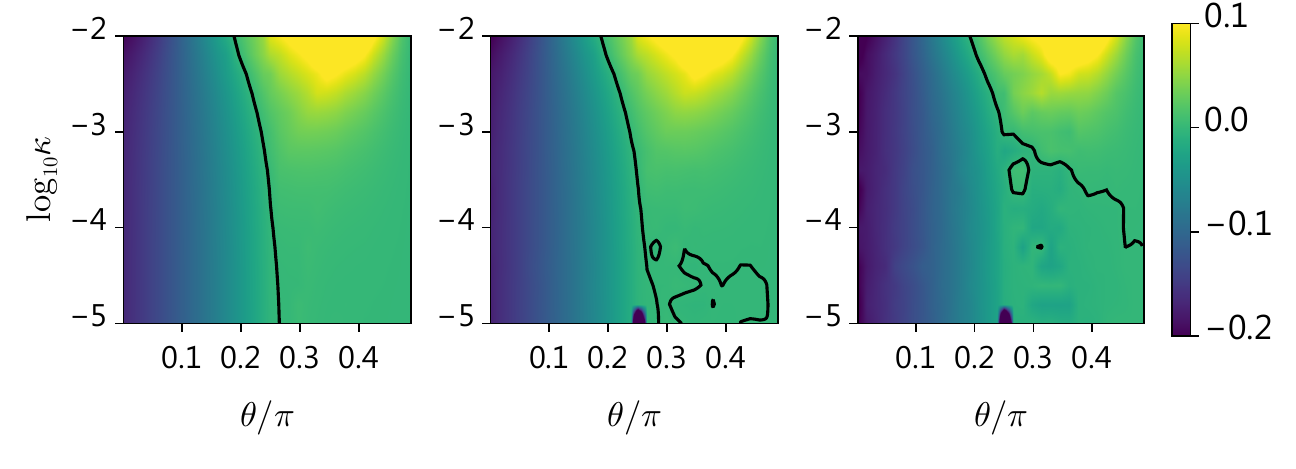}
            \put(-10,21){$N_{\rm{pars}}=12$}
            \put(18,35){$\delta=0$}
            \put(45,35){$\delta=0.02$}
            \put(73,35){$\delta=0.05$}
            \put(10,30){\color{white}\large{(a)}\color{black}}
            \put(38.5,30){\color{white}\large{(b)}\color{black}}
            \put(67,30){\color{white}\large{(c)}\color{black}}
            \put(15,21){\color{white}\large{C}\color{black}}
            \put(25,21){\color{white}\large{Q}\color{black}}
            \put(43,21){\color{white}\large{C}\color{black}}
            \put(53,21){\color{white}\large{Q}\color{black}}
            \put(55,12){\color{white}\large{C}\color{black}}
            \put(72,21){\color{white}\large{C}\color{black}}
            \put(82,27){\color{white}\large{Q}\color{black}}
        \end{overpic}
        \label{fig:12param_cool_vs_qaoa}
    \end{minipage}
    \begin{minipage}{0.9\textwidth}
        \vspace{-1.10cm}
        \centering
        \begin{overpic}[width=0.9\textwidth]{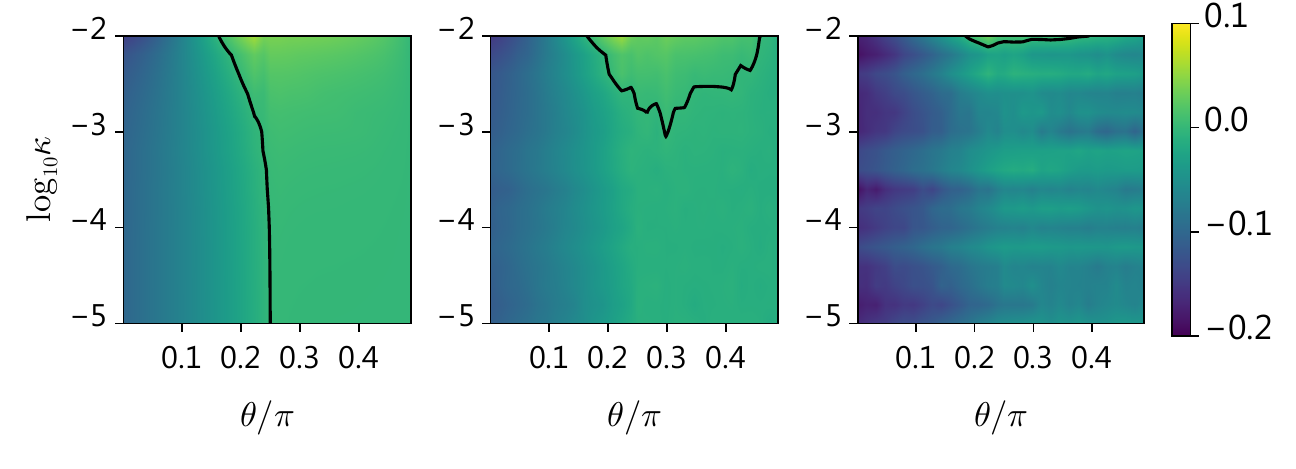}
            \put(-10,21){$N_{\rm{pars}}=20$}
            \put(10,30){\color{white}\large{(d)}\color{black}}
            \put(38.5,30){\color{white}\large{(e)}\color{black}}
            \put(67,30){\color{white}\large{(f)}\color{black}}
            \put(15,21){\color{white}\large{C}\color{black}}
            \put(25,21){\color{white}\large{Q}\color{black}}
            \put(43,21){\color{white}\large{C}\color{black}}
            \put(50,30){\color{white}\large{Q}\color{black}}
            \put(76,21){\color{white}\large{C}\color{black}}
        \end{overpic}
        \label{fig:20param_cool_vs_qaoa}
    \end{minipage}
    \vspace{-2em}
    \caption{
        Difference $e_{\rm opt}^{\rm cool}-e_{\rm opt}^{\rm QAOA}$ as a function of $\theta/\pi$ (horizontal) and $\log_{10}\kappa$ (vertical) for $N=20$. Blue: cooling gives lower energy (labeled ``C''); yellow: QAOA gives lower energy (labeled ``Q''); black contour: equal performance. Top row ($N_{\rm pars}=12$): cooling with coupling range $r_{\rm c}=0.5$ and two cycles; QAOA with $p=6$ layers. Bottom row ($N_{\rm pars}=20$): $r_{\rm c}=1.5$, $p=10$. Columns show increasing deviation of the exact parameters, ~$\delta$: (a,\,d)~$\delta=0$ (exact parameters); (b,\,e)~$\delta=0.02$; (c,\,f)~$\delta=0.05$, each averaged over 100 random perturbations. Both algorithms start from $\rho^{\rm cool}(\pi/2)$. As $\delta$ increases, the region where cooling outperforms QAOA expands, demonstrating cooling's greater robustness to parameter imperfections. The ``islands'' seen in (b, c) are caused by fluctuations in the optimal parameters in a region where both methods are almost equally effective.
    }
    \label{Fig:QAOA_vs_cooling_robustness}
\end{figure*}

\paragraph{Comparison.}
First, it should be noted that in all regions considered here, the adiabatic algorithm performs worse than both the QAOA and the cooling algorithms. Therefore, we focus on comparing the latter two. In \cref{Fig:QAOA_vs_cooling_robustness}, we show the difference between QAOA and the cooling algorithm when both are optimized over 12 and 20 free parameters, respectively. For the cooling protocol, we choose a coupling range $r_{\rm c} \in\{0.5,1.5\}$ \cite{Molpeceres2025Quantum} and allow two different cycles (each characterized by its own bath frequency~$\Delta$, cycle time~$t$, and coupling parameters), giving a total of $8(r_{\rm c}+1) \in\{12,20\}$ free parameters.
For QAOA, we use $p=4(r_{\rm c}+1) \in \{6,10\}$ layers to have the same number of parameters. As mentioned above, we choose as the initial state the state $\rho^{\rm cool}(\pi/2)$. In case the parameters can be chosen exactly, QAOA typically achieves lower energies in the trivial phase ($\theta > \pi/4$), while cooling gives lower energies in the topological phase ($\theta < \pi/4$). This is expected, as preparing an SPT state from a trivial product state is generally difficult for QAOA. Neither algorithm dominates across the entire phase diagram.

\paragraph{Robustness.}
Since the optimal parameters cannot be implemented perfectly, we also analyze the robustness of the algorithms to small deviations in the parameters (see \cref{Fig:QAOA_vs_cooling_robustness}). As shown there, the cooling algorithm is more resilient to such fluctuations than the QAOA algorithm over a wide range of parameters.

\subsection{Overall picture}
Overall, our analytical and numerical results show that the optimal strategy is strongly phase-dependent: adiabatic evolution is advantageous in the trivial phase, whereas cooling is favorable in the topological phase. Our numerical results show that QAOA performs competitively with cooling in the trivial phase but is generally outperformed by cooling in the topological phase. We further find that cooling is more robust than QAOA to parameter imperfections, reinforcing its advantage in realistic settings.

\section{Model}
\label{sec:model}

As in~\cite{Molpeceres2025Quantum}, we consider a family of translationally invariant quadratic fermionic Hamiltonians $H_S(\theta)$ parametrized by~$\theta$. Both analytic calculations and large-$N$ numerics are feasible, and all dynamics decompose into independent momentum blocks. Below we define the Hamiltonian, the bath used for cooling, and the noise channel.

\subsection{System}
\label{subsec:system}

The system is a ring of $N$ free fermions (with $N$ even) with nearest-neighbor hopping and pairing, parametrized by $\theta\in[0,\pi/2]$~\cite{Molpeceres2025Quantum}:
\begin{align}
    H_S(\theta) &= \frac{1}{2} \sin\theta \sum_{n=1}^N (a_n^\dagger a_n - a_n a_n^\dagger) \nonumber\\
                &\quad + \frac{1}{2} \cos\theta \sum_{n=1}^N \left[a_n^\dagger (a_{n+1} + i a_{n+1}^\dagger) + \text{h.c.}\right],
    \label{eq:system_hamiltonian_realspace}
\end{align}
where $a_n$ ($a_n^\dag$) are fermionic annihilation (creation) operators with $\{a_n,a_m\}=\{a_n^\dagger,a_m^\dagger\}=0$, $\{a_n,a_m^\dagger\}=\delta_{n,m}$, and we impose periodic boundary conditions $a_{N+1}=a_1$.

As mentioned before, this Hamiltonian has two phases separated by a critical point at $\theta=\pi/4$. For $\theta\to\pi/2$, the ground state is continuously connected to the trivial product vacuum (in the original fermionic basis). For $\theta\to 0$, the model approaches the Kitaev/Majorana-chain limit (topological phase).
At $\theta=\pi/4$ the gap closes at momentum $k=N/2$ and the ground state becomes degenerate. This mode is negligible for the energy density in the thermodynamic limit, but it becomes relevant for comparisons based on fidelity.

Since $H_S$ is quadratic and translation invariant, a Fourier transform decomposes it into independent $(k,-k)$ blocks. Each $(k,-k)$ block further splits into sectors of even and odd fermion parity: the even sector is spanned by $\{\ket{\Omega_S},\tilde{a}_k^\dagger \tilde{a}_{-k}^\dagger\ket{\Omega_S}\}$ and the odd sector by $\{\tilde{a}_k^\dagger\ket{\Omega_S}, \tilde{a}_{-k}^\dagger\ket{\Omega_S}\}$, where $\ket{\Omega_S}$ is the particle vacuum of the system and $\tilde{a}_k$ are the Fourier-transformed mode operators.

Because the ground state of $H_S(\pi/2)$ lies in the even-parity sector and unitary evolution generated by quadratic Hamiltonians preserves the fermion parity, both adiabatic evolution and QAOA reduce to independent effective qubits within this sector: the two-dimensional even-parity subspace of each $(k,-k)$ block defines a qubit with basis states $\ket{0_k}\equiv\ket{\Omega_S}$ and $\ket{1_k}\equiv \tilde{a}_k^\dagger \tilde{a}_{-k}^\dagger\ket{\Omega_S}$.
For the cooling algorithm, which involves a CP map obtained by tracing out or resetting the bath after joint evolution with the system, the parity of only the system is not conserved in general (it is however conserved when considering system and bath together). Nevertheless, the evolution remains Gaussian and factorizes across momentum blocks, so that the analytic and numerical treatment still reduces to independent small subsystems of at most four dimensions.
Throughout the remainder of this paper, when we refer to ``mode~$k$,'' we mean the mode pair $(k,-k)$ and the corresponding effective qubit in the even-parity sector, unless stated otherwise.

Diagonalizing $H_S$ by a Bogoliubov transformation yields the mode Hamiltonians and energies
\begin{align}
    H_S        &= \sum_{k=1}^{N/2-1} \epsilon_k (\hat a_k \hat a_k^\dag+\hat a_{-k} \hat a_{-k}^\dag)+\sum_{k=0,N/2}\epsilon_k \hat a_k \hat a_k^\dag,\\
    \epsilon_k &= \sqrt{1 + \sin2\theta \cos\frac{2\pi k}{N}}.
    \label{eq:epsilonk_main}
\end{align}
We call spectrum range the interval within which the energies of the modes vary, namely from $\epsilon_m=\sqrt{1-\sin 2\theta}$ at $k=N/2$ to $\epsilon_M=\sqrt{1+\sin 2\theta}$ at $k=0$.
The gap closes ($\epsilon_m\to 0$) at the critical point $\theta=\pi/4$.

As mentioned before, the 2--dimensional subspace of even parity is spanned by $\{\ket{0_k},\ket{1_k}\}$. The Hamiltonian block for each mode reads then
\begin{equation}
    H_{k}(\theta) = w_k(\theta) \sigma_z + r_k(\theta) \sigma_x,
    \label{eq:Hk_qubit}
\end{equation}
with $w_k(\theta)=\sin\theta+\cos\theta\cos\phi_k$, $r_k(\theta)=\cos\theta\sin\phi_k$, and $\phi_k = 2\pi k/N$.
The total relative energy can be written in terms of the energies per mode as follows
\begin{equation}
    e = \frac{\sum_{k=0}^{N/2} \epsilon_k\, e_k}{\sum_{k=0}^{N/2} \epsilon_k},
    \label{eq:relative_energy_sum}
\end{equation}
where $e_k=(E_k+\epsilon_k)/\epsilon_k$ is the relative energy of mode $k$; here $E_k$ is the energy contribution of mode $k$ and $-\epsilon_k$ is its ground-state value.
For details (Bogoliubov angles, special cases $k=0,N/2$, etc.), we refer the reader to Ref.~\cite{Molpeceres2025Quantum}.

\subsection{Bath and Coupling}
\label{subsec:bath}

For the cooling algorithm, we introduce an auxiliary bath that can be prepared (and reset) close to its ground state. As in Ref.~\cite{Molpeceres2025Quantum}, we consider $N$ independent fermionic modes with tunable onsite potential $\Delta$:
\begin{equation}
    H_B = \frac{\Delta}{2}\sum_{n=1}^N \left(b_n^\dag b_n - b_n b_n^\dag\right),
    \label{eq:bath-hamiltonian-realspace}
\end{equation}
where $b_n,b_n^\dag$ are fermionic operators acting on the bath.

The system modes ($a_n$) and bath modes ($b_n$) are coupled in the following translationally invariant way, up to a maximum range~$r_{\rm c}$:
\begin{equation}
    V_{SB} = g \sum_{n=1}^N \sum_{j=-r_{\rm c}}^{r_{\rm c}} \left[\left(\lambda_j a_n^\dag {b}_{n+j} + i \mu_j a_n {b}_{n+j}\right) + \text{h.c.}\right].
    \label{eq:system-bath-coupling_realspace}
\end{equation}
Here, $g$ is an overall coupling strength and $\lambda_j,\mu_j$ are dimensionless parameters.

In momentum space, after diagonalizing the Hamiltonian (in the Hilbert space corresponding to the system) via the Bogoliubov transformation, the joint Hamiltonian $H_{SB}=H_S+H_B+V_{SB}$ decomposes into independent $(k,-k)$ blocks.
For generic $k$, one can write $H_{SB}=\sum_k \hat{\alpha}_k^\dag h_{SB,k}\hat{\alpha}_k$ with $\hat{\alpha}_k = (\hat{a}_k,\hat{a}_{-k}^\dagger,\hat{b}_k,\hat{b}_{-k}^\dagger)^\top$, and
\begin{align}
    h_{SB,k} &=
    \begin{pmatrix}
        \epsilon_k &0           &gA_k   &gB_k\\
        0          &-\epsilon_k &gB_k   &-g A_k\\
        g A_k^*    &g B_k^*     &\Delta &0\\
        gB_k^*     &-g A_k^*    &0      &-\Delta
    \end{pmatrix},
    \label{eq:hsb_momentum_space_nn>0}
\end{align}
where $A_k,B_k$ depend on the Bogoliubov angle $\varphi_k$ and the couplings $\{\lambda_j,\mu_j\}$:
\begin{align}
    A_k &=\sum_{j=-r_{\rm c}}^{r_{\rm c}}\left(\cos(\varphi_k)\lambda_j+i\sin(\varphi_k)\mu_j\right)e^{-i\frac{2\pi jk}{N}},
    \label{eq:Ak_in_hsb_k}\\
    B_k &=\sum_{j=-r_{\rm c}}^{r_{\rm c}}\left(-\sin(\varphi_k)\lambda_j+i\cos(\varphi_k)\mu_j\right)e^{-i\frac{2\pi jk}{N}},
    \label{eq:Bk_in_hsb_k}
\end{align}
with $\varphi_k = \frac{1}{2}\tan^{-1}\left(\frac{\sin\tfrac{2\pi k}{N}}{\tan\theta + \cos\tfrac{2\pi k}{N}}\right)$.

\subsection{Noise}
\label{subsec:noise}

We model noise by a depolarizing channel acting on both system and bath during each step of coherent evolution. In Lindblad form it is described by
\begin{align}
    \mathcal{L}_E(\rho) = \kappa\sum_{n=1}^N &[ (a_n\rho a_n^\dagger + a_n^\dagger\rho a_n - \rho)\nonumber\\
                                                  &+ (b_n\rho b_n^\dagger + b_n^\dagger\rho b_n - \rho) ],
    \label{eq:LE_explicit}
\end{align}
where $\kappa$ denotes the noise strength. Since the Lindblad operators are site-local, this channel factorizes across sites, and also across modes; i.e. $e^{\mathcal{L}_Et}$ is the tensor product over all $e^{\mathcal{L}_E^k t}$, with the CPTP map~\cite{Molpeceres2025Quantum}:
\begin{equation}
    e^{\mathcal{L}_E^k t}(\rho_k)=\mathcal{D}_t(\rho_k)=e^{-2\kappa t}\rho_k+(1-e^{-2\kappa t})\frac{\mathbb{1}}{4}.
    \label{eq:Enoise_explicit}
\end{equation}
This channel commutes with any unitary generated by a quadratic fermionic Hamiltonian (see~\cite{Molpeceres2025Quantum}), which has distinct consequences for the three algorithms.

For the unitary algorithms (adiabatic and QAOA), the noise accumulated over the entire runtime can equivalently be collected into a single depolarizing channel $\mathcal{D}_T$ applied after the complete unitary circuit; the energy of any quadratic observable is then damped by a factor $e^{-2\kappa T}$.
For the cooling algorithm, the noise within each cycle can similarly be moved to act after that cycle's unitary evolution step; however, the non-unitary bath reset between successive cycles prevents combining noise from different cycles into a single channel.
As a result, increasing the total runtime of the algorithm degrades performance for all three algorithms, but the mechanism by which the optimal time balances coherent evolution against noise is algorithm-specific and constitutes the central analysis of \cref{sec:analytics_cooling_vs_adiabatic}.

\section{Quantum algorithms considered}
\label{sec:algs}

Below we describe the three algorithms compared in this work: adiabatic evolution, Quantum Approximate Optimization Algorithm (QAOA), and cooling.
For each algorithm, we also explain how depolarizing noise affects its performance.

\subsection{Adiabatic algorithm}
\label{subsec:adiabatic_alg}

We consider the adiabatic preparation of the ground state of a target Hamiltonian $H_S(\theta_f)$ starting from the ground state of an initial Hamiltonian $H_S(\theta_i)$. We choose $\theta_i = \pi/2$, corresponding to a trivial product state, and evolve towards a target $H_S(\theta_f)$.
Following the standard adiabatic protocol~\cite{Murg2004Adiabatic,Albash2018Adiabatic}, we employ a linear interpolation between the initial and target Hamiltonians:
\begin{equation}
    H(s) = (1-s) H_S(\theta_i) + s H_S(\theta_f),
    \label{eq:adiabatic_schedule}
\end{equation}
where $s=t/T \in [0,1]$ is the normalized time and $T$ is the total evolution time.

Due to the translational invariance the quadratic Hamiltonian $H(s)$ decomposes into independent $2\times 2$ blocks for each momentum mode $k$ (see Appendix~\ref{app:adiabatic_derivations} for details). The many-body dynamics thus reduces to the independent evolution of $N/2$ two-level systems.
Depending on the targeted value of~$\theta_f$, the interpolation path may cross the quantum critical point at $\theta_c=\pi/4$. We analyze how the relative energy scales in two regimes: (i)~the gapped regime, described by adiabatic perturbation theory~\cite{Albash2018Adiabatic}, where the deviation from the ground-state energy in the absence of noise scales as~$T^{-2}$, and (ii)~the gapless regime (crossing the critical point), described by the Kibble--Zurek mechanism~\cite{Kibble1976Topology,Zurek1985Cosmological,Zurek2005Dynamics,Dziarmaga2005Dynamics,Polkovnikov2005Universal,delCampo2014Universality} and Landau--Zener transitions~\cite{Landau1932Zur,Zener1932Nonadiabatic}, where it scales as~$T^{-1/2}$.

In the ideal case, the initial state is the exact ground state $\ket{0}^{\otimes N}$ at $\theta_i = \pi/2$. In the realistic scenario, the initial state is the mixed state $\rho^{\rm cool}(\pi/2)$ obtained from a cooling protocol at $\theta = \pi/2$. Since $\rho^{\rm cool}(\pi/2)$ is diagonal in the mode basis, the adiabatic evolution of each mode can be treated independently for both cases (see Appendix~\ref{subapp:realistic_adiabatic_evolution}). Moreover, as the depolarizing channel~[\cref{eq:Enoise_explicit}] commutes with the unitary evolution generated by $H(s)$, for every traceless quadratic observable $O$ and every state $\rho$ one obtains
\begin{equation}
    \tr \left[O\,e^{\mathcal{L}_E t}(\rho)\right]=e^{-2\kappa t}\tr(O\rho).
    \label{eq:noise_damping_quadratic}
\end{equation}
Therefore, the cumulative effect of noise over the total runtime $T$ is, at the level of quadratic observables such as the energy, equivalent to a single damping factor applied after the ideal unitary evolution: the total energy of the target Hamiltonian is multiplied by $e^{-2\kappa T}$\footnote{Recall that in our model, the energy of the ground state is always negative and the energy of the maximally mixed state is zero. Hence, the maximally mixed state in Eq.~\ref{eq:Enoise_explicit} does not contribute to the energy and only a damping term appears.}. The same argument applies to QAOA (see below).
This creates a trade-off: increasing~$T$ reduces deviations from the ground-state energy due to non-adiabatic transitions, but simultaneously enhances the effect of noise. The optimal runtime~$T$, which balances these competing effects, is analyzed in \cref{sec:analytics_cooling_vs_adiabatic}.

\subsection{QAOA}
\label{subsec:qaoa_alg}

The QAOA circuit consists of $p$ layers, each composed of two unitaries generated by the non-commuting Hamiltonians $H_M$ and $H_P$. We have
\begin{equation}
    \mathcal{U}(\bm{\beta},\bm{\gamma})[\rho] = U\,\rho\, U^\dagger,\quad
    U = \prod_{l=1}^p e^{-i \beta_l H_M}  e^{-i \gamma_l H_P},
    \label{eq:qaoa_ansatz}
\end{equation}
where the parameters $\bm{\beta},\bm{\gamma}\in\mathbb{R}^p$ are optimized to minimize the energy of $H_S(\theta_f)$.
We take $H_M$ and $H_P$ to be translationally invariant and quadratic, so that the circuit factorizes into independent $2\times 2$ unitaries in each $(k,-k)$ block of even parity. Specifically, within the even-parity qubit basis defined in \cref{subsec:system}, the single-mode generators read
\begin{equation}
    h_{M,k}=\sigma_z,~ h_{P,k}=\cos\phi_k \sigma_z+\sin\phi_k \sigma_x,
    \label{eq:qaoa_mode_generators}
\end{equation}
and the $l$-th QAOA layer acts on mode $k$ as
\begin{equation}
    U_{k,l}(\beta_l,\gamma_l) = e^{-i\beta_l h_{M,k}}
    e^{-i\gamma_l h_{P,k}}.
\end{equation}
The full circuit factorizes as $U = \bigotimes_{k} U_k$ with $U_k = \prod_{l=1}^p U_{k,l}$. While each $U_k$ is a product of $2\times 2$ unitaries, the parameters $\bm{\beta},\bm{\gamma}$ are shared across all modes and must be optimized jointly to minimize the total energy.

Starting from an initial state $\rho_{\rm init}=\bigotimes_k \rho_{k,\rm init}$, the variational energy is given by
\begin{equation}
    E(\bm{\beta},\bm{\gamma}) = \sum_{k=0}^{N/2}
    \tr \left[U_k\,\rho_{k,\rm init}\, U_k^\dagger\, H_{k}(\theta_f)\right],
    \label{eq:qaoa_energy}
\end{equation}
where $H_k(\theta_f)$ is the effective qubit Hamiltonian from \cref{eq:Hk_qubit}.

Using again that the depolarizing channel~[\cref{eq:Enoise_explicit}] commutes with any unitary generated by a quadratic Hamiltonian, its effect on quadratic observables is equivalent to applying a single depolarizing channel $\mathcal{D}_{t_{\rm circ}}$ after the full circuit, where the total coherent time is $t_{\rm circ}=\sum_l(|\beta_l|+|\gamma_l|)$.
Because each mode Hamiltonian~$H_k$ is traceless, the energy of the maximally mixed state vanishes, and the energy of mode $k$ reduces to
\begin{equation}
    E_\kappa = e^{-2\kappa t_{\rm circ}}\, E(\bm{\beta},\bm{\gamma}).
    \label{eq:qaoa_energy_noisy}
\end{equation}
We use this expression for the numerical comparisons in \cref{sec:numerics_qaoa_vs_cooling}.

\subsection{Cooling algorithm}

\label{subsec:cooling_alg}

A cooling algorithm extracts energy from the system by repeatedly coupling it to a bath prepared in its ground state, allowing joint evolution for a cycle time~$\tau$, and then resetting the bath. In the absence of noise, each elementary cycle is described by the completely positive trace-preserving map
\begin{equation}
    \mathcal{E}(\rho_S) = \tr_B\left[e^{-i H_{SB}\tau}(\rho_S \otimes \rho_B)e^{i H_{SB}\tau}\right],
    \label{eq:cooling_map0}
\end{equation}
where $\rho_S$ is the system density matrix, $\rho_B$ denotes the initial state of the bath, prepared in its vacuum state, and $\tr_B$ denotes the partial trace over the bath. The total Hamiltonian is $H_{SB} =H_S+H_B+V_{SB}$, as defined in \cref{sec:model}.
More sophisticated schemes consist of sequences of multiple cycles with varying parameters, such as the system--bath couplings, the cycle time $\tau$, or the bath frequency $\Delta$.

The analytical framework for our algorithm, developed in Ref.~\cite{Molpeceres2025Quantum}, yields the following effective description. The analytical results are derived in the weak coupling limit, the low noise limit and the cooling limit (see Appendix~\ref{app:analytic_cooling} for details). In the weak-coupling limit, $(g\tau)^2 \ll 1$, a Dyson expansion of the map~[\cref{eq:cooling_map0}] produces effective Lindblad dynamics for each momentum mode~$k$. Two competing processes appear: a cooling process, peaked when the bath frequency~$\Delta$ matches the mode energy~$\epsilon_k$, and a heating process, which remains small in general provided the cooling limit $\epsilon_k \tau \gg 1$ holds. A single bath frequency~$\Delta$ cools only modes near resonance, $\epsilon_k \approx \Delta$, and for a fixed cycle time, accidental resonances at $(\epsilon_k - \Delta)\tau = 2\pi n$ ($n \in \mathbb{Z}\setminus\{0\}$) lead to unwanted heating. This can be avoided using two different methods~\cite{Molpeceres2025Quantum}: the first one randomizes the cycle times and adds multiple frequencies; the second couples each bath mode to more than one of the system's modes and optimizes the cooling with respect to the corresponding couplings. The latter, in fact, does not need to operate in the weak coupling limit and gives rise to better performance for the models studied in~\cite{Molpeceres2025Quantum}, although one loses the analytical understanding of the first one. Below we detail how both methods operate.

\paragraph{Multi-time and frequency cooling}

To cool all modes simultaneously, we employ a multifrequency strategy with local couplings ($\lambda_0=\mu_0=1$ in \cref{eq:system-bath-coupling_realspace}). For each cooling cycle, we choose $\tau$ and $\Delta$ randomly according to a given distribution, $D(\tau,\Delta)=D^0_t(\tau)D^1_{R,\Delta_m,\Delta_M}(\Delta)$. The distribution is parametrized in terms of the mean cycle time, $t$, the number of frequencies, $R$, and the minimum and maximum values of the bath frequencies, $\Delta_m,\Delta_M$, respectively, which depend on $\epsilon_{m,M}$, the bandwidth of the system's single-particle spectrum. Eventually, we also scale $R\propto t$, as explained below.

\begin{figure}[ht]
    \includegraphics[width=\linewidth]{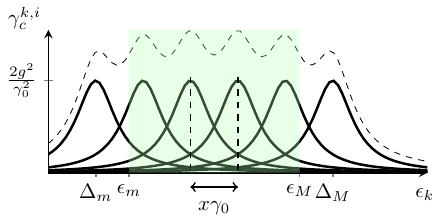}
    \caption{Individual cooling rates $\gamma^{k,i}_{\rm c}$ (Lorentzian profiles) and their sum $\tilde{\gamma}^{\rm c}_k$ (dashed curve) as a function of the mode energy $\epsilon_k$ for a set of bath frequencies $\{\Delta_r\}$ equally spaced over the interval $[\Delta_m, \Delta_M]$. The spacing between two adjacent frequencies is $x\gamma_0$, where $x$ is a dimensionless parameter and $\gamma_0 \propto 1/t$ is the Lorentzian linewidth. Each frequency generates a resonance peak of width $\gamma_0$ centered at $\epsilon_k=\Delta_r$ and height $2g^2/\gamma_0^2$. When the peaks overlap sufficiently ($x \lesssim 1$), the averaged cooling rate ${\gamma}^{\rm c}_k$ becomes approximately independent of $k$, ensuring uniform cooling of all modes in the range $[\epsilon_m,\epsilon_M]$ (green area).}
    \label{fig:multifreq_scheme}
\end{figure}

The distribution $D^0$ is uniform, namely $\tau\in [0,2t]$ with uniform probability distribution, so that $t$ is the averaged time of the cycle. The value of $t$ has to be chosen so that the weak coupling limit $(gt)^2\ll 1$ is fulfilled. The distribution $D^1$ is also uniform, but takes $R$ discrete values in the interval $[\Delta_m,\Delta_M]$, as explained previously \footnote{As mentioned before the analytical results derived in Appendix \ref{app:analytic_cooling} hold for any distribution. However, for the numerical derivations we choose here a discrete number of frequencies.}. In order to choose those quantities, we note that in the weak coupling limit the cooling rate as a function of the mode energy $\epsilon_k$ is given by a Lorentzian function centered at $\Delta$ and with a linewidth $\gamma_0\propto 1/t$ (see Eq. \eqref{eq:ratesAppA} and Fig.~\ref{fig:multifreq_scheme}). That is, if we choose one particular $\Delta$ it will cool down the modes $k$ with $\epsilon_k$ around the interval $\Delta\pm \gamma_0$. Thus, we have to choose $R$ such that for all $k$, $\epsilon_k$ lies within that interval for some $\Delta_i$ (see  Fig.~\ref{fig:multifreq_scheme} and Appendix~\ref{subapp:discrete_distrib} for details).

In order to derive analytical results, we average the map [\cref{eq:cooling_map0}] with respect to the distribution $D(\tau,\Delta)$ and then expand the result in the weak coupling limit. In Appendix~\ref{app:analytic_cooling} we show that after a sufficient number of cycles, the resulting state for each pair of modes $(k,-k)$ is ${\cal O}(g^2t^2)$ close to the one obtained by those approximations. In practice, it is simpler to first expand in the weak coupling limit and then carry out the average, as it was done in~\cite{Molpeceres2025Quantum} \footnote{Note that the result is the same as if we do it the other way around, see Appendix~\ref{subapp:weak_coupling_map}.}. The result is that for each pair of modes, we obtain a rate equation with averaged cooling and heating rates, $ \gamma^c_k$ and $ \gamma^h_k$ (see Appendix~\ref{app:analytic_cooling}). We also estimate, in Appendix~\ref{subapp:cooling_limit_expansion}, the total time needed to reach a target fidelity, finding an improvement over the bound of~\cite{Molpeceres2025Quantum}. The relative energy of the steady-state of mode $k$ in the absence of noise is
\begin{equation}
    e_k = \frac{2{\gamma}_k^{\rm h}}{{\gamma}_k^{\rm c} + {\gamma}_k^{\rm h}}.
    \label{eq:ek_cooling_noiseless}
\end{equation}

Depolarizing noise of strength~$\kappa$ modifies this expression. Assuming a low noise limit ($\kappa t\ll 1$), each elementary subcycle of duration~$\tau$ acquires a noise contribution~$\kappa\tau$ [cf.\ \cref{eq:Enoise_explicit}]; averaging over the cycle time ($\langle\tau\rangle=t$) thus adds a term~$\kappa t$ to both the cooling and heating rates, resulting in
\begin{equation}
    e_k(t) = \frac{2{\gamma}_k^{\rm h} + 2\kappa t}{{\gamma}_k^{\rm c} + {\gamma}_k^{\rm h} + 2\kappa t}.
    \label{eq:ek_noisy_cooling}
\end{equation}
When cooling dominates ($\tilde{\gamma}_k^{\rm c} \gg \kappa t$), the energy approaches the ground state ($e_k \to 0$); when noise dominates, the system depolarizes ($e_k \to 1$). The optimal trade-off between these regimes, and its scaling with~$\kappa$, is derived in \cref{subsec:analytic_cooling}. Let us note here that we could have also chosen a cooling algorithm where, at each step, $\tau$ and $\Delta$ are chosen randomly and uniformly from the interval $[0,2t]$ and $[\epsilon_m,\epsilon_M]$ resp. (see Appendix~\ref{app:analytic_cooling}). However, due to the fact that the discrete values might be experimentally more accessible, we proceed here with the discrete distribution of $\Delta$, detailed in \cref{subsec:analytic_cooling}.

\paragraph{Variationally optimized cooling.}
The second strategy we employ here uses a cooling procedure in which a cooling cycle is repeated until steady state is reached. Each global cycle is composed of a small number of cooling cycles with nonlocal couplings (coupling range~$r_{\rm c}$; see \cref{eq:system-bath-coupling_realspace}), each cycle having its own bath frequency~$\Delta$, cycle time~$\tau$, and coupling parameters $\{\lambda_j, \mu_j\}$.
All parameters are numerically optimized to minimize the total relative energy.
As shown in Ref.~\cite{Molpeceres2025Quantum}, this approach can achieve lower energies with shorter cycle times than the multifrequency strategy, at the cost of numerical fine-tuning.
When comparing this algorithm to QAOA in \cref{sec:numerics_qaoa_vs_cooling}, we assess both its performance under noise and its sensitivity to imperfections in the optimized parameters.

\section{Analytical treatment: multifrequency cooling vs.\ adiabatic algorithm}
\label{sec:analytics_cooling_vs_adiabatic}

Having described the three algorithms, we now compare the performance of the multifrequency cooling and adiabatic approaches analytically. We determine the optimal relative energy each can achieve as a function of the noise rate~$\kappa$.
For the cooling algorithm, the analytical results are derived in three limits---the weak-coupling limit $(gt)^2\ll1$, the cooling limit $\epsilon_m t\gg1$, and the low-noise limit $\kappa t\ll1$, all of which are defined and used in Appendix~\ref{app:analytic_cooling}; for these to be mutually compatible we require $\kappa\ll\epsilon_m$.

In both cases, the optimal energy scales as a power law in~$\kappa$, with an exponent that depends on the algorithm and the targeted quantum phase.
Note that a lower exponent indicates greater resilience to noise since the energy degrades more slowly as $\kappa$ increases.
We derive these exponents analytically for the multifrequency cooling and adiabatic algorithms below.

\subsection{Multifrequency cooling: \texorpdfstring{$e_{\rm opt}^{\rm cool}$}{eopt} as a function of \texorpdfstring{$\kappa$}{kappa}}
\label{subsec:analytic_cooling}

The noisy steady-state energy given in \cref{eq:ek_noisy_cooling} depends on the cycle time~$t$ through both the heating rates (which decrease with longer~$t$ via sharper Lorentzian peaks) and the noise (which increases as $\kappa t$).
In a noiseless setting, increasing~$t$ improves spectral resolution, which causes $e_k$ to decrease monotonically with $t$. In a noisy setting a more careful optimization must be carried out, since the effect of noise will also increase with $t$. We want to optimize this tradeoff to determine the best achievable energy as a function of~$\kappa$.

At fixed coupling strength $g$, the energy $e_k(t)$ in \cref{eq:ek_noisy_cooling} decreases monotonically with the cycle time $t$, but the limit $t\to\infty$ is incompatible with the weak-coupling condition $(gt)^2 \ll 1$ required for the perturbative treatment.
We therefore impose that $g t$ is constant and set $gt = \eta$, fixing the product of coupling strength and cycle time.
Numerical calculations confirm that the perturbative results remain accurate for $\eta \lesssim 0.5$; see Appendix~\ref{app:analytic_cooling} and \cref{fig:kappag2_efit} therein.
Under this constraint, the coupling strength decreases as $g = \eta/t$ and the number of bath frequencies grows as $R\simeq \sqrt{2/3} (\Delta_M-\Delta_m) t$ (implemented as $R=\lfloor\sqrt{2/3} (\Delta_M-\Delta_m) t\rfloor+1$ when the integer nature of $R$ matters), leaving $t$ as the sole free parameter.

As shown in \cref{subsec:cooling_alg}, the averaged cooling rate~${\gamma}_k^{\rm c}$ is $k$-independent, while the averaged heating rate~${\gamma}_k^{\rm h}$ varies only weakly with $k$ and can be bounded from above (see Appendix~\ref{app:analytic_cooling} and \cref{fig:multifreq_sums}). Thus, the optimal cycle time is approximately the same for all modes. We can therefore optimize the total relative energy by setting $\partial e_k / \partial t = 0$ at a representative mode and solving for the optimal time (see Appendix~\ref{subapp:energy_steady_state} for the full calculation).
With the constraint that $gt$ is constant, in particular, $gt = \eta$, the noise-induced heating grows as $\kappa t$, while the averaged cooling rate decreases with growing $t$ (since ${\gamma}_k^{\rm c} \propto g^2 t= \eta/t$) and the heating rate decreases even faster with $t$ (since ${\gamma}_k^{\rm h}\propto g^2=\eta^2/ t^{2}$).
The energy therefore first decreases and then increases with~$t$, reaching a minimum at an intermediate optimal time $t_{\rm opt}^{\rm cool} \sim (\kappa \eta^{-2})^{-1/3}$; substituting back into \cref{eq:ek_noisy_cooling} gives the minimum relative energy:
\begin{gather}
    e_{{\rm opt}}^{\rm cool} \approx \frac{3}{C}(A^{2}\eta^{-2}\kappa)^{1/3}, \mbox{ with}\label{eq:e_cool_scaling}\\
    A =\frac{1}{(\epsilon_{N/4}+\Delta_m)(\epsilon_{N/4}+\Delta_M)},\quad C=\frac{\sqrt{2}\pi }{\sqrt{3}(\Delta_M-\Delta_m)},\nonumber
\end{gather}
where $\Delta_{M,m}$ are the largest and smallest bath frequencies, and $A$ is evaluated at $k=N/4$, i.e. in the middle of the spectrum (the weak $k$-dependence is discussed in Appendix~\ref{app:analytic_cooling}).

\subsection{Adiabatic algorithm: \texorpdfstring{$e_{\rm opt}^{\rm adiab}$}{eopt} as a function of \texorpdfstring{$\kappa$}{kappa}}

To compare the performance of the cooling algorithm with the adiabatic algorithm, we now derive the relative energy as a function of the noise rate for the latter. Starting from $\theta_i=\pi/2$ and targeting~$\theta_f$, two qualitatively different regimes arise depending on whether the interpolation path~[\cref{eq:adiabatic_schedule}] crosses the quantum critical point at $\theta_c=\pi/4$ or not.
Since $H(s)$ is quadratic and translationally invariant, the many-body dynamics factorizes into independent two-level problems, one for each momentum mode~$k$.

While each individual mode undergoes standard two-level dynamics, the nontrivial step is summing over all $N/2$ modes---each with a different instantaneous gap and transition amplitude---to obtain the total energy and its scaling with~$T$.
Recall that, as the depolarizing channel commutes with the unitary evolution (see \cref{subsec:noise}), its effect on the energy is given by an overall damping factor $e^{-2\kappa T}$. Non-adiabatic effects increase the deviation from the ground-state energy for faster processes, whereas depolarizing noise leads to higher energies for longer evolution times. The competition between these two mechanisms gives rise to an optimal runtime that minimizes the achievable energy.

We sketch the derivations below (full details are given in Appendix~\ref{app:adiabatic_derivations}).
Throughout, we use the superscripts ``prod'' (product/trivial phase) and ``ent'' (entangled/topological phase) to distinguish the results for the two phases, as they give different scalings. In Table~\ref{tab:scaling_summary} we summarize the scalings of the relative energy as a function of $\kappa$ for both the cooling and adiabatic algorithms.

\paragraph{Trivial phase, no crossing ($\theta_f>\pi/4$):}
When the adiabatic path does not cross the critical point, the spectral gap remains of order~$\mathcal{O}(1)$ for all modes throughout the evolution.
Applying first-order adiabatic perturbation theory \cite{Sakurai2020Modern} to each $2\times 2$ mode block, we expand the time-evolved state in the instantaneous eigenbasis and find that
the transition amplitude to the excited state is given by an oscillatory integral whose rapid phase oscillations (at rate $\propto T$) suppress contributions from the bulk of the integration domain; only the boundary terms survive. This yields transition probability $|c_1(k)|^{2}\propto T^{-2}$
for each mode (see Appendix~\ref{app:adiabatic_derivations}).
Summing over all modes, the final energy in the absence of noise takes the form
\begin{equation}
    E\approx E_{\rm GS} \left(1-\frac{\tilde{A}}{T^{2}}\right),
    \quad \mbox{ with }\tilde{A}\approx\frac{\cos^{2}\theta_f}{8}\,.
    \label{E_adiab_prod_noiseless}
\end{equation}
Including the damping factor due to noise, $e^{-2\kappa T}$, and expanding for $\kappa T\ll 1$, the total energy becomes $E\approx E_{\rm GS}(1-\tilde{A}/T^{2}-2\kappa T)$. The non-adiabatic contribution ($\propto T^{-2}$) is suppressed at large~$T$, while the noise contribution ($\propto \kappa T$) is small at short~$T$.
Optimizing over~$T$ under the constraint $\kappa T\ll 1$ gives
\begin{equation}
    T_{\rm opt}^{\rm prod} \approx\left(\frac{\tilde{A}}{\kappa}\right)^{ 1/3} ,\mbox{ and }
    e_{\rm opt}^{\rm prod} \approx 3\tilde{A}^{1/3}\kappa^{2/3}.
    \label{eq:e_adiab_prod}
\end{equation}

\paragraph{Topological phase, crossing the critical point ($\theta_f<\pi/4$):}

When the adiabatic path crosses the critical point, the spectrum of mode $k = N/2$ becomes gapless, while nearby modes also develop very small gaps. As a result, the system is particularly susceptible to excitations in this region.

Close to the critical point, the dynamics of modes near $k = N/2$ can be approximated by the well-known Landau--Zener model~\cite{Landau1932Zur,Zener1932Nonadiabatic}. This model calculates the excitation probability of a two-level system evolution that starts in the ground state and goes through an avoided crossing at some speed (for details, see Appendix~\ref{app:adiabatic_derivations}). We use this model to analyze each mode $k$ in our system. The energy levels approach the avoided crossing with a ``velocity'' set by $v_{\rm LZ} = 2/T$, while the minimum gap in this crossing depends on how far the mode is from $k = N/2$, which is captured by the parameter $\delta_k = 2\pi k/N - \pi$. Within this approximation, the excitation probability takes a Gaussian form,
\begin{equation}
    P_{\rm LZ}^{k} = \exp\left(-\pi G\, T\, \delta_k^2\right),
\end{equation}
where $G=\cos^{2}\theta_f/(1+\cos\theta_f-\sin\theta_f)^{3}$ depends only on the final parameter $\theta_f$. For large evolution time $T$, this Gaussian becomes sharply peaked around $k = N/2$, meaning that only a narrow range of modes is excited. The width of this region scales as $\sim 1/\sqrt{T}$, which is the characteristic behavior predicted by the Kibble--Zurek mechanism~\cite{Zurek2005Dynamics,Dziarmaga2005Dynamics,Polkovnikov2005Universal}.

By approximating the sum over modes as an integral, one finds that the energy deviation from the ground state decreases as
\begin{equation}
    E \approx E_{\rm GS}\left(1 - \frac{\tilde{B}}{\sqrt{T}}\right),
\end{equation}
where $\tilde{B}$ is a constant of order one that depends on $\theta_f$ and on the behavior of the spectrum near the critical point.

Including the effect of noise and optimizing over the total evolution time $T$ yields
\begin{equation}
    T_{\rm opt}^{\rm ent} \approx \left(\frac{\tilde{B}}{4\kappa}\right)^{2/3},
    \quad
    e_{\text{opt}}^{\rm ent} \approx \frac{3}{2^{1/3}} \tilde{B}^{2/3} \kappa^{1/3}.
    \label{eq:e_adiab_ent}
\end{equation}

Compared to the trivial phase, the energy approaches the ground state more slowly with increasing $T$ (as $T^{-1/2}$ rather than $T^{-2}$). Even though we find a more favorable scaling with $\kappa$ than in the trivial phase, the relative energy is higher due to the prefactors.

Finally, at very low noise, the energy cannot be reduced arbitrarily. This is because the mode $k = N/2$ has exactly zero gap at the critical point and is therefore always excited, no matter how slow the evolution is. This sets a lower bound on the energy that is independent of $\kappa$. The crossover between this finite-size limit and the $\kappa^{1/3}$ scaling occurs when $\kappa^{1/3} \sim \epsilon_{N/2}/|E_{\rm GS}|$, and shifts to smaller values of $\kappa$ as the system size $N$ increases (see Appendix~\ref{app:adiabatic_derivations} and \cref{fig:finite_N_crossover}).

\begin{figure}
    \begin{minipage}{0.23\textwidth}
        \centering
        \begin{overpic}[width = 1\textwidth]{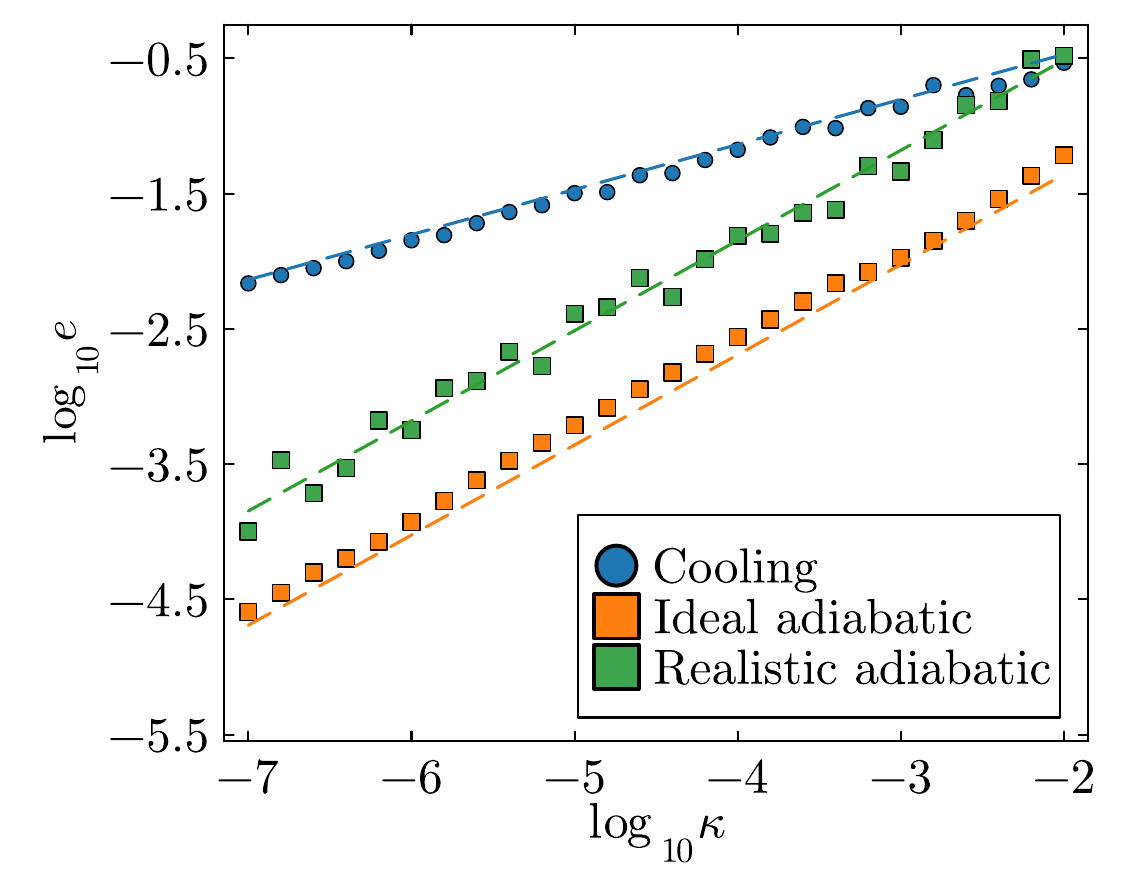}
            \put(-2,70){{(a)}}
            \put(38,80){\small$\pi/2\to\pi/3$}
        \end{overpic}
        \label{fig:pi3_cool_adiab}
    \end{minipage}
    \begin{minipage}{0.23\textwidth}
        \centering
        \begin{overpic}[width = 1\textwidth]{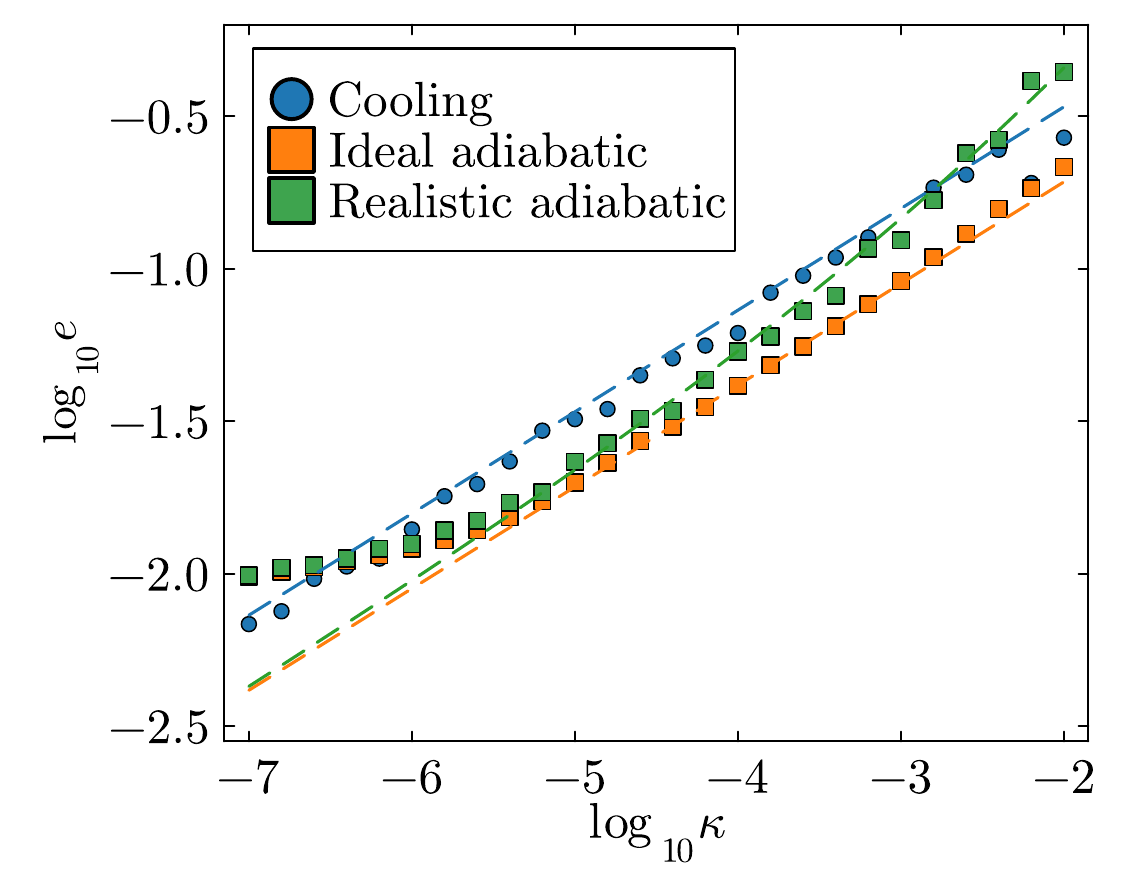}
            \put(-2,70){{(b)}}
            \put(38,80){\small$\pi/2\to\pi/6$}
        \end{overpic}
        \label{fig:pi6_cool_adiab}
    \end{minipage}
    \begin{minipage}{0.23\textwidth}
        \centering
        \begin{overpic}[width = 1\textwidth]{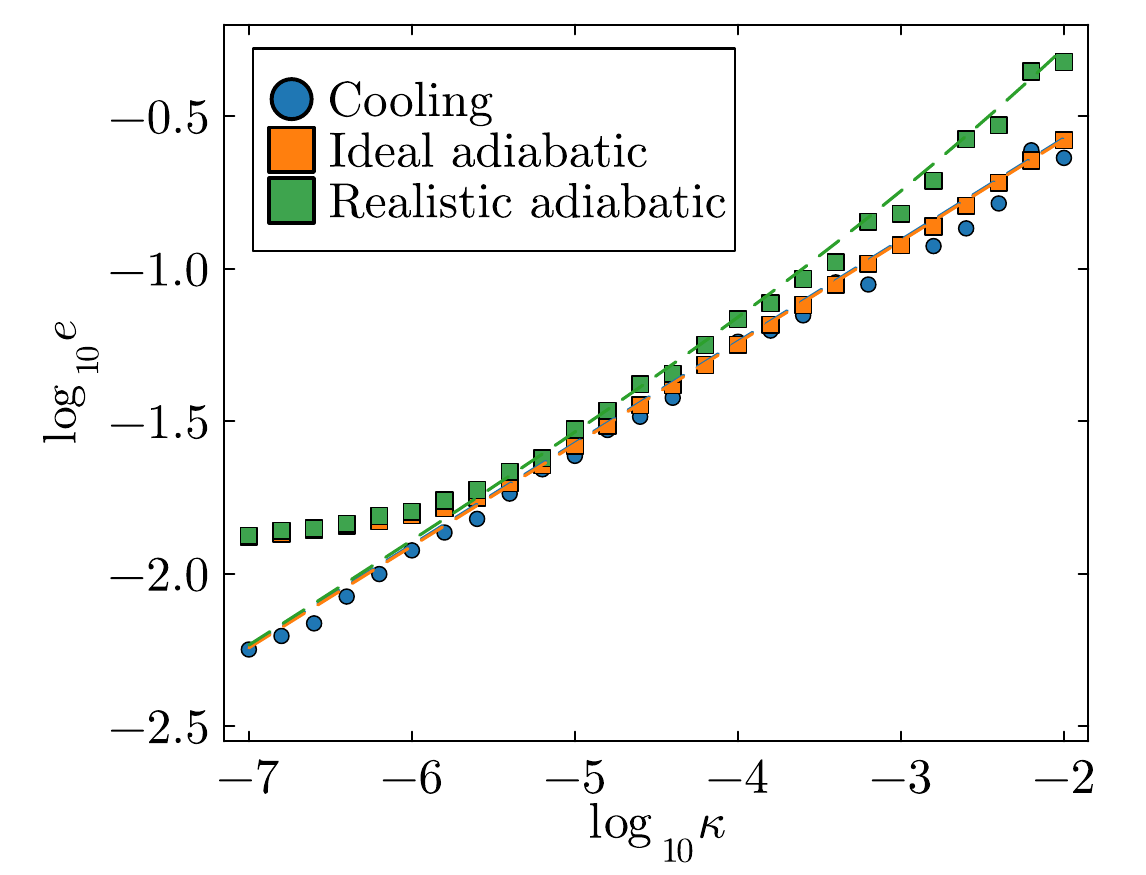}
            \put(-2,70){{(c)}}
            \put(38,80){\small$\pi/2\to\pi/8$}
        \end{overpic}
        \label{fig:pi8_cool_adiab}
    \end{minipage}
    \begin{minipage}{0.23\textwidth}
        \centering
        \begin{overpic}[width = 1\textwidth]{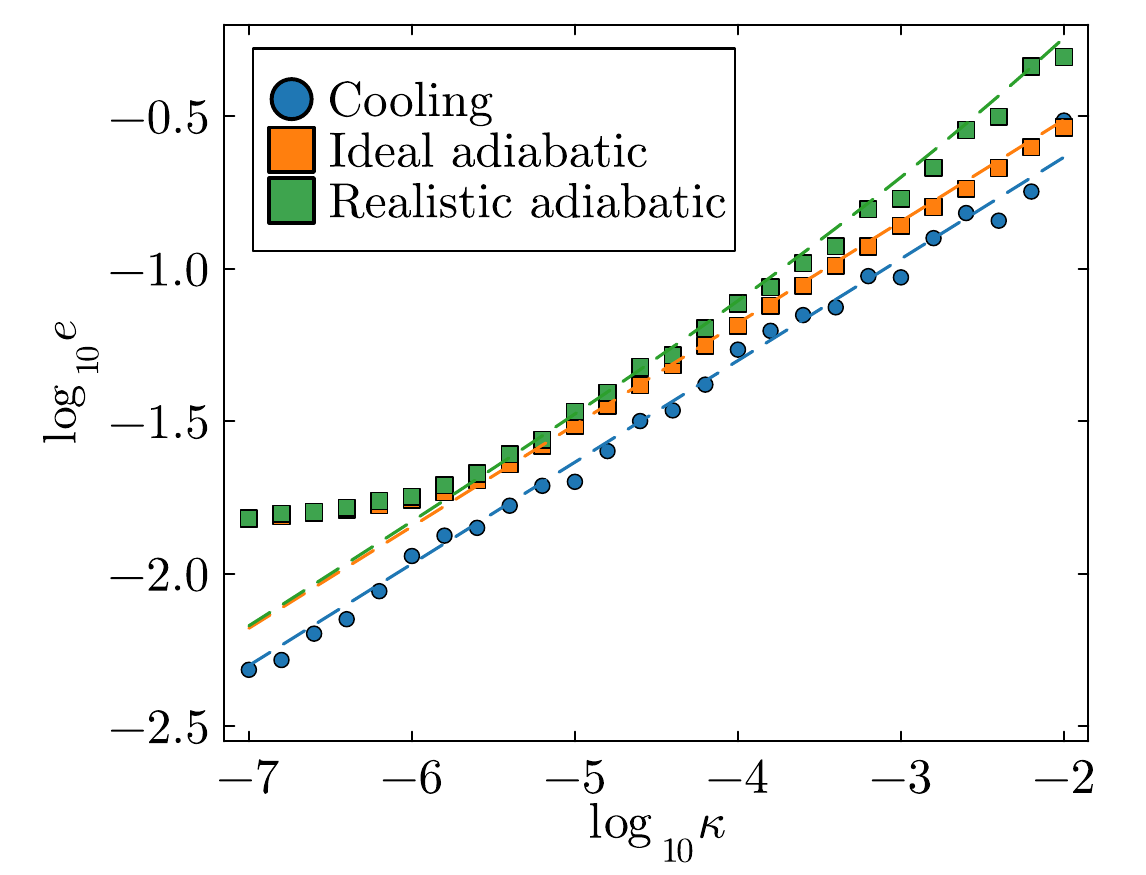}
            \put(-2,70){{(d)}}
            \put(36,80){\small$\pi/2\to\pi/10$}
        \end{overpic}
        \label{fig:pi10_cool_adiab}
    \end{minipage}
    \caption{Optimal relative energy $e_{\text{opt}}$ (log scale) as a function of the noise strength $\kappa$ (log scale) for $N=100$ sites, comparing three protocols: analytic cooling with $gt=0.5$ and $\Delta_{M(m)}=\epsilon_{M(m)}\pm 0.4$ (blue circles), ideal adiabatic with initial state  $\ket{0}^{\otimes N}$ (orange squares), and realistic adiabatic with initial state $\rho^{\rm cool}(\pi/2)$ (green squares). Dashed lines show the analytic predictions from \cref{eq:e_cool_scaling} (blue, $\kappa^{1/3}$) and Appendix~\ref{app:adiabatic_derivations} (orange, green).
        (a)~$\theta_f=\pi/3$ (trivial phase): the adiabatic $\kappa^{2/3}$ scaling gives lower energy than the achievable rate of the cooling algorithm, which is $\kappa^{1/3}$.
        (b)~$\theta_f=\pi/6$ (topological phase): both scale as $\kappa^{1/3}$; the adiabatic algorithm gives slightly lower energy.
        (c)~$\theta_f=\pi/8$: the two algorithms become comparable.
        (d)~$\theta_f=\pi/10$: cooling gives lower energy than both adiabatic protocols.
        In panels (b)--(d), the adiabatic curves flatten at small $\kappa$ due to the unavoidable excitation of the zero-gap mode $k=N/2$.
    }
    \label{fig:cooling_vs_adiabatic_scaling}
\end{figure}
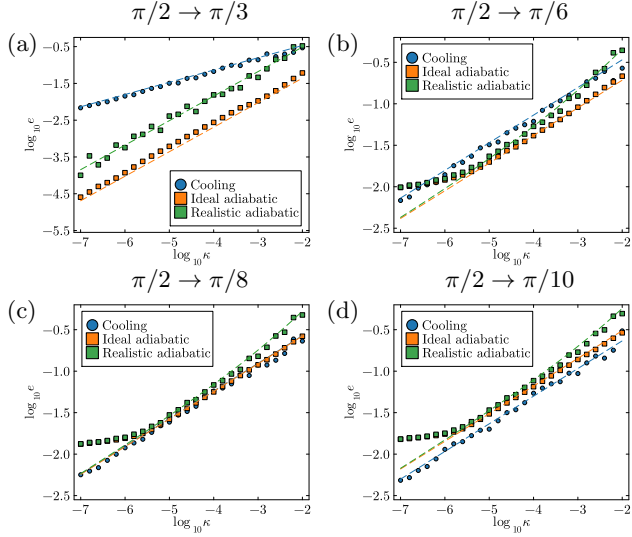

\Cref{fig:cooling_vs_adiabatic_scaling} compares the analytic cooling and adiabatic algorithms across four representative target values of~$\theta_f$, all starting from $\theta_i=\pi/2$ on a system of $N=100$ sites.
For each value of~$\kappa$, both the total adiabatic evolution time~$T$ and the cooling cycle time~$t$ are independently optimized to minimize the relative energy. One of the key findings is that the results are phase-dependent. In the trivial phase [panel~(a), $\theta_f = \pi/3$], the adiabatic $\kappa^{2/3}$ scaling gives lower energy than cooling's $\kappa^{1/3}$ across the entire range of~$\kappa$ shown, because the small prefactor $\tilde{A} \approx \cos^2\theta_f/8$ keeps the adiabatic energy well below the cooling curve.
In the topological phase [panels~(b)--(d)], Landau--Zener excitations near the gap-closing momentum reduce the noiseless convergence to $T^{-1/2}$, so after optimization both algorithms scale as $e \propto \kappa^{1/3}$. The relative performance is then determined by prefactors: as $\theta_f$ decreases deeper into the topological phase, the adiabatic prefactor~$\tilde{B}$ grows while the cooling prefactor stays roughly constant. The ideal adiabatic algorithm gives slightly lower energy at $\theta_f = \pi/6$ [panel~(b)], the two are comparable at $\theta_f = \pi/8$ [panel~(c)], and cooling gives lower energy at $\theta_f = \pi/10$ [panel~(d)].

At small~$\kappa$ in panels~(b)--(d), the adiabatic curves flatten to a noise-independent floor: the mode $k = N/2$ has exactly zero gap at the critical point and is unavoidably excited regardless of sweep time. This floor shifts to smaller~$\kappa$ with increasing~$N$ (\cref{fig:finite_N_crossover}).

\Cref{fig:cooling_vs_adiabatic_heatmaps} maps out the difference $e_{\rm cool}-e_{\rm adiab}$ across the full $(\theta,\kappa)$ plane in the limit $N\gg 1$.
Dark regions indicate that cooling achieves lower energy (``C''); bright regions (``A'') indicate that the adiabatic algorithm does; the black contour marks equal performance.
In the topological phase [panels~(a,\,c)], cooling outperforms the adiabatic algorithm throughout most of the plane. In the trivial phase [panel~(b)], the adiabatic algorithm, with scaling $\kappa^{2/3}$ leads to lower energies. Panels~(c,\,d) show the realistic scenario (initial state $\rho^{\rm cool}(\pi/2)$); the imperfect initialization adds a $\kappa^{2/3}$ contribution that shifts the equal-performance contour in favor of cooling, especially at high noise [panel~(d), $\kappa \gtrsim 10^{-3}$].

These power-law behaviors found analytically rely on the assumption of large-$N$; for sufficiently small $\kappa$, the adiabatic curves can cross over to a different regime due to finite-size effects (see \cref{fig:finite_N_crossover}), so the $\kappa^{2/3}$ and $\kappa^{1/3}$ laws found analytically hold within a finite window $\kappa\in[\kappa_{\min},\kappa_{\max}]$, as shown in Appendix~\ref{app:adiabatic_derivations}.
\begin{figure}
    \centering
    \includegraphics[width=0.7\linewidth]{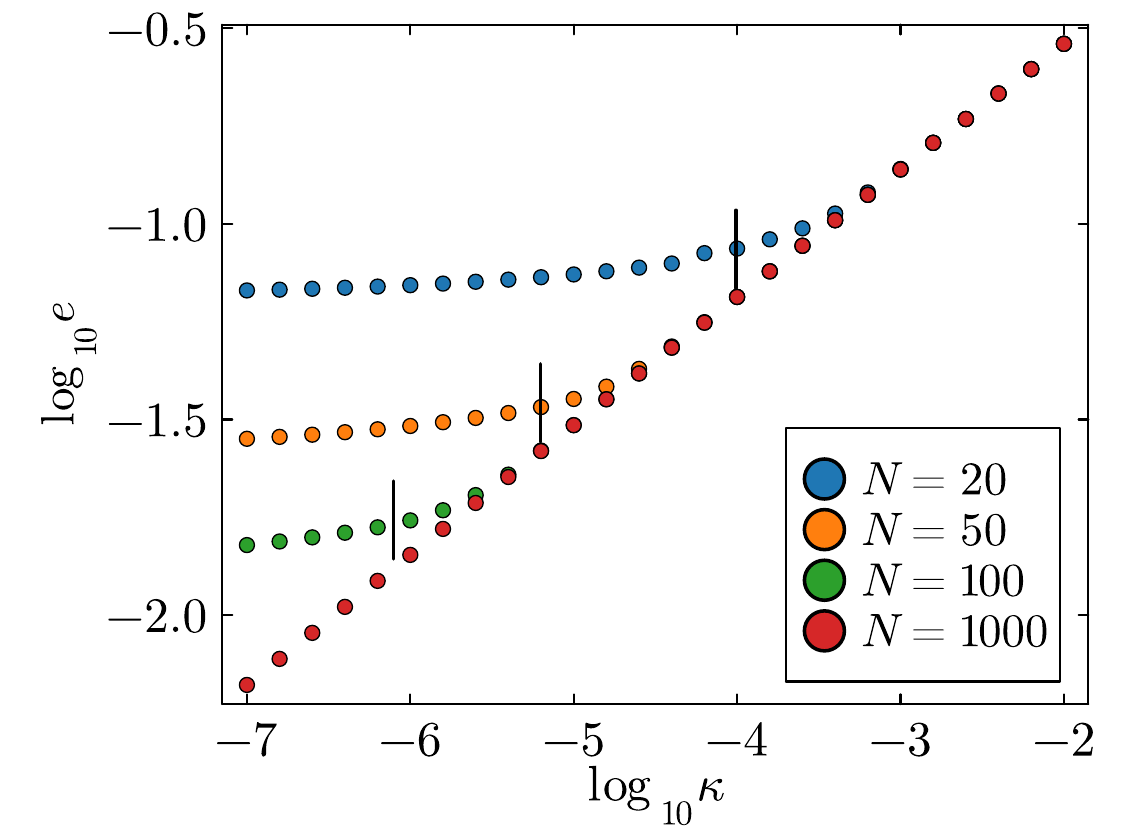}
    \caption{Optimal relative energy $e_{\text{opt}}^{\text{adiab}}$ as a function of the noise rate $\kappa$ for different system sizes $N$ and target $\theta_f=\pi/10$ for the ideal adiabatic algorithm. The smaller the system size, the more prominent the effect of the $k=N/2$ mode is, and the larger the $\kappa_{\rm min}$ where the crossover (predicted by the vertical lines) happens.
        The predicted plateau values of $\log_{10}e$ are determined by the finite-size floor $2\epsilon_{N/2}/|E_{\rm GS}|$.
    }
    \label{fig:finite_N_crossover}
\end{figure}

\section{Numerical optimization: QAOA vs optimized cooling}
\label{sec:numerics_qaoa_vs_cooling}

The analytical expressions for the relative energy as a function of the noise rate, derived in \cref{sec:analytics_cooling_vs_adiabatic}, enable a fair comparison between the multifrequency cooling and adiabatic algorithms. Here, we study how the variational approach, QAOA, performs relative to those state-preparation protocols. We perform a direct numerical optimization over the full parameter space $(\theta,\kappa)$.

For the variational protocols studied in Section VI, i.e. QAOA and optimized cooling, we find that both outperform the particular linear-ramp adiabatic protocol of \cref{eq:adiabatic_schedule}. Although alternative adiabatic paths could yield improved performance, optimizing such paths is beyond the scope of this work. Consequently, the adiabatic algorithm is excluded from the final comparison performed here.

Our numerical comparison proceeds in two steps. First, we compare the relative energies achieved by cooling and QAOA using optimally tuned parameters. Since such precise parameter settings are typically not accessible in experiments, we then analyze the robustness of both approaches under perturbations of the optimized parameters.

Before we begin with the comparison, let us make the following general remark. While QAOA performs well in the trivial phase, it struggles to prepare the ground state near $\theta=0$. The limit $\theta\to 0$ corresponds to a Kitaev/Majorana chain, which realizes a symmetry-protected topological (SPT) phase~\cite{Chen2012SymmetryProtected}. Preparing an SPT state from a trivial product state using a local unitary circuit such as in QAOA requires a circuit depth that scales linearly with~$N$~\cite{Bravyi2006LiebRobinson,Chen2010Local}. Since the QAOA circuits considered here have constant depth in the sense that the number of layers $p$ does not scale with~$N$, and the initial state is close to a product state, QAOA is expected to face difficulties in efficiently preparing a state with high fidelity with respect to the ground state. This result gives some intuition as to what energies we should expect to observe in both phases for QAOA. In contrast to that, the cooling algorithm is non-unitary and thus not subject to these circuit-depth constraints. Since the performance of the cooling algorithm mostly depends on the energy spectrum of the system, and this spectrum is symmetric with respect to $\theta=\pi/4$, we expect cooling to work roughly equally well in both phases.

\subsection{Optimal energies with exact optimal parameters}

This subsection compares the best energies obtained when QAOA and optimized cooling are implemented with their numerically optimized parameters.

\begin{figure}[t]
    \begin{minipage}{0.215\textwidth}
        \centering
        \includegraphics[width = \textwidth]{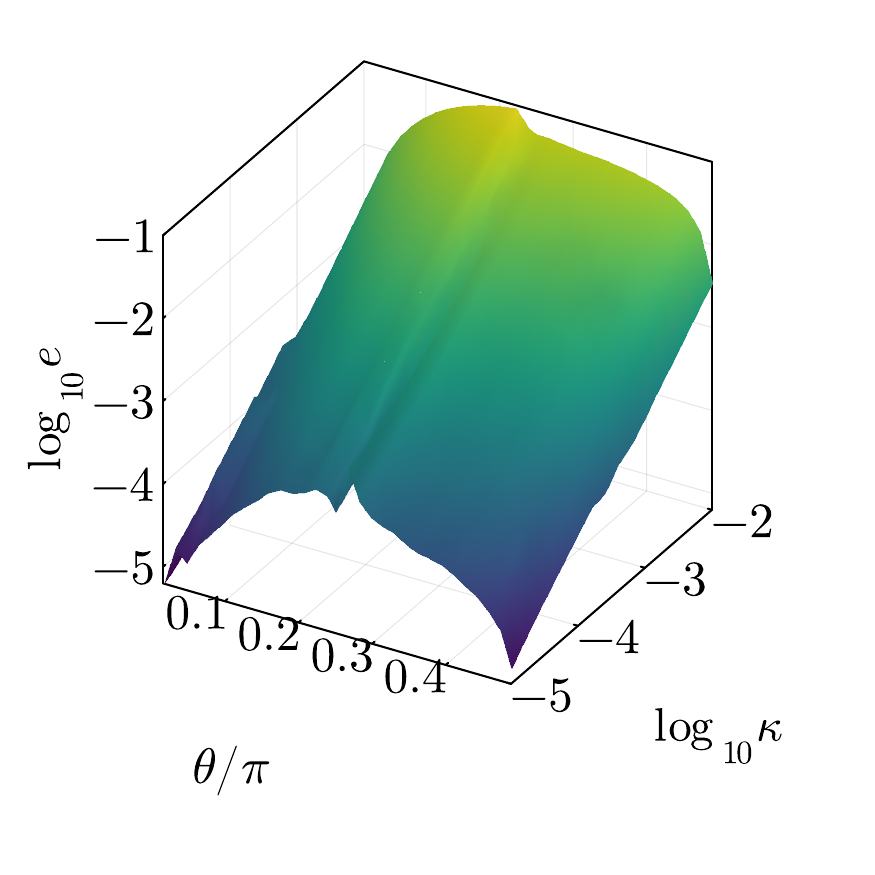}
        \label{fig:left}
    \end{minipage}
    \hfill
    \begin{minipage}{0.26\textwidth}
        \centering
        \includegraphics[width = \textwidth]{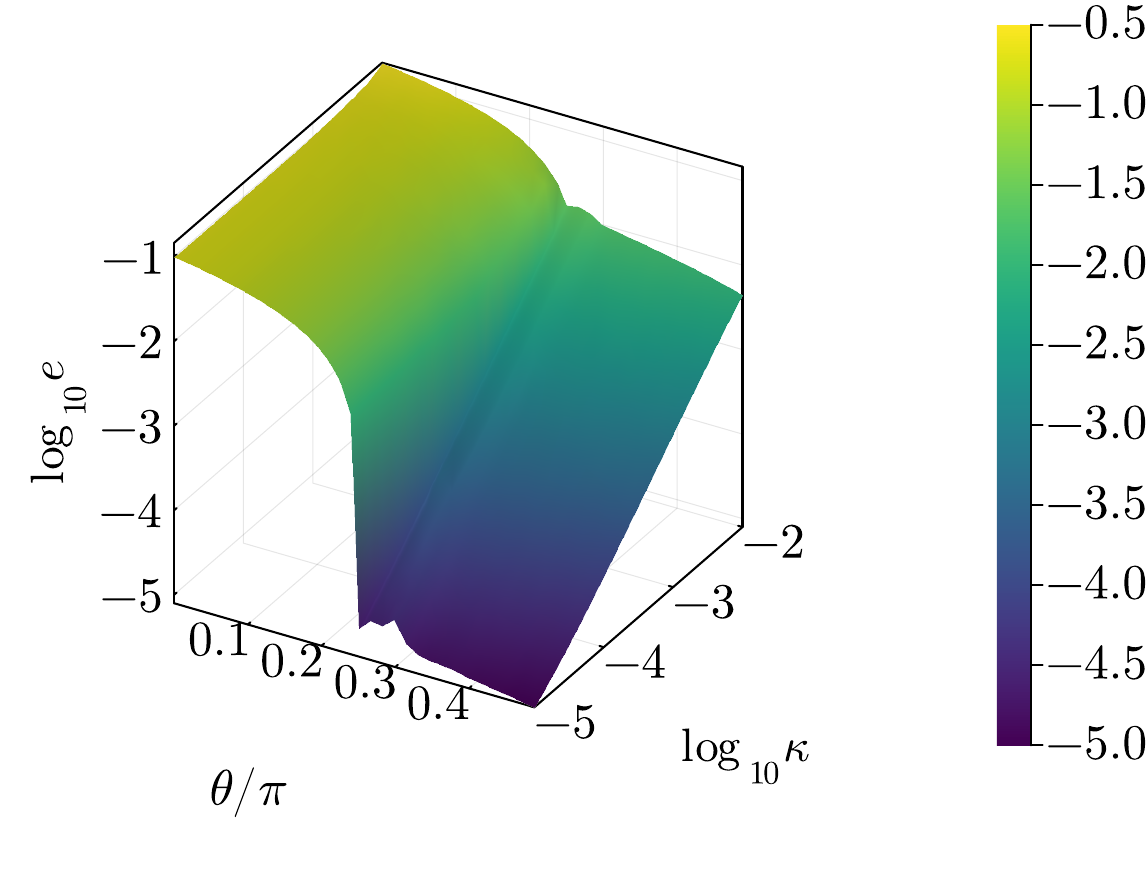}
        \label{fig:right}
    \end{minipage}
    \caption{Optimal relative energies $e_{\text{opt}}^{\text{cool}}$ and $e_{\text{opt}}^{\text{QAOA}}$
        as a function of $\theta/\pi$ and $\log_{10}\kappa$ for $N=20$ with 20 free parameters.
        Left: optimized cooling with coupling range $r_{\rm c}=1.5$ and two cooling cycles.
        Right: QAOA with $p=10$ layers.
        Both algorithms use as initial state $\rho^{\rm cool}(\pi/2)$. The color bar applies to both panels.
        Cooling achieves lower energies in the topological phase ($\theta\lesssim\pi/4$), while QAOA performs comparably or better in the trivial phase ($\theta\gtrsim\pi/4$).
        Panel~(d) of \cref{Fig:QAOA_vs_cooling_robustness} shows the same optimized data as a two-dimensional heatmap.
    }
    \label{fig:qaoa_cooling_surfaceplots}
\end{figure}

We proceed with the comparison as follows. For the cooling protocol, we choose a coupling range $r_{\rm c}$ and allow two different cooling cycles. Each of them is characterized by its own bath frequency~$\Delta$, cycle time~$t$, and coupling parameters $\{\lambda_j,\mu_j\}$, giving a total of $N_{\rm pars}=8(r_{\rm c}+1)$ free parameters.
For QAOA, we use $p=4(r_{\rm c}+1)$ layers, giving the same total number of free parameters, $2p$, as we have two parameters $\beta_l, \gamma_l$ for each layer. For the numerics, we chose $r_{\rm c}=0.5$ and $r_{\rm c}=1.5$, which amounts to 12 and 20 free parameters, respectively.
The half-integer coupling ranges follow the convention of Ref.~\cite{Molpeceres2025Quantum}, where $r_{\rm c} = 0.5$ means only the $j = +1$ neighbor is coupled and not the $j=-1$.

Since the exact ground state $\ket{0}^{\otimes N}$ at $\theta = \pi/2$ cannot be prepared perfectly in the presence of noise, we use the same realistic initial state for both algorithms: the steady state obtained from cooling at $\theta=\pi/2$, denoted $\rho^{\rm cool}(\pi/2)$. Note that this steady state may, in general, differ from the one derived in Appendix~\ref{subapp:few-frequency_corrections_cooling} and utilized before, since the cooling protocol used here is the variationally optimized one rather than the analytic single-frequency scheme. To avoid inconsistencies, we compute this ``realistic initial state'' numerically for each parameter set.

All parameters are optimized using the BFGS algorithm as implemented in the \texttt{Optim.jl} package for Julia.
For each $(\theta,\kappa)$ grid point, we perform three independent optimization runs from random initial parameter vectors and retain the set yielding the lowest energy.

\Cref{fig:qaoa_cooling_surfaceplots} shows the optimal relative energies as three-dimensional surface plots over the $(\theta/\pi,\log_{10}\kappa)$ plane for $N=20$ sites. Both surfaces decrease (i.e., improve) as $\kappa$ decreases and as $\theta$ approaches $\pi/2$. The main difference arises near $\theta=0$ (topological phase). There, QAOA faces limitations, which might be related to the SPT depth constraint mentioned above, whereas cooling attains lower energies. In the trivial phase ($\theta\to\pi/2$), the performance of both algorithms is comparable. Therefore, in the SPT phase, there is a clear advantage of cooling with respect to QAOA.

\subsection{Parameter resilience}
\label{subsec:param_resilience}

In practice, control parameters cannot be implemented exactly. To quantify the sensitivity of each algorithm to such imperfections, we introduce random relative perturbations to the optimized parameters. Specifically, for an optimized parameter vector $\bm{\vartheta}=(\vartheta_1,\dots,\vartheta_M)$, we sample i.i.d.\ variables $u_j\sim\mathrm{Unif}[-\delta,\delta]$ and replace each parameter as $\vartheta_j \mapsto \vartheta_j(1+u_j)$. For each perturbation strength~$\delta$, we compute the resulting relative energy and average over 100 independent realizations.

\Cref{Fig:QAOA_vs_cooling_robustness} shows the difference of the relative energies, $e_{\text{opt}}^{\text{cool}}-e_{\text{opt}}^{\text{QAOA}}$, across the $(\theta,\kappa)$ plane. Panels~(a,\,d) show the comparison with exact optimal parameters ($\delta=0$): QAOA gives lower energies in the trivial phase ($\theta>\pi/4$), while cooling gives lower energy in the topological phase ($\theta<\pi/4$). Panels~(b,\,e) show the same comparison with $2\%$ relative deviations ($\delta=0.02$), and panels~(c,\,f) with $5\%$ deviations ($\delta=0.05$). As the deviations grow, the region in which cooling gives lower energy expands noticeably, demonstrating that the cooling protocol is more robust to parameter imperfections than QAOA.

This greater robustness of cooling can be attributed to its convergence to a fixed point: the steady state is an attractor of the cooling map, so small perturbations in the coupling parameters shift the attractor only slightly. In contrast, QAOA produces a single time-evolved state whose energy is directly sensitive to each parameter value, which does not have any stabilizing effect from a fixed-point structure.

\section{Conclusion and Outlook}
\label{sec:conclusion}

We have developed a unified framework for comparing ground-state preparation algorithms---multifrequency cooling, adiabatic evolution, and QAOA---in the presence of depolarizing noise, applied to an exactly solvable family of quadratic fermionic Hamiltonians.

Our central analytical result is the derivation of scaling laws for the achievable relative energy as a function of the noise rate~$\kappa$ (\cref{tab:scaling_summary}). In the trivial (gapped) phase, the adiabatic algorithm achieves a $\kappa^{2/3}$ scaling which, together with its small prefactor in the gapped regime, yields lower relative energies than the $\kappa^{1/3}$ cooling result in the parameter range studied here. In the topological phase, where the adiabatic path crosses a quantum critical point, both algorithms scale as $\kappa^{1/3}$. However, cooling achieves lower prefactors for the targeted values of $\theta_f$ deep in the topological phase. No single algorithm dominates across the full phase diagram. That is, the optimal strategy is phase dependent.

Numerical optimization of QAOA and numerically optimized cooling algorithms lead to a similar picture. QAOA performs competitively with cooling in the trivial phase but is outperformed in the topological regime. This is consistent with the circuit-depth constraints on preparing symmetry-protected topological states from product states. The cooling protocol also exhibits greater robustness to parameter imperfections (\cref{Fig:QAOA_vs_cooling_robustness}), a practical advantage for realistic implementations.

Several open questions remain. Most importantly, future work should investigate whether the phase-dependent advantage of cooling persists for interacting, non-integrable Hamiltonians, where the structure of the spectrum and the nature of the phase transition change qualitatively. Optimizing the adiabatic schedule beyond the linear ramp (e.g., via counterdiabatic driving~\cite{GueryOdelin2019Shortcuts}) and combining cooling with variational optimization of bath parameters, or using machine learning methods to dynamically optimize the cooling cycle, are additional directions worth exploring.

Let us also mention that our multifrequency cooling strategy shares conceptual similarities with the filtered Lindbladian approach of Zhan et al.~\cite{Zhan2026Rapid}, though the mechanism (cooling vs dissipative state preparation), and also the implementations differ: static couplings with randomized cycle times and frequencies (ours) versus time-dependent couplings (theirs).
A systematic comparison of these complementary approaches in terms of convergence time and time to solution would be valuable, especially in a noisy setting.

On the experimental side, incorporating Trotterization errors~\cite{Childs2021Theory} for digital implementations and gate-level noise models would connect our continuous-time analysis to near-term superconducting~\cite{Mi2024Stable} and trapped-ion~\cite{Barreiro2011Opensystem} platforms. More broadly, the analytical approach developed here---deriving closed-form noise scaling laws and comparing them across quantum phases---provides a template for benchmarking ground-state preparation algorithms beyond the specific model and noise channel studied in this work.

\begin{acknowledgments}
    This research is part of the Munich Quantum Valley, which is supported by the Bavarian state government with funds from the Hightech Agenda Bayern Plus.
    The research is partly funded by THEQUCO as part of the Munich Quantum Valley, which is supported by the Bavarian state government with funds from the Hightech Agenda Bayern Plus.
    The work is partially supported by the Deutsche Forschungsgemeinschaft (DFG, German Research Foundation) under Germany's Excellence Strategy - EXC-2111 - 390814868.
    The work at MPQ is supported from the German Federal Ministry of Education and Research (BMBF) through the funded project ALMANAQC, grant number 13N17236 within the research program ``Quantum Systems''.
    Furthermore, we acknowledge the BMW endowment fund.
\end{acknowledgments}

\FloatBarrier

\clearpage
\appendix

\section{Multifrequency cooling: analytical derivations}
\label{app:analytic_cooling}

In this Appendix, we derive the analytical scaling laws for the multifrequency cooling protocol under depolarizing noise.
We proceed in three stages: we first review the multifrequency cooling algorithms in the weak coupling regime, where we focus on the specific choice of times and frequencies presented in \cref{sec:analytics_cooling_vs_adiabatic} (Appendix~\ref{subapp:weak_coupling_map}); we then derive the steady-state relative energy of each mode in the presence of noise (Appendix~\ref{subapp:energy_steady_state}) and optimize the cycle time to obtain the lowest achievable energy as a function of the noise rate (Appendix~\ref{subapp:optimal_cycle_time}). Finally, in Appendix~\ref{subapp:few-frequency_corrections_cooling}, we treat a single frequency protocol and show that it is advantageous in the edges of the spectrum. We also bridge the region close to the edges of the spectrum with a few-frequency protocol.

\subsection{Weak-coupling and cooling limits}
\label{subapp:weak_coupling_map}

In Section~\ref{sec:analytics_cooling_vs_adiabatic} we provided some analytical expressions for the energy in steady state after the cooling procedure in the presence of noise. Those are valid in the weak coupling and cooling limits, and for low noise rate. Furthermore, they will be obtained after averaging over different realizations of the cooling maps. In this subsection we show that such average faithfully describes the cooling process, and derive the analytical expressions for the steady state in both the weak coupling and cooling limits. In the following subsection, we will also include noise in this derivation.

We consider the map $\mathcal{E}_{k,\tau,\Delta}$ describing a cooling cycle for the density operator in a sector including the modes labeled by $(k,-k)$\footnote{Recall that the evolution of the mode pairs $(k,-k)$ decouples and, therefore, we can consider them separately.} where $\tau$ and $\Delta$ are the cycle time and bath frequency, respectively. They are randomly chosen from flat distributions $\tau\in[0,2t]$ and $\Delta\in[\Delta_m,\Delta_M]$ with
\[
    \Delta_M > \epsilon_M \ge \epsilon_m > \Delta_m>0,
\]
where $\epsilon_{m,M}$ are the minimum and maximum values of the mode energy $\epsilon$. This ensures that all modes are cooled, as described in \cite{Molpeceres2025Quantum}. Note that we have assumed that the Hamiltonian is gapped, i.e.\ $\epsilon_m>0$.

For a given realization we can write the action of the next cooling cycles as a sequence
\begin{equation}
    \rho_{\ell+1}= {\cal E}_{k,\tau_{\ell+1},\Delta_{\ell+1}} (\rho_\ell)
    \label{eq:mapk1}
\end{equation}
where $\rho_\ell$ is the density operator corresponding to the modes with momenta $\pm k$ at cycle
$\ell$, and $\tau_\ell,\Delta_\ell$ the corresponding values chosen for the map. In the following, for the sake of simplicity we will omit all the subscripts whenever they are not needed. We are interested in the state $\rho_\ell$ for $\ell\gg 1$. Although no steady state will be reached by applying all those maps, since at each time step we apply a different random channel, we will demonstrate that the final state (for $\ell\gg 1$) will be close to the steady state of the map that is obtained by averaging ${\cal E}_{k,\tau,\Delta}$ with respect to $\tau$ and $\Delta$, as long as we restrict ourselves to the weak-coupling limit $gt\ll 1$. Furthermore, we will use the cooling limit $\epsilon_m t\gg 1$ in order to obtain simple analytical expressions for that steady state.

We have divided the remainder of this subappendix into five parts. In the first, we set up the method used to compute both the steady state of the averaged map and its standard deviation. From there it will be clear that averaging the map and expanding it in the weak coupling and cooling limits can be performed in any order to approximate the steady state of such a map. In the second, we explicitly compute that steady state in the weak-coupling limit. In the third, we evaluate the standard deviation and show that, in this limit, each individual realization---the state obtained after applying $\ell\gg1$ cooling maps with random $\tau$ and $\Delta$---is close to the stationary state of the averaged map. In Section~\ref{subapp:cooling_limit_expansion} we further expand the resulting state, and the cooling and heating rates, in the cooling limit. Finally, in Section~\ref{subapp:discrete_distrib} we consider a discrete distribution for the values of $\Delta$, which is needed to analyze the dependence of the final energy on the number of frequencies, as discussed in \cref{subsec:analytic_cooling}.

\subsubsection{Averaged map and steady state}

Given the fermion parity superselection rule, all the information about $\rho$ is contained in the following expectation values:
\begin{subequations}
    \label{eq:ndpq}
    \begin{eqnarray}
        n &=&\frac{1}{2}\left(\langle a^\dagger_k a_k\rangle + \langle a^\dagger_{-k} a_{-k}\rangle\right),\\
        d &=&\langle a^\dagger_k a_k\rangle - \langle a^\dagger_{-k} a_{-k}\rangle,\\
        p&=& \langle a^\dagger_k a_{-k}^\dagger \rangle\\
        q&=& \langle a^\dagger_k a_{-k}\rangle.
    \end{eqnarray}
\end{subequations}
We can define a vector $P=(n,d,p,p^\ast,q,q^\ast)$, so that the sequence in \cref{eq:mapk1} can be equally written as
\begin{equation}
    P_{\ell+1} = X_{\ell+1} P_\ell + Y_{\ell+1}
    \label{eq:mapk2}
\end{equation}
where $X_\ell,Y_\ell$ are a matrix and a vector, respectively, which depend on both $\tau_\ell$ and $\Delta_\ell$.

We are interested in $P_\ell$ in the limit $\ell\to \infty$, since this contains the information about
the final energy and therefore about the success of the cooling process. For each realization, we
will obtain a different result since it contains different choices of the cycle times and bath
frequencies, and thus $P_\ell$ follows a distribution itself. In order to characterize it, we may
consider its mean value and the standard deviation (or, more generally, the covariance matrix).
In case the latter is small, the cooling will be
well characterized by the first. In the next subsection we will show that the standard deviation is ${\cal O}((gt)^2)$, so
that in the weak-coupling limit we can confidently extract the conclusions about the success of the
cooling process from the averaged values.

We define $\bar P_\ell$ as the vector obtained when averaging over $P_\ell$ with respect to all previous choices of $\tau$ and
$\Delta$, following the distributions mentioned above. Then we have,
\begin{equation}
    \bar P_{\ell+1} = \bar X \bar P_\ell + \bar Y
    \label{eq:mapk3}
\end{equation}
where the bar indicates the average with respect to the $\ell+1$ random values of $\tau$ and $\Delta$
chosen in the $\ell+1$ different steps. Note that the average of $X_{\ell+1} P_\ell$ can be written as the
product of averages since the first only depends on the choice of $\tau$ and $\Delta$ in step
$\ell+1$, while the second depends on all previous steps. In the limit  $\ell\to \infty$, we obtain
\begin{equation}
    \bar P = \lim_{\ell\to\infty}\bar P_\ell=  (\Id - \bar X )^{-1} \bar Y
    \label{eq:mapk4}
\end{equation}

In order to compute the standard deviation of the distribution, we consider
\begin{equation}
    P_{\ell+1}\otimes  P_{\ell+1} = (X_{\ell+1} P_\ell + Y_{\ell+1})\otimes (X_{\ell+1} P_\ell + Y_{\ell+1})
\end{equation}
Then, we average with respect to all realizations, and take the limit $\ell\to \infty$ to obtain
\begin{equation}
    Q= \overline{P\otimes  P}- \bar P\otimes \bar P
\end{equation}
which contains all the information about the corresponding covariance matrix. Once we obtain the results for $\bar P$ and $Q$, we can perform the expansions in the weak coupling and cooling limit, which involve expanding $\bar X$, $\overline{X\otimes X}$, etc.\ in terms of $gt$ and $\epsilon_m t$. Crucially it holds that, for all those quantities, the average of the expansion coincides with the expansion of the average. Thus, we can first do the expansion in $gt$, then take the average, and finally the expansion in $\epsilon_m t$. Indeed, we will do that in the subsequent  subsections since this heavily simplifies the derivations.

\subsubsection{Average in the weak-coupling limit}

We will first expand the map ${\cal E}_{k,\tau,\Delta}$ up to second order in $\eta=gt$ and then average it with respect to $\tau$ and $\Delta$. Both steps were carried out in~\cite{Molpeceres2025Quantum}, so we will take some of the results from there. Note, however, that in that reference we averaged over a discrete distribution in $\Delta$, while here we will consider a continuous distribution in order to get simpler formulas. At the end of this section we will also consider a discrete distribution in which the discretization depends on $t$, since most of the computations in the main text have been carried out for such a distribution.

In the weak coupling limit, $\eta^2=(gt)^2 \ll 1$ we have that the cooling map can be transformed via a Dyson expansion into the following Lindbladian form~\cite{Molpeceres2025Quantum}:
\begin{align}
    \mathcal{E}(\rho) &= e^{-i h_S \tau} \left(\rho + \eta^2 (\mathcal{L}_{l_1}+\mathcal{L}_{l_2})(\rho)\right) e^{i h_S \tau}+\mathcal{O}(\eta^4).
\end{align}
Here, $\mathcal{L}_{O}$ denotes the Lindblad superoperator $\mathcal{L}_{O}(\rho)=O\rho O^\dag-\frac{1}{2}\{O^\dag O,\rho\}$, and
\begin{align}
    h_S &= \epsilon(\hat{a}_k^\dag\hat{a}_k+\hat{a}_{-k}^\dag\hat{a}_{-k}),\\
    l_1 &= x\hat{a}_k-iy\hat{a}_{-k}^\dag,\\
    l_2 &= x\hat{a}_{-k}-iy\hat{a}_{k}^\dag,
\end{align}
with $\epsilon$ the energy of the fermionic modes and coefficients $x$ and $y$ given by
\begin{align}
    x &= \frac{1}{\tau}\int_0^\tau e^{i(\epsilon-\Delta)\tau'} \dd\tau' = \frac{1 - e^{i (\Delta-\epsilon) \tau}}{i (\Delta-\epsilon)\tau}, \label{eq:xT_def:app}\\
    y &= -\frac{1}{\tau}\int_0^\tau e^{i(\epsilon+\Delta)\tau'} \dd\tau' = \frac{1 - e^{i (\Delta+\epsilon) \tau}}{i (\Delta+\epsilon)\tau}. \label{eq:yT_def:app}
\end{align}

We have that $|x|,|y|\leq 1$, since $\tau\in[0,2t]$ and the bound is reached for the resonances $(\Delta\pm\epsilon)\tau\ll1$. As explained in \cite{Molpeceres2025Quantum}, the interpretation of this map is that apart from the system evolution, $e^{-i h_S \tau}$, we have dissipative terms resulting from the weak coupling to the bath.

We can now use this equation to obtain the change of the parameters in \cref{eq:ndpq} in one cooling cycle,
\begin{subequations}
    \label{eq:ndpq_gt}
    \begin{eqnarray}
        n_{\ell +1} &=&  n_{\ell} + \eta^2 \Big[|y|^2 -  |z|^2 n_{\ell}\nonumber\\
            &&+ i(x^\ast y-x y^\ast) (p_\ell +p_\ell^\ast) \Big],\\
        d_{\ell +1} &=&  (1-2 \eta^2 |z|^2) d_{\ell},\\
        p_{\ell +1} &=&  e^{2i\epsilon\tau} \left[(1-2 \eta^2 |z|^2) p_{\ell}+2i\eta^2 xy^\ast d_\ell\right],\\
        q_{\ell +1} &=&  (1-2 \eta^2 |z|^2) q_{\ell} ,
    \end{eqnarray}
\end{subequations}
where we have defined $|z|^2=|x|^2+|y|^2$, and we have dropped the subscript $\ell$ in $x$, $y$ and $z$. From these equations, we can easily retrieve the matrix $X$ and the vector $Y$ defined in \cref{eq:mapk2} and then compute the corresponding averages in the limit $\ell\to \infty$. However, before doing that, note that for any realization we have that $|d_\ell|,|p_\ell|,|q_\ell|\to 0$ in that limit \footnote{ The reason for that is that both coefficients in front of $d_\ell$ and $q_\ell$ are smaller than 1 (for all but possibly a zero measure set of choices of the pairs $(\tau, \Delta)$). Hence, in the limit $\ell \to \infty$ both, $d_\ell$ and $q_\ell$, will vanish. Setting $d_\ell=0$ in the equation for $p_{\ell+1}$ we also have that the absolute value of $p$ decreases to $0$ and therefore $p_\ell$ vanishes too. }. Since we are neglecting terms ${\cal O}(\eta^4)$, we conclude that all those quantities are of that order in the limit $\ell\to\infty$. In particular, we can replace those values of $p_{\ell}$ in the equation for $n_{\ell+1}$ so that for sufficiently large $\ell$ we have
\begin{equation}
    n_{\ell +1} =  n_{\ell} + \eta^2 \left[|y|^2 - |z|^2 n_{\ell}\right]+{\cal O}(\eta^4),
    \label{eq:nl+1_no_avg}
\end{equation}
We can now average \cref{eq:nl+1_no_avg} with respect to $\tau$ and $\Delta$, obtaining
\begin{equation}
    \bar n_{\ell +1} - \bar n_{\ell} = - (\gamma^{\rm c}+\gamma^{\rm h}) \bar n_{\ell} + \gamma^{\rm h} +{\cal O}(\eta^4),
    \label{eq:nl+1}
\end{equation}
where
\begin{equation}
    \gamma^{\rm c,h} =  \frac{g^2}{\Delta_M-\Delta_m} \left[F(\Delta_M\mp \epsilon,t)-F(\Delta_m\mp \epsilon,t)\right],
    \label{eq:gammach}
\end{equation}
with
\[
    F(\epsilon,t)=\frac{\sin(2\epsilon t)}
    {2\epsilon^2t}+\frac{\cos\!\left(2\epsilon t\right)-2}
    {\epsilon} +2t\,\mathrm{Si}\!\left(2\epsilon t\right)
\]
and $\mathrm{Si}(y)$ is the sine integral function. We can thus write
\begin{equation}
    \bar n_k =\lim_{\ell\to\infty} \bar n_\ell = \frac{\gamma^{\rm h}}{\gamma^{\rm h}+\gamma^{\rm c}},
\end{equation}
which coincides with the averaged map obtained in \cite{Molpeceres2025Quantum}.
Note that, $\bar n$ together with $d,p,q=0$ contains all the information about the final state (for the average) in the limit $\ell \to \infty$. Moreover, we have $\langle a_k^\dagger a_k \rangle =\langle a_{-k}^\dagger a_{-k} \rangle$ (since $d=0$), which implies in particular that the final state has a $k \leftrightarrow -k$ symmetry and, clearly, that there is a unique stationary state for the average map.
From the averaged rates we can also obtain the steady-state energy
\begin{equation}
    \label{eq:steady_state_energy}
    E_{\infty}=-\epsilon(1-2\bar{n})=\epsilon\frac{-{\gamma}^{\rm c}+{\gamma}^{\rm h}}{{\gamma}^{\rm c}+{\gamma}^{\rm h}}+\mathcal{O}(g^2t^2).
\end{equation}
This result coincides with the one found in \cite{Molpeceres2025Quantum}, but now we show its validity for any chosen distribution of times and frequencies.

\subsubsection{Standard Deviation}

We can also compute the standard deviation of the distribution of $n_\ell$ by squaring \cref{eq:nl+1}, obtaining
\begin{equation}
    n_{\ell +1}^2 =  n_{\ell}^2 + 2 \eta^2 \left[|y|^2 - |z|^2 n_{\ell}\right] n_\ell+{\cal O}(\eta^4),
    \label{eq:nl+1squared}
\end{equation}
Averaging over realizations and taking the limit $\ell\to \infty$ we obtain the following variance:
\begin{equation}
    \lim_{\ell\to\infty} \overline{n_{\ell}^2}-\overline{n_\ell}^2 =  {\cal O}(\eta^2).
\end{equation}
Note that the same is true of the other variables in \cref{eq:ndpq_gt}.
Thus, as announced, in the weak-coupling limit each realization, i.e.\ the state obtained after applying $\ell \gg1$ random cooling maps, will converge to a state close to the one obtained with the averaged map.

We emphasize that, independent of the weak-coupling limit, the standard deviation is upper bounded by $\bar n$, since $0\le n_\ell\le 1$ \footnote{This follows trivially from $\overline{n_{\ell}^2}-\overline{n_\ell}^2\leq \overline{n_{\ell}^2} \leq \overline{n_{\ell}}$} . Thus, $\bar n\ll 1$ is sufficient in order to ensure that cooling will take place for most of the realizations. Additionally, since the average map converges to the steady state in constant time, each of the realizations will also provide cooling in constant time.

\subsubsection{Cooling limit}
\label{subapp:cooling_limit_expansion}

We consider now the {\em cooling limit}, defined as $\epsilon_m t\gg 1$. Let us first concentrate on the heating rate, $\gamma^{\rm h}$. Since $|\Delta_{m,M}+\epsilon_M| t \ge \epsilon_m t\gg 1$, we can expand \cref{eq:gammach} and obtain
\begin{equation}
    \label{eq:gamma_h_continuous}
    \gamma^{\rm h} = \frac{2g^2} {(\Delta_M+\epsilon)(\Delta_m+\epsilon)} +{\cal O}\!\left(t^{-2}\right).
\end{equation}
Here, we have taken into account that the arguments of $F$ are both positive. For $\gamma^{\rm c}$, however, one is positive and the other negative. Assuming $(\Delta_M-\epsilon)t, (\epsilon-\Delta_m)t \gg 1$ we obtain
\begin{equation}
    \label{eq:gamma_c_continuous}
    \gamma^{\rm c} = \frac{2\pi g^2 t}{\Delta_M-\Delta_m} +{\cal O}\!\left(1\right).
\end{equation}
Note that the conditions above are implied by the cooling limit if we choose
\begin{equation}
    \label{eq:Delta_limits}
    \Delta_m=\epsilon_m/2, \quad \Delta_M=\epsilon_M+\epsilon_m/2.
\end{equation}
With all that, we obtain the following simple expression for $\bar n$ and therefore for the final state,
\begin{equation}
    \label{eq:barn}
    \bar n =
    \frac{2(\Delta_M-\Delta_m)} {\pi\,t\,(\Delta_M+\epsilon)(\Delta_m+\epsilon)}.
\end{equation}

With these results, we can also compute the time required to cool the system. For that, we can use Eq. (\ref{eq:nl+1}) combined with the cooling limit derived here. In case we want to obtain a value of $n_\ell = O(\delta)$, where $\delta$ is a small parameter, we need to choose, according to (\ref{eq:barn}), $t=O(1/\delta)$. Furthermore, the total time required to reach that values is $Mt$, where $M$ is the number of cooling steps. In order to obtain that, we define $m_\ell=n_\ell - \bar n$, which fulfills the recursion equation $m_{\ell +1}= (1-\gamma^c-\gamma^h) m_\ell$. Thus imposing that $m_\ell=O(\delta)$ we need $M=O(\log(\delta^{-1})/\gamma^c)$. Furthermore, since $\gamma^c=O(g^2 t)$, and $g=O(1/t)$ in order to respect the weak coupling limit, we have that the total time
\begin{equation}
    T = O(M t) = O ( t^2 \log(\delta^{-1})) = O (\delta^{-2}\log(\delta^{-1})).
\end{equation}
In case we are interested in reaching a fidelity $F=1-O(\delta_F)$ we need to choose $\delta=\delta_F/N$, so that the total number of excitations is $\mathcal{O}(\delta_F)$. Thus, the total time scales as $T_F =  O (N^2\delta_F^{-2}\log(N/\delta_F))$. This improves over the value obtained in \cite{Molpeceres2025Quantum}.

\subsubsection{Discrete distribution}
\label{subapp:discrete_distrib}

\begin{figure}[ht!]
    \begin{minipage}{0.43\textwidth}
        \centering
        \begin{overpic}[width = 1\textwidth]{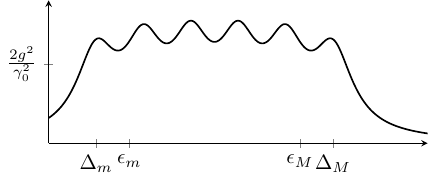}
            \put(1,40){{(a)}}
        \end{overpic}
        \label{fig:freqsum1}
    \end{minipage}
    \begin{minipage}{0.43\textwidth}
        \centering
        \begin{overpic}[width = 1\textwidth]{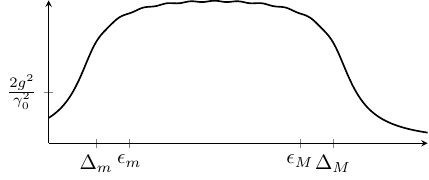}
            \put(1,40){{(b)}}
        \end{overpic}
        \label{fig:freqsum2}
    \end{minipage}
    \begin{minipage}{0.43\textwidth}
        \centering
        \begin{overpic}[width = 1\textwidth]{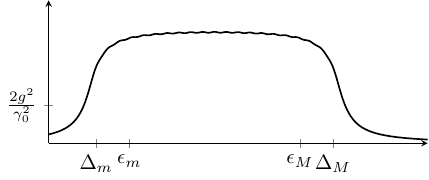}
            \put(1,40){{(c)}}
        \end{overpic}
        \label{fig:freqsum3}
    \end{minipage}
    \caption{Rescaled averaged cooling rate $R{\gamma}^{\rm c}_k$ versus mode energy $\epsilon_k$ for three configurations of bath frequencies. The interval $[\epsilon_m, \epsilon_M]$ marks the system's spectral range; $[\Delta_m, \Delta_M]$ marks the bath frequency range.
        (a)~Too few frequencies: pronounced peaks and valleys lead to strongly non-uniform cooling.
        (b)~Sufficiently many frequencies but short cycle times (broad linewidth $\gamma_0$): resonance peaks extend well beyond the spectrum.
        (c)~Many frequencies with long cycle times (narrow linewidth): the averaged rate is approximately flat across $[\epsilon_m, \epsilon_M]$ and decays sharply outside, yielding the desired uniform cooling profile.}
    \label{fig:multifreq_sums}
\end{figure}

\begin{figure}[ht]
    \includegraphics[width=0.7\linewidth]{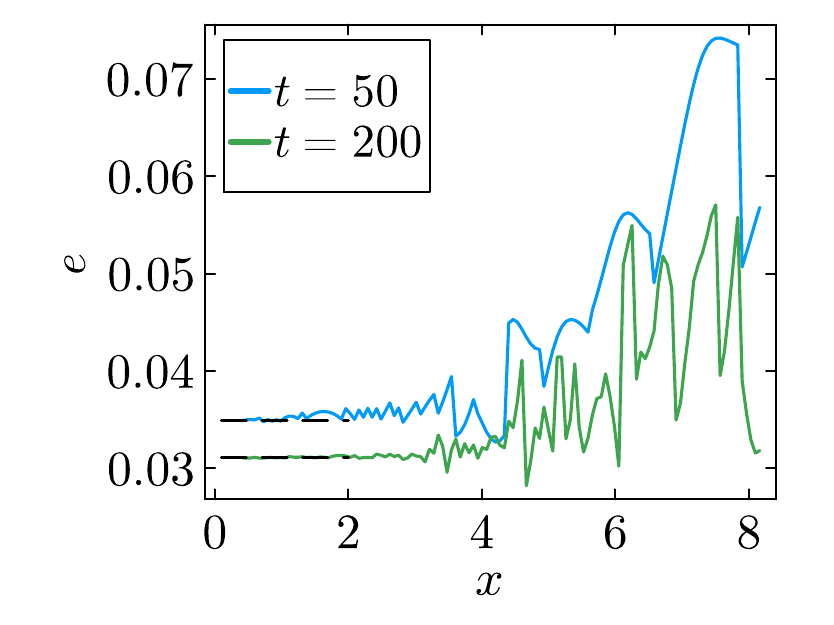}
    \caption{Relative energy~$e$ as a function of the spacing parameter~$\mathrm{x}$ for cycle times $t\in\{50,200\}$, computed by exact numerical simulation ($N=20$, $g=10^{-3}$, $\kappa=0.1g^2$, $\theta=\pi/8$, $\Delta_{M(m)}=\epsilon_{M(m)}\pm 0.1$).
        For $x\lesssim 1$, all curves reach a plateau consistent with the prediction of \cref{eq:ek_x_approx}, represented by the black dashed lines in the figure.
        The jagged features at larger~$x$ arise because the number of frequencies $R$ changes discretely: each unit increase in~$R$ produces a jump in the averaged cooling rate.
    }
    \label{fig:lowx_check}
\end{figure}

We now consider a discrete distribution for $\Delta$ composed of $R\propto t$ frequencies, equally distributed in the interval $[\Delta_m,\Delta_M]$ with a spacing $\propto t^{-1}$. We have $\Delta_{M(m)}=\epsilon_{M(m)}+(-)\delta$, with $\delta$ a constant. In the terminology used in the main text, the regime $R\gg1$ is the many-frequency regime; in the limit $R\to\infty$ the discrete sum approaches the continuum case, derived above.
We proceed as follows. Firstly, we average with respect to time. This average can then be approximated by a Lorentzian function, yielding
\begin{equation}
    \label{eq:ratesAppA}
    \gamma^{\rm c}_\Delta=\frac{2g^2}{(\Delta-\epsilon)^2+\gamma_0^2},\quad \gamma^{\rm h}_\Delta=\frac{2g^2}{(\Delta+\epsilon)^2+\gamma_0^2},
\end{equation}
where  $2\gamma_0=\sqrt{6}/t$ is the effective linewidth. In the expression for $\gamma^{\rm h}_\Delta$, this linewidth term can be neglected as we do not consider negative frequencies, thus $(\Delta+\epsilon)t\gg1$ due to the cooling limit and $\Delta+\epsilon\gg\gamma_0$. Choosing the spacing $\mathrm{x}\gamma_0$, we have
\begin{equation}
    R=\frac{\Delta_M-\Delta_m}{\mathrm{x}\gamma_0},
\end{equation}
and we find bounds for our averaged cooling rate at any given $k$:
\begin{align}
    {\gamma}^{\rm c}_{k_1}        &=\frac{2g^2}{R\gamma_0^2}\left(1+2\sum_{n=1}^\infty\frac{1}{n^2\mathrm{x}^2+1}\right)\nonumber\\
                                  &=\frac{2\sqrt{6}\pi g^2t}{3(\Delta_M-\Delta_m)}\coth\left(\frac{\pi}{\mathrm{x}}\right),\\
    {\gamma}^{\rm c}_{k_2}        &=\frac{2g^2}{R\gamma_0^2}2\sum_{n=1}^\infty\frac{1}{(n-\frac{1}{2})^2\mathrm{x}^2+1}\nonumber\\
                                  &=\frac{2\sqrt{6}\pi g^2t}{3(\Delta_M-\Delta_m)}\tanh\left(\frac{\pi}{\mathrm{x}}\right),\\
    \min\{{\gamma}^{\rm c}_{k_1}, &{\gamma}^{\rm c}_{k_2}  \}\leq\gamma^{\rm c}_{k}\leq \max\{{\gamma}^{\rm c}_{k_1},{\gamma}^{\rm c}_{k_2}  \}=\mathcal{O}(g^2t),
    \label{eq:bounds_gammac}
\end{align}
where $k_1$ is a mode located exactly in resonance with some $\Delta_i$ and $k_2$ is located exactly between two resonance peaks. The difference ${\gamma}^{\rm c}_{k_1}-{\gamma}^{\rm c}_{k_2}$ tends to zero for $\mathrm{x}/\pi\ll 1$: since $\coth(y)-\tanh(y) = 2/\sinh(2y)$, the two rates become equal for $\mathrm{x}\to 0$. Therefore, choosing $\mathrm{x}/\pi\ll 1$ makes the bounds in \eqref{eq:bounds_gammac} equal and ${\gamma}^{\rm c}_k$ independent of $k$ (see \cref{fig:multifreq_sums,fig:lowx_check}). We choose $\mathrm{x}=1$ as it fulfills this approximation while at the same time keeping the number of frequencies $R$ small, yielding
\begin{equation}
    \gamma^c\approx\frac{2\sqrt{6}\pi g^2t}{3(\Delta_M-\Delta_m)}.
    \label{eq:noiseless_cooling_rate}
\end{equation}

Note that one should obtain the same result as in \cref{eq:gamma_c_continuous}, found in the continuum limit, but the prefactor is slightly different due to the Lorentzian approximation done in \cref{eq:ratesAppA}. For the averaged heating rate [Eq. \eqref{eq:gamma_h_continuous}], the result does depend on $k$, but can be upper-bounded by ${\gamma}^{\rm h}_{-N/2}$ or roughly approximated by ${\gamma}^{\rm h}_{-N/4}$.

Let us make a few remarks. Firstly, we have chosen  $\Delta_{M,m}$ such that the modes at the edges of the spectrum can still be treated as modes in the middle of the spectrum. Choosing $\Delta_{M(m)}=\epsilon_{M(m)}+(-)\delta$, as we do for the discrete distribution, we require $\delta t\gg 1$ to remain in the cooling limit and for the analytical results to hold. However, it can be favorable to choose less frequencies, even if the edges of the spectrum are cooled unevenly and $\delta t\gg 1$ is not fulfilled. This is the case, for example, in the presence of noise, where unneeded frequencies are detrimental.

We also remark that cooling will become less effective near the phase transition. This is due to two crucial details. The first is that the choice of $\Delta_{M,m}$ needs to be such that the edges of the spectrum $\epsilon_{M,m}$ are surrounded by resonant frequencies on both sides. Near the phase transition, where $\epsilon_m\rightarrow 0$, this cannot be achieved, since adding frequencies below $\epsilon_m$ (possibly negative) would imply reheating other modes. The cooling limit $\epsilon_m t\gg1$ also indicates that cycle time must be inversely proportional to the gap, which will ask for an unfeasible $t$.

Finally, as mentioned before, a similar flat-rate profile was obtained in Ref.~\cite{Lloyd2025Quantum}. However, in contrast to the choice of random cycle times and multiple frequencies proposed here, the authors of Ref.~\cite{Lloyd2025Quantum} achieve a similar profile with time-dependent couplings.

\subsection{Steady-state energy in the low-noise limit}
\label{subapp:energy_steady_state}

We now include depolarizing noise and derive the steady-state relative energy of mode~$k$ in the low-noise limit $\kappa t \ll 1$. In the main text the noise is written as a channel~[\cref{eq:Enoise_explicit}], equivalently generated by the depolarizing master equation~[\cref{eq:LE_explicit}]. Acting on a mode pair during one cooling subcycle of duration~$\tau$ it reads $\mathcal{D}_\tau(\rho)=e^{-2\kappa\tau}\rho+(1-e^{-2\kappa\tau})\tfrac{\mathbb{1}}{4}$ and drives the average occupation towards that of the maximally mixed state, $\bar n=1/2$, as $\bar n(\tau)=e^{-2\kappa\tau}\bar n(0)+(1-e^{-2\kappa\tau})/2$, i.e.\ $\dot{\bar n}=-2\kappa(\bar n-\tfrac12)$. In the low-noise limit $\kappa\tau\ll1$ we expand to first order in $\kappa\tau$ and, since the depolarizing generator commutes with both the coherent evolution and the cooling Lindbladian, add this contribution directly to the single-realization recursion~[\cref{eq:nl+1_no_avg}],
\begin{equation}
    n_{\ell+1}=n_{\ell}+\eta^2\!\left(|y|^2-|z|^2 n_{\ell}\right)-2\kappa\tau\!\left(n_{\ell}-\tfrac12\right)
    \label{eq:nl+1_noisy_no_avg}
\end{equation}
with corrections of order ${\cal O}[\eta^4,/\kappa t)^2,\eta^2\kappa]$. Averaging with respect to $\tau$ and $\Delta$ (recall $\langle\tau\rangle=t$) we obtain
\begin{align}
    \bar n_{\ell+1}-\bar n_{\ell}
     &=-(\gamma^{\rm c}+\gamma^{\rm h})\bar n_{\ell}+\gamma^{\rm h}-2\kappa t(\bar n_{\ell}-\tfrac12)\nonumber\\
     &=-(\gamma^{\rm c}+\gamma^{\rm h}+2\kappa t)\bar n_{\ell}+(\gamma^{\rm h}+\kappa t).
\end{align}
This is precisely the noiseless recursion~[\cref{eq:nl+1}] with the replacement $\gamma^{\rm c,h}\to\gamma^{\rm c,h}+\kappa t$, which establishes
\begin{equation}
    \label{eq:noisy_rates}
    \gamma^{c,h}_{\rm noisy}=\gamma^{c,h}+\kappa t.
\end{equation}

We can now combine these noisy rates with the results of the previous subsection to obtain the steady-state relative energy of mode~$k$. From the relative energies of the individual modes one obtains the total relative energy, $e=(\sum_k \epsilon_k  e_k)/\sum_k \epsilon_k$, as a function of the cycle time~$t$ and the noise rate~$\kappa$; the optimization of $e$ over $t$ for any given $\kappa$ is carried out in Appendix~\ref{subapp:optimal_cycle_time}.

The relative energy of mode $k$ is defined as
\begin{equation}
    e_k = \left|\frac{E_k - E_{k,\mathrm{GS}}}{E_{k,\mathrm{GS}}}\right|,
    \label{eq:ek_def_app}
\end{equation}
where $E_{k,\mathrm{GS}}=-\epsilon_k$ is the ground-state energy of the mode. Inserting the noiseless steady-state energy of the averaged map~[\cref{eq:steady_state_energy}]  gives
\begin{equation}
    e_k = 2\bar{n}_k+\mathcal{O}(g^2t^2),
    \qquad
    \bar{n}_k = \frac{{\gamma}^{\rm h}_k}{{\gamma}^{\rm h}_k+{\gamma}^{\rm c}_k}.
    \label{eq:ek_2nbar}
\end{equation}
Replacing ${\gamma}^{\rm c,h}_k\to{\gamma}^{\rm c,h}_k+\kappa t$ according to \cref{eq:noisy_rates} gives
\begin{equation}
    e_k =\frac{2{\gamma}^{\rm h}_{k}+2\kappa t}{{\gamma}^{\rm h}_{k}+{\gamma}^{\rm c}_{k}+2\kappa t},
    \label{eq:ek_noisy_rates_app}
\end{equation}
which is the expression quoted in the main text [cf.\ \cref{eq:ek_noisy_cooling}]. Now, we can choose any distribution of frequencies and calculate the rates.

\subsubsection{Discrete distribution}

We now insert the explicit averaged heating and cooling rates derived in the weak-coupling and cooling limits~[\cref{eq:gamma_h_continuous,eq:noiseless_cooling_rate}] for the discrete distribution (see \cref{subapp:discrete_distrib}). Dividing the numerator and denominator of \cref{eq:ek_noisy_rates_app} by $2g^2$, we obtain
\begin{align}
    e_k &=\frac{\frac{2}{(\epsilon_k+\Delta_m)(\epsilon_k+\Delta_M)}+\kappa t/g^2}{\frac{1}{(\epsilon_k+\Delta_m)(\epsilon_k+\Delta_M)}+\frac{\sqrt{2}\pi t}{\sqrt{3}(\Delta_M-\Delta_m)}+\kappa t/g^2}.
    \label{eq:ek_x_approx}
\end{align}

To make the structure of \cref{eq:ek_x_approx} transparent, we collect the three contributions into the coefficients
\begin{align}
    A_k &=\frac{1}{(\epsilon_k+\Delta_m)(\epsilon_k+\Delta_M)},\nonumber\\
    B   &=\frac{\kappa}{g^2},\quad C=\frac{\sqrt{2}\pi }{\sqrt{3}(\Delta_M-\Delta_m)},
    \label{eq:ABC_definitions}
\end{align}
where $A_k$ originates from the heating rate~[\cref{eq:gamma_h_continuous}], $Ct$ from the cooling rate~[\cref{eq:noiseless_cooling_rate}], and $Bt$ from the noise term~[\cref{eq:noisy_rates}]. The relative energy in \cref{eq:ek_x_approx} then can be written as
\begin{align}
    e_{k}(t) &=\frac{2A_k+Bt}{A_k+(B+C)t}.
    \label{eq:ek_closed_form}
\end{align}
From \cref{eq:ek_closed_form} one clearly sees that cooling ($e_k\to0$) is obtained when $Ct\gg A_k,\,Bt$, which translates into $t\gg(\Delta_M-\Delta_m)/(\Delta_m\Delta_M)$ and $\kappa/g^2\ll 1/(\Delta_M-\Delta_m)$. Taking $\Delta_m,\Delta_M$ as in the choice $\Delta_m=\epsilon_m/2,\ \Delta_M=\epsilon_M+\epsilon_m/2$ used above, the first condition is nothing but the cooling limit $\epsilon_m t\gg1$, whereas the second, combined with $\kappa\ll\epsilon_m$, implies $\kappa\ll g$. Thus we can only expect cooling under this condition. In addition, if we merely impose $Bt\gg A_k$, i.e.\ $\kappa t/g^2\gg 1/(\Delta_m\Delta_M)$, we obtain a simple expression for the final energy:
\begin{equation}
    e^{\rm cool}\approx\frac{1}{1+\frac{\sqrt{2}\pi g^2}{\sqrt{3}\kappa(\Delta_M-\Delta_m)}}.
    \label{eq:e_longt_approx_app}
\end{equation}

\begin{figure}[ht!]
    \begin{overpic}[width=\linewidth]{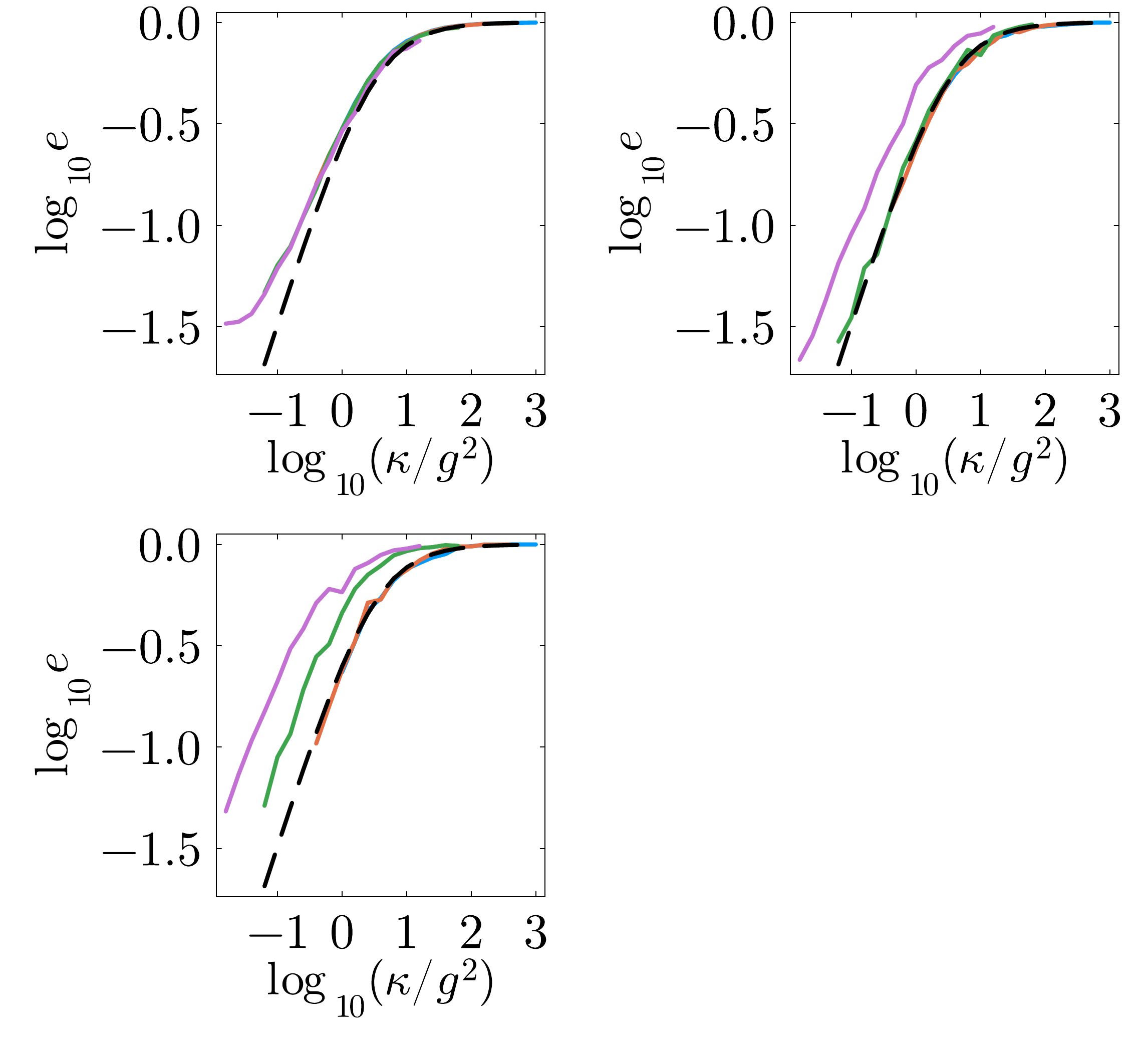}
        \put(3, 87){(a)}
        \put(53, 87){(b)}
        \put(3, 42){(c)}
    \end{overpic}
    \caption{Relative energy~$e$ versus the ratio $\kappa/g^2$ for cooling with three cycle times: (a)~$t=10$, (b)~$t=50$, (c)~$t=100$, and four coupling strengths $g=0.003$ (blue), $0.005$ (orange), $0.013$ (green), $0.025$ (purple).
        The dashed black line shows the analytic prediction from \cref{eq:e_longt_approx_app}.
        Agreement is good when $\kappa/g^2 \gtrsim 1$ and $gt \lesssim 0.5$; deviations at smaller $\kappa/g^2$ or larger $gt$ indicate breakdown of the long-time or weak-coupling approximations.
        Parameters: $N=20$, $\theta=\pi/8$, $\Delta_{M(m)}=\epsilon_{M(m)}\pm 0.1$.}
    \label{fig:kappag2_efit}
\end{figure}

\subsection{Optimal cycle time and energy}
\label{subapp:optimal_cycle_time}

Having identified the regime in which cooling occurs, we now optimize the average cycle time $t$ at fixed noise rate $\kappa$ to obtain the lowest achievable relative energy.
The parameter regime in which \cref{eq:e_longt_approx_app} approximates the relative energy well is illustrated in \cref{fig:kappag2_efit}: we can see that the weak-coupling condition prevents us from choosing $t$ arbitrarily large.
Thus, we will now consider a constrained optimization of $e_k$ in which the product $gt$ is kept fixed and denote, as before,
\begin{align}
    gt = \eta.
    \label{eq:gt_constraint}
\end{align}
Under this constraint $g=\eta/t$, so the noise term in \cref{eq:ek_closed_form} grows as $Bt=\kappa\eta^{-2}t^3$. Substituting into \cref{eq:ek_x_approx} gives
\begin{align}
    e_k(t) &=\frac{2A_k+\kappa\eta^{-2} t^3}{A_k+Ct+\kappa\eta^{-2} t^3}.
    \label{eq:ek_constrained}
\end{align}
Unlike the unconstrained case, the noise now grows as $t^3$ while the cooling grows linearly in $t$, so $e_k(t)$ has a minimum at finite $t$. Setting $\partial e_k/\partial t=0$ yields
\begin{align}
    \frac{\partial e_k}{\partial t} &=0 \iff 3A_kt^2-2Ct^3=\frac{-2A_kC\eta^{2}}{\kappa}.
    \label{eq:stationarity_condition}
\end{align}
This can be solved exactly, but the scaling is clearer under a mild approximation: since $A_k,C=\mathcal{O}(1)$ and $Ct\gg A_k$ in the cooling limit, the cubic term dominates the left-hand side, giving $t^3\approx A_k\eta^2/\kappa$. The optimal cycle time and energy of mode~$k$ are then
\begin{align}
    t_{{\rm opt}, k}^{\rm cool} &\approx\left(\frac{A_k\eta^2}{\kappa}\right)^{1/3},\\
    e_{{\rm opt}, k}^{\rm cool} &\approx \frac{3}{C}(A_k^2\eta^{-2}\kappa)^{1/3}.
    \label{eq:cooling_e_analytic}
\end{align}
These expressions hold provided $\kappa t_{{\rm opt},k}^{\rm cool}\ll1$, consistent with the low-noise assumption.
This scaling is expected to hold for the whole system, even though the actual optimal times differ slightly from mode to mode (since $A_k$ depends on $k$).
We can also observe that, since the cooling rate increases with the value of $gt$, it is advisable to choose the highest possible value of $\eta$ that still satisfies the weak coupling condition in order to achieve the best efficiency.
For all numerical results and figures, we use $\eta=0.5$, as it numerically proves to be close to the limit where the weak-coupling assumption does not hold anymore, but still holds.
We further approximate $A_k \approx A_{N/4}$, which acts as a representative value across all $A_k$, to obtain an approximation for the total relative energy $e_{\rm opt}^{\rm cool}$. This approximation is numerically verified to be close to the exact value of $e_{\rm opt}^{\rm cool}$ for the whole spectrum and range of $\kappa$ studied (see \cref{fig:cooling_vs_adiabatic_scaling}).

As a final note, in the analysis above we have treated $R\gg1$ as a continuous variable to make the functions well-behaved, but in actual simulations $R\in\mathbb{N}$, which produces discrete jumps whenever a new frequency is added. In special cases like vanishing spectrum or high noise rates, using only one or a few frequencies might be advantageous over the $R\gg1$ regime. We discuss this in the next subsection.

\subsection{The few frequency case}
\label{subapp:few-frequency_corrections_cooling}
In the previous subsections, we have analyzed the many-frequency regime, $R\gg1$, which allowed us to cool all modes in the system evenly, while at the same time keeping the effect of noise low.
We now consider two alternatives: the single-frequency ($R=1$) and few-frequency ($R\gtrsim1$) regimes.
These regimes are advantageous when the spectrum is narrow enough that one or only a few resonant frequencies cool all modes, or when the noise rate is large enough that the optimized cycle time is short.
Adding unnecessary frequencies should be avoided, because it increases the runtime of the algorithm and makes the impact of noise larger. We will analyze in this Section where the $R=1$ regime can prove advantageous, and what corrections can be made to our algorithm in order to bridge the gap between the $R\gg1$ and the $R=1$ regimes. Finally, we find a formula for the scaling with noise rate $\kappa$ that can be applied in the intermediate regime.

We will begin by analyzing an important region of our model: the edge of the trivial phase, $\theta = \pi/2$. In this case, only a single frequency is required, since all $k$-mode energies $\epsilon_k=1$ coincide. Therefore, we have $R=1$ independent of the cycle time $t$ and of the noise rate $\kappa$. This causes the dependence of the steady state energy on $\kappa$ to differ from that observed in the many-frequency regime. Following similar steps as in the previous subsections, we obtain
\begin{gather}
    e                      =\frac{1+2\kappa\eta^{-2}t^3}{\frac{1}{2}+\frac{4}{3}t^2+2\kappa\eta^{-2}t^3},\\
    t_{\rm opt}^{\rm cool} \approx \left(\frac{\eta^2}{\kappa}\right)^{1/3},\\
    e_{\rm opt}^{\rm cool} \approx D\kappa^{2/3},\qquad D=\frac{9}{4\eta^{4/3}}.
    \label{eq:D_definition}
\end{gather}

This protocol yields a steady state $\rho_k=\rho_{00}\ket{0}\bra{0}+(1-\rho_{00})\ket{1}\bra{1}$\footnote{This state is then used as the starting state for adiabatic evolution, as detailed in Appendix~\ref{subapp:realistic_adiabatic_evolution}.} that is the same for every mode, with
\begin{equation}
    \rho_{00} \approx 1-\frac{D}{2}\kappa^{2/3}.
\end{equation}

The scaling in the single-frequency scenario is $e\sim\kappa^{2/3}$, clearly different from the many-frequency case detailed previously. Now, we will show a way to bridge the area between the two regimes. To that end, we introduce the following corrections to the protocol:
\begin{itemize}
    \item $R$ is taken to be a natural number: $R=\lfloor\sqrt{2/3}(\Delta_M-\Delta_m) t\rfloor+1$, so that when $\Delta_M-\Delta_m\to 0$ the single-frequency result is recovered, and in the limit $R\gg 1$ one extra frequency has a negligible effect.
    \item We take $\Delta_{M(m)}=\epsilon_{M(m)}+(-) f(\theta)$, where $f(\theta)$ is a positive function that is roughly constant for most values of $\theta$ (ensuring that the full spectrum is still covered as explained in previous subsections) but goes to zero in the limits $\theta\to 0$ and $\theta\to\pi/2$. We can choose, for example, the following approximation to a square pulse:
          \begin{equation}
              \label{eq:f_theta}
              f(\theta)=\frac{\tanh(20\theta-1)-\tanh(20\theta-10\pi+1)}{20}.
          \end{equation}
          In general, for this choice of $f(\theta)$, $\Delta_{M(m)}\approx\epsilon_{M(m)}\pm 0.1$ for most values of $\theta$. But for $\theta\in[0,0.1]$ (and analogously for $\theta\in[\pi/2-0.1,\pi/2]$) the function behaves as a linear decay: $f(\theta)\approx\theta$.
\end{itemize}

In the regime $\theta \in [0,0.1]$, $R$ contains both a term linear in~$t$ and a constant contribution (the additional $+1$), neither of which can be neglected. Furthermore, the dependence on $t$ is not smooth, as $R$ exhibits discrete jumps due to its integer nature; consequently, there exists a range of cycle times that yield the same value of $R$. In addition, the coefficients $\gamma_k^{\rm c,h}$ depend on the number of frequencies and therefore also display discontinuous behavior. Taking these features into account, the determination of the optimal cycle time and steady-state energy in this regime is most reliably carried out numerically.

\begin{figure}[ht!]
    \begin{minipage}{0.23\textwidth}
        \centering
        \begin{overpic}[width = 1\textwidth]{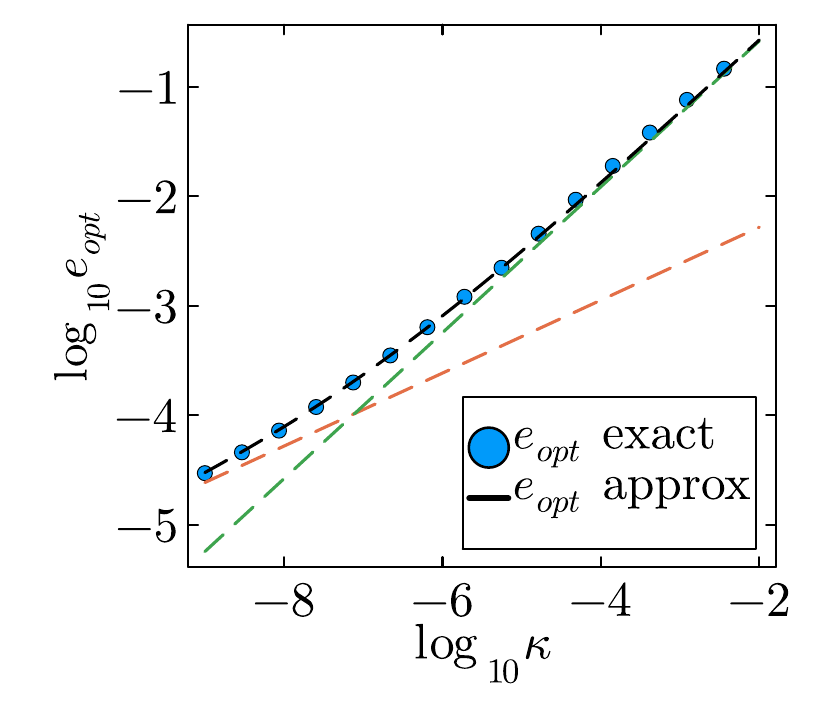}
            \put(-2,70){{(a)}}
        \end{overpic}
        \label{fig:cool_kappa_pi1000}
    \end{minipage}
    \begin{minipage}{0.23\textwidth}
        \centering
        \begin{overpic}[width = 1\textwidth]{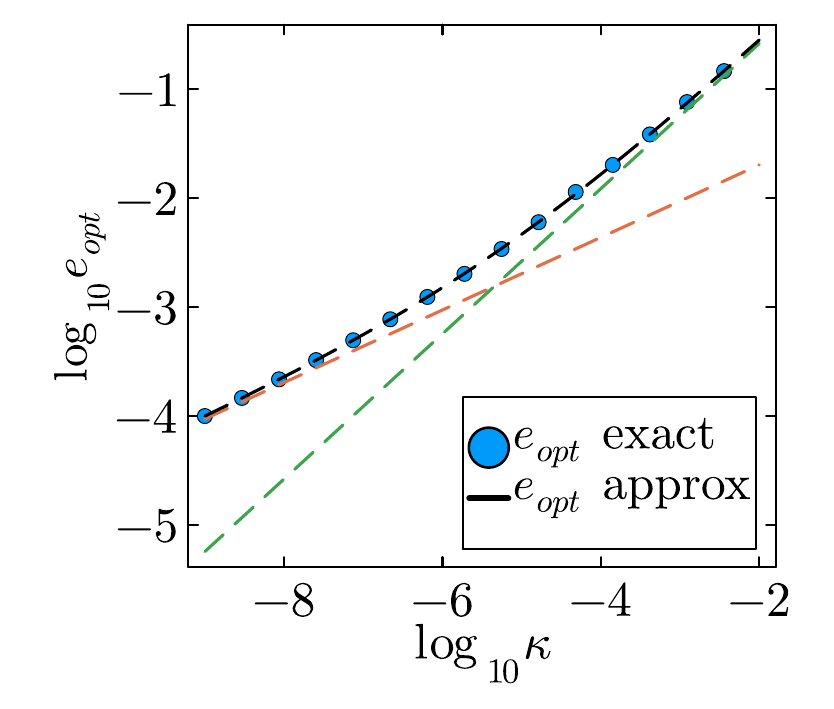}
            \put(-2,70){{(b)}}
        \end{overpic}
        \label{fig:cool_kappa_pi100}
    \end{minipage}
    \begin{minipage}{0.23\textwidth}
        \centering
        \begin{overpic}[width = 1\textwidth]{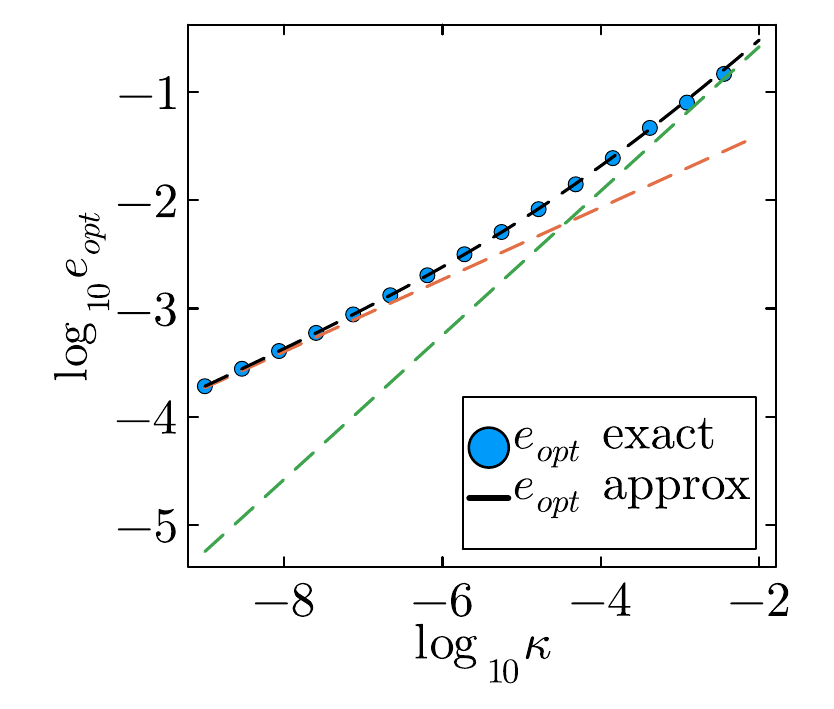}
            \put(-2,70){{(c)}}
        \end{overpic}
        \label{fig:cool_kappa_pi50}
    \end{minipage}
    \begin{minipage}{0.23\textwidth}
        \centering
        \begin{overpic}[width = 1\textwidth]{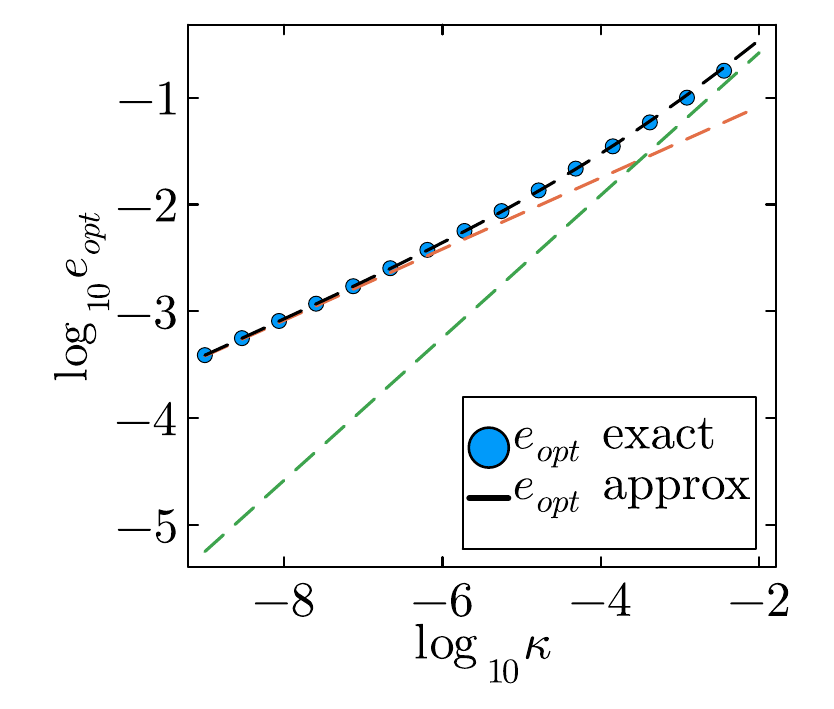}
            \put(-2,70){{(d)}}
        \end{overpic}
        \label{fig:cool_kappa_pi20}
    \end{minipage}
    \caption{Optimal relative energy $e_{\rm opt}^{\rm cool}$ versus noise strength~$\kappa$ (both log scale) for $N=100$, $gt=0.5$, and $\Delta_{M(m)}=\epsilon_{M(m)}\pm f(\theta)$ in our cooling protocol.
        Blue circles: exact numerical optimization over~$t$.
        Dashed black lines: combined approximation from \cref{eq:combined_cooling_approx}.
        Dashed green lines: multifrequency scaling $\propto\kappa^{1/3}$.
        Dashed orange lines: single-frequency scaling $\propto\kappa^{2/3}$.
        (a)~$\theta=\pi/1000$: the single-frequency term dominates.
        (b)~$\theta=\pi/100$: both terms contribute.
        (c)~$\theta=\pi/50$: multifrequency term begins to dominate.
        (d)~$\theta=\pi/20$: the $\kappa^{1/3}$ many-frequency scaling holds over most of the range.}
    \label{fig:cooling_kappa_transition}
\end{figure}

Moreover, since the optimal value of~$t$ decreases as $\kappa$ increases, a crossover from $\kappa^{1/3}$ to $\kappa^{2/3}$ scaling can occur for sufficiently large $\kappa$ across the spectrum, as illustrated in \cref{fig:cooling_kappa_transition}. Finding the true analytical expression for the few-frequency regime between the $\mathcal{O}(\kappa^{1/3})$ and $\mathcal{O}(\kappa^{2/3})$ regimes would require summing over all different $k$ modes, each with their own $k$-dependent cooling and heating rates (see \cref{fig:multifreq_sums}b). However, the total relative energy can be well approximated by simply combining the contributions from both analytically tractable limits:
\begin{equation}
    e_{\rm opt}^{\rm cool}\approx D\kappa^{2/3}+\frac{3}{C}(A^2\eta^{-2}\kappa)^{1/3}.
    \label{eq:combined_cooling_approx}
\end{equation}
This result is supported by the exact numerics shown in \cref{fig:cooling_kappa_transition}.

Note that the number of additional frequencies could, in principle, also be optimized, suggesting the existence of an optimal functional form of $f(\theta)$; however, identifying this form lies beyond the scope of the present work.

\section{Analytic derivation of the adiabatic evolution}
\label{app:adiabatic_derivations}

In this Appendix we derive how the relative energy of the adiabatic algorithm scales with the noise strength~$\kappa$.
We use here the well-known single-mode adiabatic perturbation theory and Landau--Zener formulas~\cite{Albash2018Adiabatic,Landau1932Zur,Zener1932Nonadiabatic} and apply them to our specific model. We then sum over all momentum modes to obtain the total relative energy, and optimize over the evolution time~$T$ in the presence of depolarizing noise.
We find formulas for the relative energy in the trivial phase (Appendix~\ref{subapp:trivial_phase_adiabatic}) and the topological phase (Appendix~\ref{subapp:topological_phase_LZ}), then optimize the relative energy in both phases taking into account the effect of noise (Appendix~\ref{subapp:adding_noise_adiabatic}), and finish by modifying our formulas in the case of a realistic initial state (Appendix~\ref{subapp:realistic_adiabatic_evolution}).

\subsection{Trivial phase: first-order adiabatic perturbation theory}
\label{subapp:trivial_phase_adiabatic}
Let us first review the adiabatic approximation and then apply it to our protocol. We start with the Schr\"odinger equation for a time-dependent two-level system corresponding to a single momentum mode~$k$. For clarity we drop the $k$ subscript and write
\begin{equation}
    i\frac{d}{dt}\ket{\psi(t)}=H(t)\ket{\psi(t)},
\end{equation}
with
\begin{equation}
    H(t)=\begin{pmatrix}
        f(t) &r(t)\\
        r(t) &-f(t)
    \end{pmatrix},
\end{equation}
and $t\in[0,T]$. We use the rescaled variable $s=t/T\in[0,1]$. The eigenvalues and eigenvectors (in terms of $s$) are:
\begin{align}
    E_0(s)     &=-\sqrt{f^2+r^2},\\
    E_1(s)     &=\sqrt{f^2+r^2}=-E_0,\\
    \ket{0(s)} &=\begin{pmatrix}
                     f+E_0\\
                     r
                 \end{pmatrix}\frac{1}{\sqrt{r^2+(f+E_0)^2}},\\
    \ket{1(s)} &=\begin{pmatrix}
                     -r\\
                     f+E_0
                 \end{pmatrix}\frac{1}{\sqrt{r^2+(f+E_0)^2}}.
\end{align}
We expand $\ket{\psi(t)}=c_0(t)\ket{0(t)}+c_1(t)\ket{1(t)}$ in the instantaneous eigenbasis. Substituting into the Schr\"odinger equation and projecting onto the instantaneous eigenbasis gives:
\begin{align}
    \dot{c}_0(t)+c_0(t)\braket{0(t)|\dot{0}(t)}+c_1(t)\braket{0(t)|\dot{1}(t)} &=-ic_0(t)E_0(t),\\
    \dot{c}_1(t)+c_1(t)\braket{1(t)|\dot{1}(t)}+c_0(t)\braket{1(t)|\dot{0}(t)} &=-ic_1(t)E_1(t).
\end{align}

The adiabatic approximation consists of neglecting the off-diagonal coupling $c_1\braket{0|\dot{1}}$ in the first equation, which is valid when the spectral gap is large. Solving for $c_0(0)=1$ we obtain
\begin{align}
    c_0(t)= e^{-\int_0^t\left(\braket{0(t')|\dot{0}(t')}+iE_0(t')\right)\dd t'}.
\end{align}
Substituting back into the second equation and solving for $c_1(0)=0$ yields
\begin{align}
    c_1(t)=\int_0^t &\left[e^{\int_0^{t'}\braket{1(t'')|\dot{1}(t'')}-\braket{0(t'')|\dot{0}(t'')}\dd t''}\right.\nonumber\\
                         &\left.\times e^{i\int_0^{t'}E_1(t'')-E_0(t'')\dd t''}\braket{1(t')|\dot{0}(t')}\right]\dd t'.
\end{align}
Using $s=t/T$ and denoting $v=1/T$ we have
\begin{equation}
    c_1(s)=\int_0^s e^{-i\gamma_{10}(s')}e^{\frac{i}{v}\omega_{10}(s')}\braket{1(s')|\dot{0}(s')}\dd s',
    \label{eq:c1_exact}
\end{equation}
with Berry and dynamical phases:
\begin{align}
    \gamma_{nm}(s) &=i\int_0^s \dd s'\Bigl(\braket{n(s')|\frac{d}{\dd s'}|n(s')}\nonumber\\
                   &\qquad -\braket{m(s')|\frac{d}{\dd s'}|m(s')}\Bigr),\\
    \omega_{nm}(s) &=\int_0^s \dd s'(E_n(s')-E_m(s')),
\end{align}
where $n,m$ take the values $0,1$. To extract the leading behavior for large~$T$ (small $v$) in \cref{eq:c1_exact}, we use the following identity for integrals of the form
\begin{equation}
    I=\int_0^s\dd s'B(s')e^{\frac{1}{v}\int_0^{s'}\dd s''C(s'')},
    \label{eq:I_integral}
\end{equation}
where we can expand them in growing terms of $v$ by iteratively applying the identity
\begin{align}
    B(s')e^{\frac{1}{v}\int_0^{s'}\dd s''C(s'')}=
     &v\frac{d}{\dd s'}\left(\frac{B(s')}{C(s')}e^{\frac{1}{v}\int_0^{s'}\dd s''C(s'')}\right)\nonumber\\
     &-ve^{\frac{1}{v}\int_0^{s'}\dd s''C(s'')}\frac{d}{\dd s'}\left(\frac{B(s')}{C(s')}\right).
\end{align}
The first term gives us the first order in $v$; the second gives us an integral of the same type as $I$ but with a factor of $v$ already in front and $B$ replaced by $\frac{d}{\dd s'}\left(\frac{B(s')}{C(s')}\right)$, therefore contributing only to higher orders of $v$. We keep only the first order in $v$.

Let us now apply this to our choice of $H(s)$ (see \cref{eq:adiabatic_schedule}). The Berry phase vanishes: $\gamma_{10}(s)=0$, so \cref{eq:c1_exact} has the shape of \cref{eq:I_integral}. The resulting transition probability to the excited state after the complete evolution is, up to first order in $v$,
\begin{equation}
    |c_1(1)|^2\approx\frac{1}{T^2}\left|e^{iT\omega_{10}(1)}\frac{\braket{1(1)|\dot{0}(1)}}{2E_1(1)}-\frac{\braket{1(0)|\dot{0}(0)}}{2E_1(0)}\right|^2.
    \label{eq:c1k_squared}
\end{equation}

We now map this to our specific Hamiltonian. Starting from $\theta_i=\pi/2$ and targeting $\theta_f$ via the linear interpolation $H(s)=H_0(1-s)+H_f s$, the matrix elements are
\begin{align}
    f(s) &=1+s(\sin\theta_f+\cos\theta_f\cos\phi_k-1),\\
    r(s) &=s\cos\theta_f\sin\phi_k,
\end{align}
where $\phi_k=2\pi k/N$.

The dynamical phase integral evaluates to
\begin{align}
     &\omega_{10}(1)=\int_0^1\sqrt{ax^2+bx+1} \dd x\nonumber\\
     &=\frac{1}{8a^{3/2}}\left[(4a-b^2)\log(2\sqrt{a}\sqrt{ax^2+bx+1}+2ax+b)\right.\nonumber\\
                             &\left.+2\sqrt{a}\sqrt{ax^2+bx+1}(2ax+b)\right]_0^1
\end{align}
with
\begin{align}
    a &=(\sin\theta_f+\cos\theta_f\cos\phi_k-1)^2+(\cos\theta_f\sin\phi_k)^2,\\
    b &=2(\sin\theta_f+\cos\theta_f\cos\phi_k-1).
\end{align}
The matrix elements in \cref{eq:c1k_squared} for $s>0$ are given by:
\begin{equation}
    \braket{1(s)|\dot{0}(s)}=\frac{r}{s}\left(1+\frac{f}{\sqrt{f^2+r^2}}\right)\frac{1}{r^2+(f+\sqrt{f^2+r^2})^2},
\end{equation}
whereas for $s=0$ we have
\begin{equation}
    \braket{1(0)|\dot{0}(0)}=\frac{1}{2}\cos\theta_f\sin\phi_k.
\end{equation}

Since we are interested in scalings rather than exact prefactors, we do not evaluate these expressions further. Summing over all modes, the total energy after the adiabatic evolution in the absence of noise is
\begin{equation}
    E=\sum_k -\epsilon_k\left(1-|c_1(k)|^2\right)=E_{\rm GS}\left(1-\frac{\tilde{A}}{T^2}\right).
    \label{eq:noiseless_adiab_product_energy}
\end{equation}
The prefactor $\tilde{A}$ can be calculated numerically. We have assumed that the time dependence inside the absolute value in \cref{eq:c1k_squared} can be neglected, since it is only a phase and does not significantly alter the value of $|c_1(k)|^2$; consequently, $\tilde{A}$ is approximately time-independent. One can further approximate $\tilde{A}\approx T^2|c_1(N/4)|^2\approx \left(\cos^{2}\theta_f\right)/8$, which can be verified numerically.

\subsection{Topological phase: Landau--Zener transition}
\label{subapp:topological_phase_LZ}

When the adiabatic path crosses the phase transition, the perturbative treatment fails for modes in the vicinity of the gap-closing point. In this regime, the energy can instead be determined using the Landau--Zener formalism~\cite{Landau1932Zur,Zener1932Nonadiabatic,Albash2018Adiabatic}. We apply the standard Landau--Zener formulas to our adiabatic evolution.

This approximation is most accurate when the energy gaps in the initial and final configurations are large. To extract an analytic approximation, we follow a path starting from $\theta=\pi/2$ and going to $\theta=0$. We redefine our linear path such that $H_k(s)=H^{\pi/2}_k(0.5-s)+H^{0}_k(0.5+s)$, where $s=t/T$ and we run $t$ from $-T/2$ to $T/2$. Close to the transition point $s\approx0$ and close to the mode where the gap closes, $\delta_k=\phi_k-\pi\approx0$ (with $k\in[0,N]$), the eigenenergies of the instantaneous Hamiltonian can be approximated by
\begin{equation}
    \epsilon_k(\delta_k,t)\approx\sqrt{\frac{4t^2}{T^2}+\frac{\delta_k^2}{4}}.
\end{equation}
This dispersion relation can be identified with that of the standard Landau--Zener Hamiltonian
\begin{align}
    H_{\rm LZ}(t)         &=v_{\rm LZ}t\sigma_z+\Delta\sigma_x,\\
    \epsilon_{\rm LZ}^\pm &=\pm\sqrt{(v_{\rm LZ}t)^2+\Delta^2},
\end{align}
which, by comparison, gives
\begin{align}
    v_{\rm LZ}      &=\frac{2}{T},\\
    \Delta_{\rm LZ} &=\frac{|\delta_k|}{2}.
\end{align}
From the standard Landau--Zener result, the transition probability is (with $\hbar=1$):
\begin{equation}
    P_{\rm LZ}^k=\exp\left(-\frac{\pi\Delta_{\rm LZ}^2}{v_{\rm LZ}}\right)=\exp\left(-\frac{\pi}{8}T\delta_k^2\right).
\end{equation}

This gives a relative energy per mode $e_k=2P_{\rm LZ}^k$.
As expected, the smaller the gap of the mode, the longer the evolution time required to achieve a low excitation probability. A true closing of the gap occurs at $k=N/2$ (i.e., $\delta_k=0$), but in the limit $N\gg1$ this single-mode contribution becomes negligible.

This approximation does not depend on the target angle $\theta_f$ and was derived for the specific path to $\theta_f=0$ as an illustration.

A result for general~$\theta_f$ can be extracted in a similar fashion:
\begin{align}
    P_{\rm LZ}^k &=\exp\left(-\pi GT\delta_k^2\right),\\
    G            &=\frac{\cos^2\theta_f}{(1+\cos\theta_f-\sin\theta_f)^3}.
\end{align}
We recover $G(\theta_f=0)=1/8$, but for other values of $\theta_f$ in the topological phase, $G(\theta_f)$ changes significantly, e.g., $G(\pi/8)\approx 0.23$.

The derivation of the total energy is more involved and, in general, does not yield a closed-form expression. Following the Kibble--Zurek argument~\cite{Kibble1976Topology,Zurek1985Cosmological,Zurek2005Dynamics,Dziarmaga2005Dynamics,Polkovnikov2005Universal}, for a large enough system size the energy can be approximated by
\begin{align}
    E           &\approx\int_{0}^{N} E_k\, \dd k=E_{\rm GS}+2\int_{0}^{N}e^{-\pi GT\delta_k^2}\epsilon_k\, \dd k\nonumber\\
                &\approx E_{\rm GS}+\frac{N}{\pi}\int_{-\infty}^\infty e^{-\pi GT(x-\pi)^2}\epsilon(x) \dd x,\\
    \epsilon(x) &=\sqrt{1+\sin(2\theta_f)\cos x},\nonumber
\end{align}
where the integral has been extended to infinity since the contribution of modes with large gaps is negligible. The value of the integral depends on the target angle~$\theta_f$. For $\theta_f=0$ or $\theta_f=\pi/2$, one has $\epsilon_k=1$ for all $k$ and a closed form exists; in the general case, further approximation is needed. For large $T$, the Gaussian factor is sharply peaked in $x$ and the energy factor is approximately constant over the range where the Gaussian is concentrated. Evaluating the energy at the peak of the Gaussian and taking it out of the integral gives
\begin{align}
    E &\approx E_{\rm GS}+\frac{N}{\pi}\epsilon(\pi)\int_{-\infty}^\infty e^{-\pi GT(x-\pi)^2}\dd x\nonumber\\
      &=E_{\rm GS}+\frac{N}{\pi\sqrt{GT}}\sqrt{1-\sin(2\theta_f)}.
    \label{eq:noiseless-LZ-energy}
\end{align}

To obtain a closed equation for the relative energy $e$, an approximation for $E_{\rm GS}$ is required, which we derive below.

\paragraph{Approximating $E_{\rm GS}$.}
 In the large system size limit, the ground state energy is given by the following integral:
\begin{equation}
    E_{\rm GS}(\theta)=-\frac{N}{\pi}\int_0^\pi \sqrt{1+\sin(2\theta)\cos x}\, \dd x.
\end{equation}
To approximate this integral, we construct a function that has the same expansion as the $|\sin(2\theta)|\ll 1$ approximation of the integral, is symmetric with respect to $\theta=\pi/4$, and lies close to the exact value for $\theta=\pi/4$:
\begin{equation}
    E_{\rm GS}(\theta)\approx-N\sqrt{1-\frac{\sin^2(2\theta)}{8}-\frac{\sin^4(2\theta)}{17}}.
    \label{eq:Egs-approximation}
\end{equation}
This approximation can be numerically verified to be within $1\%$ of the true value of the integral, and allows us to approximate the ground-state energy for any given $\theta$ in the large $N$ limit. It can also be verified to lie close to the Pad\'e--Chebyshev approximant for the same function.
\subsection{Adding noise}
\label{subapp:adding_noise_adiabatic}

\begin{figure}[t]
    \centering
    \begin{minipage}{0.23\textwidth}
        \centering
        \begin{overpic}[width = 1\textwidth]{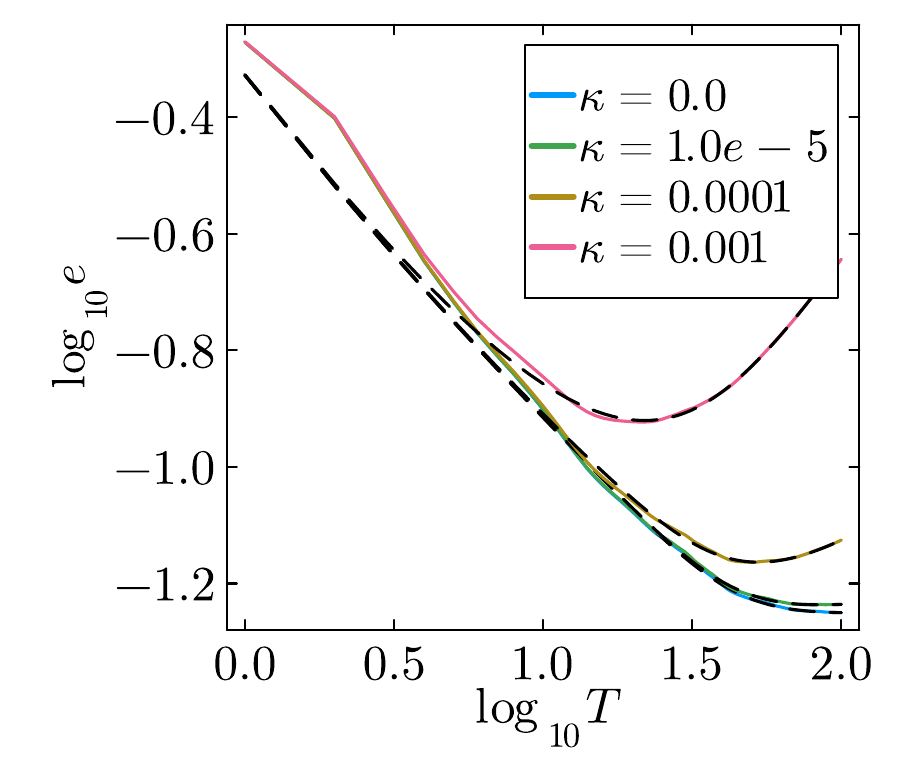}
            \put(-2,70){{(a)}}
        \end{overpic}
    \end{minipage}\hspace{0.01\textwidth}%
    \begin{minipage}{0.23\textwidth}
        \centering
        \begin{overpic}[width = 1\textwidth]{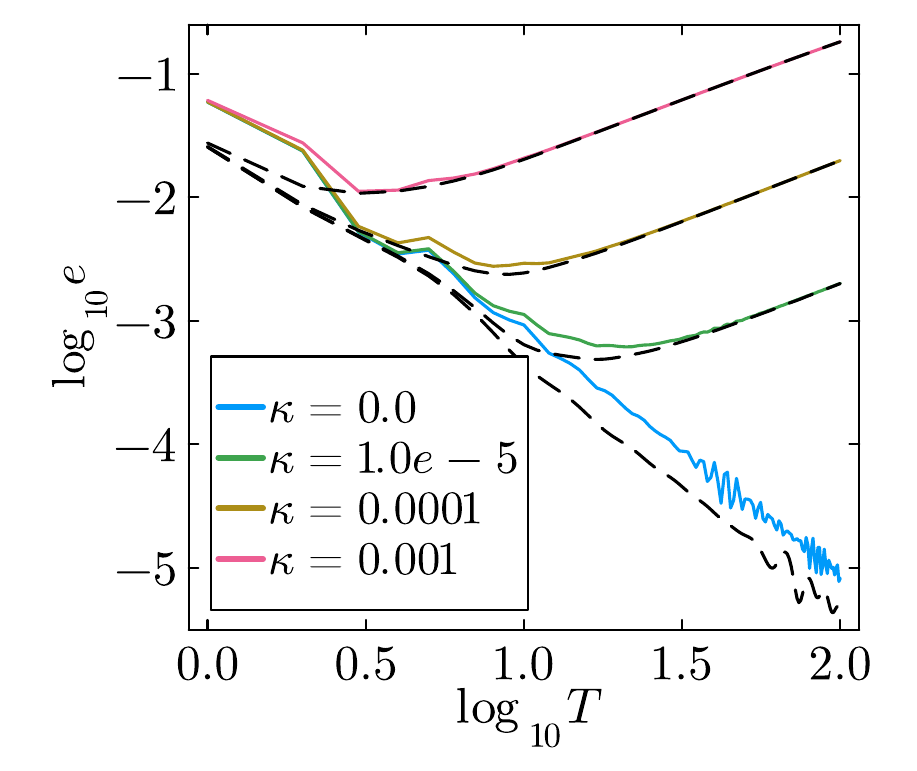}
            \put(-2,70){{(b)}}
        \end{overpic}
    \end{minipage}
    \caption{Relative energy~$e$ versus total evolution time~$T$ (log-log scale) for adiabatic evolution targeting (a) $\theta_f=\pi/8$  and (b) $\theta_f=\pi/3$  at several noise rates~$\kappa$, for system size $N=20$.
        Solid lines: exact numerical simulation. Dashed lines: analytic predictions.
        In the presence of noise, each curve exhibits a minimum at an optimal~$T$, reflecting the trade-off between non-adiabatic excitations and depolarizing noise. For larger~$T$, the energy increases as noise effects dominate. The analytical predictions accurately capture both the scaling behavior and the position of these minima.}
    \label{fig:adiab_analytic_vs_numeric}
\end{figure}

For depolarizing noise of strength~$\kappa$, the noise channel commutes with the unitary evolution (see Ref.~\cite{Molpeceres2025Quantum} and \cref{subsec:noise}). For our model, the maximally mixed state has $E=0$. Then, the energy in the presence of noise is related to the energy in the noiseless setting by an overall damping factor as
\begin{equation}
    E_{\text{noisy}}=e^{-2\kappa T}E_{\text{ideal}}.
\end{equation}

\Cref{fig:adiab_analytic_vs_numeric} illustrates the effect of noise on the adiabatic evolution targeting $\theta_f=\pi/3$ (within the trivial phase, no gap closing). It shows that the analytical predictions accurately capture both the scaling behavior and the position of these minima. For short~$T$, the energy decreases as~$T^{-2}$ and is dominated by non-adiabatic excitations present even in the absence of noise. For large~$T$, depolarization effects become dominant. The optimal~$T$ arises from balancing these two contributions.

We can now determine the scalings for the adiabatic algorithm in the presence of noise, for both cases (crossing or not crossing the phase transition).

\paragraph{Gapped regime (no gap closing, $\theta_f>\pi/4$).}
When the path does not cross the phase transition and for sufficiently small noise rate $\kappa$, the total energy can be easily determined by multiplying \cref{eq:noiseless_adiab_product_energy} with the exponential damping factor resulting from the depolarizing noise. We obtain

\begin{equation}
    E=e^{-2\kappa T}\left(1-\frac{\tilde{A}}{T^2}\right)E_{\rm GS}\approx E_{\rm GS}\left(1-2\kappa T-\frac{\tilde{A}}{T^2}\right),
\end{equation}
where the approximation results from an expansion around small $\kappa T$. \cref{fig:adiab_analytic_vs_numeric} illustrates the agreement between analytics and numerics in this scenario. Differentiating with respect to $T$ and setting the derivative to zero yields the optimal time and energy:
\begin{align}
    T_{\rm opt} &\approx\left(\frac{\tilde{A}}{\kappa}\right)^{1/3},\\
    E_{\rm opt} &\approx E_{\rm GS}(1-3\tilde{A}^{1/3}\kappa^{2/3}),\\
    e_{\rm opt} &\approx 3\tilde{A}^{1/3}\kappa^{2/3},
\end{align}
where $e_{\rm opt}$ is calculated using \cref{eq:e_definition}.

This result is consistent with the expectation that, in the absence of noise, the adiabatic algorithm in the limit $T\to\infty$ produces the exact ground state.

\paragraph{Gapless regime (gap closing, $\theta_f<\pi/4$).}
The analysis when the path crosses the phase transition requires more care, since each mode contributes differently to the total energy, yet all modes evolve for the same time $T$. We assume large $N$ and large $T$, multiply \cref{eq:noiseless-LZ-energy} with the exponential damping factor resulting from the depolarizing noise, and use the same approximation as before to obtain:

\begin{align}
    E &\approx e^{-2\kappa T}\left(E_{\rm GS}+\frac{N}{\pi\sqrt{GT}}\sqrt{1-\sin(2\theta_f)}\right)\nonumber\\
      &\approx E_{\rm GS}\left(1-2\kappa T-\frac{\tilde{B}}{\sqrt{T}}+2\kappa \tilde{B}\sqrt{T}\right),
    \label{eq:E_LZ_1}
\end{align}
with
\begin{align}
    \tilde{B} &= \frac{N\sqrt{1-\sin(2\theta_f)}}{\pi\sqrt{G} |E_{\rm GS}|}\nonumber\\
              &\approx \frac{\sqrt{1-\sin(2\theta_f)}}{\pi\sqrt{G}\sqrt{1-\frac{\sin^2(2\theta_f)}{8}-\frac{\sin^4(2\theta_f)}{17}} }.
\end{align}

In order to ease the determination of the optimal $T$ we approximate this expression further. Using that $\tilde{B}=\mathcal{O}(1)$ and $\kappa T\ll 1$, we have that the term $2\kappa \tilde{B}\sqrt{T}$ is negligible compared to $\tilde{B}/\sqrt{T}$ and we can simplify \cref{eq:E_LZ_1}:
\begin{align}
    E &\approx E_{\rm GS}\left(1-2\kappa T-\frac{\tilde{B}}{\sqrt{T}}\right).
\end{align}
Differentiating with respect to $T$ and optimizing yields:
\begin{align}
    T_{\rm opt} &\approx\left(\frac{\tilde{B}}{4\kappa}\right)^{2/3},\\
    E_{\rm opt} &\approx E_{\rm GS}\left(1-\frac{3}{2^{1/3}}\tilde{B}^{2/3}\kappa^{1/3}\right),\\
    e_{\rm opt} &\approx \frac{3}{2^{1/3}}\tilde{B}^{2/3}\kappa^{1/3}.
    \label{eq:adiabatic_LZ_e_analytic}
\end{align}
These scalings are consistent with the numerical results presented in the main text (\cref{fig:cooling_vs_adiabatic_scaling}). The prefactor $\tilde{B}$ is monotonically decreasing from $\theta=0$, where $\tilde{B}\approx 0.9$, to $\theta=\pi/4$, where $\tilde{B}=0$. This is consistent with the intuition that the deeper into the topological phase the target lies, the more difficult it becomes to prepare the ground state via the adiabatic algorithm, as the effects of finite evolution time cause some modes to have non-zero transition probability to their excited states.

\paragraph{Effects of finite system size.}

In \cref{fig:cooling_vs_adiabatic_scaling}, for adiabatic trajectories that cross the critical point, the relative energy exhibits a crossover from $\kappa^{1/3}$ scaling to a constant value independent of $\kappa$ for sufficiently small $\kappa$. This constant contribution arises from the mode $k=N/2$, whose gap closes at the phase transition. The excitation of this mode is unavoidable when crossing the critical point, irrespective of~$T$.

As a result, in the limit $\kappa = 0$, the optimal evolution time $T \to \infty$ yields an energy $E = E_{\rm GS} + 2\epsilon_{N/2}$. The crossover between the two regimes occurs at $\kappa^{1/3} \sim \epsilon_{N/2}/|E_{\rm GS}|$, as confirmed numerically in \cref{fig:finite_N_crossover}. For larger system sizes, this crossover shifts to smaller values of $\kappa$, and vanishes in the thermodynamic limit $N \gg 1$, where the contribution of a single mode becomes negligible.

\subsection{Realistic adiabatic evolution}
\label{subapp:realistic_adiabatic_evolution}

Instead of starting from the ideal product state $\ket{0}^{\otimes N}$, we consider a more realistic scenario in which the initial state is already affected by noise. To this end, we apply the cooling procedure at $\theta=\pi/2$ for a given noise rate $\kappa$ (see \cref{subapp:few-frequency_corrections_cooling}).
This yields a state $\rho^{\rm cool}(\pi/2)$ that is diagonal in the mode basis, with each mode in a mixture $\rho_k = \rho_{00}\ket{0}\bra{0} + (1-\rho_{00})\ket{1}\bra{1}$ and $\rho_{00} \approx 1 - (D/2)\kappa^{2/3}$.

Since all derivations in the previous subsections for an initial state $\ket{0}$ apply analogously to an initial state $\ket{1}$ (due to the symmetry of the spectrum), we can apply the derivations independently to both cases, weight the resulting energies by $\rho_{00}$ and $1-\rho_{00}$, respectively, and sum them to obtain the total energy:
\begin{align}
    E_{\rm realistic}=E_{\rm ideal}(1-D\kappa^{2/3}),
\end{align}
where the factor $D$ has been defined in \cref{eq:D_definition} (see \cref{subapp:few-frequency_corrections_cooling} for the full derivation). Crucially, this does not alter the optimal times, so the optimizations performed for the ideal initial state remain valid. The final energies change in different ways depending on the target phase. In the trivial phase, which already has a $\kappa^{2/3}$ scaling, the correction merely modifies the prefactor. In the topological phase, the imperfect initialization contributes only a subleading additive term leading to
\begin{align}
    e^{\rm ent}_{\rm realistic}  &\approx \frac{3}{2^{1/3}}\tilde{B}^{2/3}\kappa^{1/3}+D\kappa^{2/3},\\
    e^{\rm prod}_{\rm realistic} &\approx (3\tilde{A}^{1/3}+D)\kappa^{2/3}.
\end{align}

In the topological phase, the correction from the imperfect initial state ($D\kappa^{2/3}$) becomes negligible for sufficiently small~$\kappa$: this is consistent with the fact that, at low noise rates, cooling can prepare the initial state very close to $\ket{0}^{\otimes N}$. For sufficiently large $\kappa$, however, both contributions are significant and the curve deviates from the ideal $\kappa^{1/3}$ behavior.

\FloatBarrier

\bibliography{library.bib}
\end{document}